\documentclass[aps,preprint,nofootinbib,superscriptaddress]{revtex4}

\usepackage{graphicx}
\usepackage{amsmath}
\usepackage{amsfonts}
\usepackage{bm}
\usepackage{xcolor}
\usepackage{ulem}
\usepackage{appendix}

\usepackage{verbatim} 

\newcommand{\simge}{\hspace*{0.2em}\raisebox{0.5ex}{$>$}
     \hspace{-0.8em}\raisebox{-0.3em}{$\sim$}\hspace*{0.2em}}
\newcommand{\simle}{\hspace*{0.2em}\raisebox{0.5ex}{$<$}
     \hspace{-0.8em}\raisebox{-0.3em}{$\sim$}\hspace*{0.2em}}

\newcommand{\bea}{\begin{eqnarray}}
\newcommand{\eea}{\end{eqnarray}}
\newcommand{\beq}{\begin{equation}}
\newcommand{\eeq}{\end{equation}}
\newcommand{\bqa}{\begin{eqnarray}}
\newcommand{\eqa}{\end{eqnarray}}

\newcommand{\nn}{\nonumber \\}
\def\mqo2{{\!\!\!}}

\begin{document}

\title{Effective Field Theory Analysis of Three-Boson Systems at
Next-To-Next-To-Leading Order\footnote{Dedicated to Professor Henryk Witala at the occasion of his 60th birthday.}}
\author{Chen Ji}\email{jichen@triumf.ca}
\affiliation{Department of Physics and Astronomy and Institute of Nuclear and Particle Physics, Ohio University, Athens, OH, 45701, USA\\}
\affiliation{TRIUMF,  4004 Wesbrook Mall, Vancouver, B.C. V6T 2A3, Canada\footnote{Permanent address}}
\author{Daniel R. Phillips$^1$}\email{phillips@phy.ohiou.edu}

\begin{abstract}
We use an effective field theory for short-range forces (SREFT) to analyze systems of three identical bosons interacting via a two-body potential that  generates a scattering length, $a$, which is large compared to the range of the interaction, $\ell$. The amplitude for the scattering of one boson off a bound state of the other two is computed to next-to-next-to-leading order (N$^2$LO) in the $\ell/a$ expansion. At this order, two pieces of three-body data are required as input in order to renormalize the amplitude (for fixed $a$). We apply our results to a model system of three Helium-4 atoms, which are assumed to interact via the TTY potential. We generate N$^2$LO predictions for atom-dimer scattering below the dimer breakup threshold using the bound-state energy of the shallow Helium-4 trimer and the atom-dimer scattering length as our two pieces of three-body input. Based on the convergence pattern of the SREFT expansion, as well as differences in the predictions of two renormalization schemes, we 
conclude that our N$^2$LO 
phase-
shift predictions will receive higher-order corrections of $< 0.2$\%. In contrast, the prediction of SREFT for the binding energy of the ``deep" trimer of Helium-4 atoms displays poor convergence.
\end{abstract}

\smallskip
\maketitle

\section{Introduction}
\label{sec:introduction}
Few-body systems share universal features at low energies, for which details of their short-distance structure and interactions are not essential. Studies of such systems' universal behavior normally involve separation of a large-distance scale and a short-distance scale. The large-distance scale is related to the two-body scattering length, $a$, which determines the zero-energy total cross section in S-wave elastic scattering of two particles:
\begin{equation}
\sigma = 4\pi a^2~,
\end{equation}
where $\sigma =2\pi a^2$ if the two particles are identical.
The short-distance scale is represented by the range of the interparticle short-distance interactions, $\ell$. Alternatively, the separation of scales is also reflected in a hierarchy of momentum scales. The low-momentum scale, $Q\sim1/a$, is the typical momentum scale of such few-body systems (e.g. the binding momentum). A high-momentum scale, $\sim1/\ell$, sets the breakdown scale for the description of universal physics: high-momentum degrees of freedom above $1/\ell$ ($\simeq$ short-distance physics corresponding to interactions of range $\simle \ell$) are integrated out.

Universal behavior thus occurs in few-body systems where the two-body scattering length is much larger than the range of the interaction, $|a|\gg\ell$. For example, in systems with three equal-mass particles, Efimov proved the existence of an infinite tower of three-body bound states ({\it trimers}) in the unitary limit $|a|\rightarrow\infty$~\cite{Efimov70}. These trimer states have a geometric spectrum: the ratio between two consecutive binding energies is always $515$. This behavior, along with many other universal features, has been studied in systems in atomic, nuclear and particle physics which obey $|a|\gg\ell$.
In ultracold atomic gases,
the atom-atom scattering length is controlled by an external magnetic field and can be tuned to arbitrarily large values through Feshbach resonances~\cite{Chin:2008}. A large scattering length occurs without manipulation in the interactions of Helium-4 atoms. Experimental data~\cite{Grisenti:2000zz} suggests that the ratio $a/\ell$ is of order 10 in this system.
 In few-nucleon systems, the nucleon-nucleon (e.g. $np$ and $nn$) scattering length is naturally about 3 times the range, $a/\ell\sim 3$~\cite{Mathelitsch:1984hq}. These systems, together with halo nuclei~\cite{Jensen-04,Zhukov-93} and exotic charmonium~\cite{Canham:2009zq}, offer testing grounds for the study of universal physics. For reviews of the application of this notion of universality in few-body physics, see Refs.~\cite{Braaten:2004rn, Hammer:2010kp}.

One powerful tool to describe universal physics is Effective Field Theory (EFT). It provides a systematic expansion in the ratio of a low- and a high-momentum scale. It thereby permits study of not only the universal features at low energies, but also corrections to these features from dynamics at (or below) the length scale $\ell$, which can be evaluated in perturbation theory. For example, the binding of nucleon-nucleon systems is much smaller than the pion mass, $m_{\pi}$. Such systems' low-energy behavior is thus insensitive to physics with a momentum scale at or above $m_{\pi}$, e.g. pion exchanges. It can be described by an effective theory, called pionless EFT, in which nucleons are treated as point-like particles and the nucleon-nucleon potential becomes a string of contact interactions with increasing powers of momentum. These contact terms parametrize short-distance physics in nucleon-nucleon systems, and all those with two or more derivatives are investigated perturbatively when higher-order 
corrections are examined in this EFT.

Pionless EFT has been successfully applied to nucleon-nucleon systems and is able to describe the low-energy component of nuclear forces~\cite{Kaplan:1998tg, vanKolck:1998bw,vanKolck:1999mw,Beane:2000fx,Bedaque:2002mn,Rho:2002sh}. Using renormalization-group methods, Birse {\it et al.} showed that, if $|a| \gg \ell$, this EFT is equivalent to the effective-range expansion for scattering by a short-ranged potential~\cite{Birse:1998dk}. Therefore, this EFT can be understood as an expansion in the ratio of the effective range, $r_0$, to the scattering length, $a$. These are defined from the effective-range expansion of the two-body S-wave phase shift:
\begin{equation}
k\cot \delta = -\frac{1}{a} + \frac{1}{2}r_0 k^2 + \cdots.
\label{eq:ere}
\end{equation}
We refer hereafter to this EFT as short-range EFT (SREFT). If spin and isospin degrees of freedom are excluded, SREFT can be used to describe low-energy boson-boson systems with a short-range interaction. 

The SREFT expansion is valid for $k\sim1/|a| \ll 1/\ell$, and $r_0$ is usually of the same order as $\ell$. When $|a| \gg \ell$, leading-order (LO) calculations in SREFT are computed in the zero-range limit $\ell=0$, and $a$ is treated as an input quantity fixed by extrapolation from scattering data or by model calculations with an underlying interaction beyond EFT. Effects of the short-distance physics are systematically included in SREFT as higher-order corrections to the zero-range limit. At these orders $r_0$ is also an input, to be obtained in the same way as $a$. Once $a$ and $r_0$ are fixed, results from these EFT calculations are insensitive to short-distance details.

The universal features of three-body systems, which were predicted by Efimov, have recently been rederived in SREFT by Bedaque {\it et al.}~\cite{Bedaque:1998kg, Bedaque:1998km}. The leading-order calculation corresponds to the limit $\ell =0$ and large $a$. In SREFT at LO, diagrams involving loops still contribute to the same EFT order in powers of momentum as do tree diagrams, which thereby leads to a non-perturbative calculation for three-body observables in this EFT at LO. A three-body counterterm is needed at LO for consistent renormalization, i.e. to cancel notable dependence of the non-perturbative result on the momentum cutoff, $\Lambda$. This counterterm is introduced to describe short-distance physics in the three-body system, and is tuned to reproduce one three-body observable. 

Beyond LO in SREFT, three-body observables are calculated as a perturbative expansion in powers of $r_0/a$. The next-to-leading-order (NLO) range corrections, $\sim r_0/a$, have been calculated in~\cite{Hammer:2001gh, Bedaque:2002yg} for systems with a fixed scattering length, and in~\cite{Platter:2008cx, Ji:2010su, Ji:2011qg} for systems with a variable scattering length. The latter is associated with needs in the experimental study of cold atoms, where the atom-atom scattering length is manipulated near a Feshbach resonance. The existence of a scattering-length-dependent parameter in the NLO three-body counterterm was reported in~\cite{Ji:2010su, Ji:2011qg}. 

Meanwhile, few-body systems with a large, fixed scattering length are realized in nuclear and molecular physics. In the nuclear case $r_0/a\sim1/3$, and so we expect NLO predictions to have errors $\sim 10$\%, which is above the desired accuracy when comparing to experiments. (For example, even the thirty-year-old measurement of the neutron-deutron scattering length by Dilg {et al.} has an accuracy of about $6\%$~\cite{Dilg71}.) Therefore, next-to-next-to-leading-order (N$^2$LO) effective-range effects, $\sim r_0^2/a^2$, need to be considered in nuclear systems.

Such an N$^2$LO calculation was carried out by Bedaque {\it et al.}~\cite{Bedaque:2002yg}, who calculated the S-wave neutron-deuteron phase shift at N$^2$LO and showed that an additional, energy-dependent, three-body counterterm is needed in the renormalization. This conclusion is supported by analytic arguments based on a renormalization-group treatment of the three-body problem in hyperspherical co-ordinates~\cite{BB05}.  
However, the calculation of Bedaque {\it et al.} is not done according to a strict expansion in $r_0/a$. Instead, range corrections up to $\mathcal{O}(r_0^2/a^2)$ are included in the two-body scattering amplitude, which is then used as an input in a non-perturbative three-body calculation, thereby arbitrarily including higher-order corrections above N$^2$LO. Bedaque {\it et al.} argue that these higher-order corrections should be small, as long as the cutoff $\Lambda$ is kept below $1/\ell$. 

Platter and Phillips analyzed the three-boson system up to N$^2$LO by using a similar partial resummation as in Ref.~\cite{Bedaque:2002yg}, and concluded that an energy-dependent three-body counterterm is not needed for renormalization in the limit $\Lambda \gg 1/r_0$~\cite{Platter:2006ev}. This contradicts Bedaque {\it et al.}'s result~\cite{Bedaque:2002yg}. However, it must be emphasized that Ref.~\cite{Platter:2006ev} only reached this conclusion for the case that the cutoff $\Lambda \rightarrow \infty$. 
Ji {\it et al.}~\cite{Ji:2011qg} subsequently showed that the partial resummation of NLO range corrections carried out in Ref.~\cite{Platter:2006ev} is only consistent with a rigorous perturbative expansion when $\ell \ll r_0$, with $\Lambda \sim 1/\ell$. This condition should also apply to calculations at N$^2$LO. 

In order to solve the controversy raised by these works, an N$^2$LO calculation in a rigorous perturbative expansion of three-body observables is needed. In this paper, we calculate the $r_0^2/a^2$ corrections to three-body bound-state energies and scattering phase shifts perturbatively, and compare our findings with those two previous works~\cite{Bedaque:2002yg,Platter:2006ev}. Our derivation is an extension of the NLO calculation reported in~\cite{Ji:2011qg}. In the rigorous perturbative expansion, we find that an additional energy-dependent three-body counterterm needs to be included for a consistent renormalization at N$^2$LO. This result thus supports Bedaque {et al.}'s conclusion, and shows that Platter and Phillips' findings do not apply to the generic situation where $r_0 \sim \ell$. 

In Sec.~\ref{sec:pionfree-eft}, we will briefly review the SREFT in a
three-boson system, and summarize the formalism at LO in~\cite{Bedaque:1998kg}, which is based
on a modified Skorniakov-Ter-Martirosian integral equation, as well as the formalism at NLO, which is based on a perturbative expansion~\cite{Hammer:2001gh,Ji:2011qg}. We then discuss in Sec.~\ref{sec:n2lo-amplitude} the calculation of the N$^2$LO three-body scattering t-matrix and relate it to the real K-matrix amplitude. 
Following that, in
Sec.~\ref{sec:n2lo-observables}, we relate that K-matrix to an N$^2$LO calculation of three-body
observables: three-body binding energies and the S-wave phase shift. Sec.~\ref{sec:result-lo} shows the asymptotic
behavior of the LO three-body amplitude and defines several relevant integrals. This section serves as background for Sec.~\ref{sec:renormalization}, which discusses renormalization at N$^2$LO, via an analysis of the regularized results' cutoff dependence and the counterterms needed to cancel it. We also show our numerical result for the N$^2$LO three-body force as a function of cutoff and energy, and compare it with analytic expressions. In Sec.~\ref{sec:he-trimer}, we apply our formalism to observables in Helium trimers, comparing our results with precise calculations done with both SREFT and model potentials in Refs.~\cite{Platter:2006ev,Roudnev2000, Roudnev2003}. We conclude with a summary, followed by appendices.

\section{The SREFT in three-body systems}
\label{sec:pionfree-eft}
In this paper, we simplify our consideration to a non-relativistic system including three identical bosons. In particular, the simplifications in bosonic cases do not result in a loss of generality, due to the universal behavior in non-relativistic three-body systems. As shown by Bedaque {\it et al.}, the formalism for three identical bosons can be straightforwardly adapted (with appropriate modifications for the inclusion of spin and isopsin) to the study of three-nucleon systems~\cite{Bedaque:2000}. Therefore our N$^2$LO study in bosonic systems can be extended to three-nucleon systems, such as the neutron-deuteron system (including $^3$H) and the proton-deuteron system (including $^3$He). (Although, in the latter case, the Coulomb potential between two protons must be accounted for~\cite{Rupak:2001ci, Ando:2010wq, Konig:2011yq}.) For terminological simplicity in our three-identical-boson system, we from now on refer to a single boson as a (bosonic) atom, a two-boson state as a dimer and a three-boson 
state as a trimer.

Because we are considering only short-range interactions, we write the Lagrangian in the SREFT to describe non-relativistic three-atom systems, so it includes an atom field ($\psi$), a dimer field ($T$), an atom-atom contact interaction and an atom-dimer contact interaction:
\begin{equation}
\label{eq:Lagrangian}
\ensuremath{\mathcal{L}}=\psi^\dagger\left(i\partial_0 +
  \frac{\nabla^2}{2m}\right)\psi +\sigma T^\dagger\left(i\partial_0 +
  \frac{\nabla^2}{4m}-\Delta \right)T
-\frac{g}{\sqrt{2}}\left(T^\dagger
  \psi\psi+\textrm{h.c}\right)+hT^\dagger T \psi^\dagger \psi +
\cdots.
\end{equation}
The ellipses represent interactions above LO in powers of momentum, that are suppressed at low momenta. As demonstrated by Kaplan~\cite{Kaplan:1996nv}, a positive effective range can be described by the theory with $\sigma=-1$, which describes an atom and an auxiliary dimer. 

In the following, we will employ Eq.~\eqref{eq:Lagrangian} and expand each quantity described by this Lagrangian in powers of $r_0$. In the $r_0$ expansion, this Lagrangian is order-by-order equivalent to one containing only a single atom field (for detailed proofs of this equivalence, see Refs.~\cite{Bedaque:1998kg,Bedaque:1998km,Bedaque:1999vb}). The atom-dimer contact interactions in our SREFT are equivalent to the three-atom counterterm in Refs.~\cite{Bedaque:1998kg,Bedaque:1998km}, which is an equivalent description of the short-distance physics in a three-body system. Thus, from now on, we will refer to the atom-dimer contact interaction as a three-body counterterm (or three-body force). 

One might be concerned that the van der Waals' potential which governs the longest-range part of the atom-atom interaction is not truly short-ranged. It is indeed the case that the effective-range expansion for atom-atom scattering via a van der Waals' potential contains a term $\sim k^3$, i.e. a term that is non-analytic in $k^2$. Gao~\cite{Gao1998} showed that, upon including the van der Waals interaction, the two-body S-wave effective-range expansion is modified to:
\begin{equation}
\label{eq:kcot-vdw}
k\cot \delta_0 = -\frac{1}{a} + \frac{1}{2}r_0 k^2 -P r_0^3 k^4 +\mathcal{O}\left(\frac{r_0^4 k^3}{a^2} \right).
\end{equation}
However, the $k^3$ term is also suppressed by $r_0/a$ relative to the shape-parameter term $\sim k^4$, and so is less important than that piece of the effective-range expansion
 in either of the regimes $k\sim 1/a$ or $k~\simle~1/r_0$. Since the shape-parameter term does not enter SREFT calculations until N$^3$LO we conclude that 
SREFT can describe atom-atom interactions up to at least N$^2$LO.

The atom propagator is formulated as
\begin{equation}
iS(p_0,\mathbf{p})=\frac{i}{p_0-\frac{\mathbf{p}^2}{2m}+i\epsilon},
\end{equation}
where $p_0$ and $\mathbf{p}$ denote, respectively, the time component and the space
component of 4-momentum. The dressed dimer
propagator is renormalized to reproduce $r_0$ and $a$ in the effective-range expansion for the atom-atom scattering amplitude (see Eq.~(\ref{eq:ere})), and is expressed as
\begin{equation}
\label{eq:2}
i\mathcal{D}(p_0,\mathbf{p})=\frac{-i4\pi/mg^2}{-\gamma+\frac{1}{2}
r_0(\gamma^2+mp_0-\mathbf{p}^2/4)+\sqrt{-mp_0+\mathbf{p}^2/4-i\epsilon}+i\epsilon},
\end{equation}
where $\gamma$ is the ``typical momentum" of the atom-atom system, which is, from Eq.~(\ref{eq:ere}), related to $a$ and $r_0$ by:
\begin{equation}
\label{eq:a-gamma}
\frac{1}{a}=\gamma-\frac{1}{2}r_0\gamma^2 + \mathcal{O}\left(\frac{r_0^3}{a^4}\right).
\end{equation}
Therefore $\gamma\sim 1/a$ for $r_0\ll |a|$. The dimer is thus a real (virtual) bound state for $\gamma>0$ ($\gamma<0$), and in either case it is close to being a zero-energy resonance. 

$\mathcal{D}(p_0,\mathbf{p})$ has a spurious pole at high momenta ($p=|\mathbf{p}| \sim 1/r_0$), which is beyond the valid regime of SREFT and introduces unphysical short-distance effects. Here, we consider the physical dimer to propagate with a low momentum $p\sim \gamma$, and so expand the dimer propagator in powers of $r_0\gamma$ and $r_0 p$:
\begin{equation}
i\mathcal{D}(p_0,\mathbf{p})=\sum_n \frac{-i4\pi/mg^2}{-\gamma+\sqrt{-mp_0+\mathbf{p}^2/4-i\epsilon}}
\left(\frac{r_0}{2}\right)^n \left(\gamma+\sqrt{-mp_0 +\mathbf{p}^2/4}\right)^n.
\end{equation}

The $n$-th order dimer propagator is thus
\begin{equation}
\label{eq:dimer-expan}
i\mathcal{D}^{(n)}(p,\mathbf{p})=-i\frac{4\pi}{mg^2} \times \left(\frac{r_0}{2}\right)^n
\frac{\left(\gamma+\sqrt{-mp_0+\mathbf{p}^2/4}\right)^n}{-\gamma+\sqrt{-mp_0+\mathbf{p
}^2/4-i\epsilon}}.
\end{equation}
Here we will consider up to $n=2$ in our N$^2$LO analysis.

\begin{figure}[tbpic:stm]
\centerline{\includegraphics[width=14cm,angle=0,clip=true]{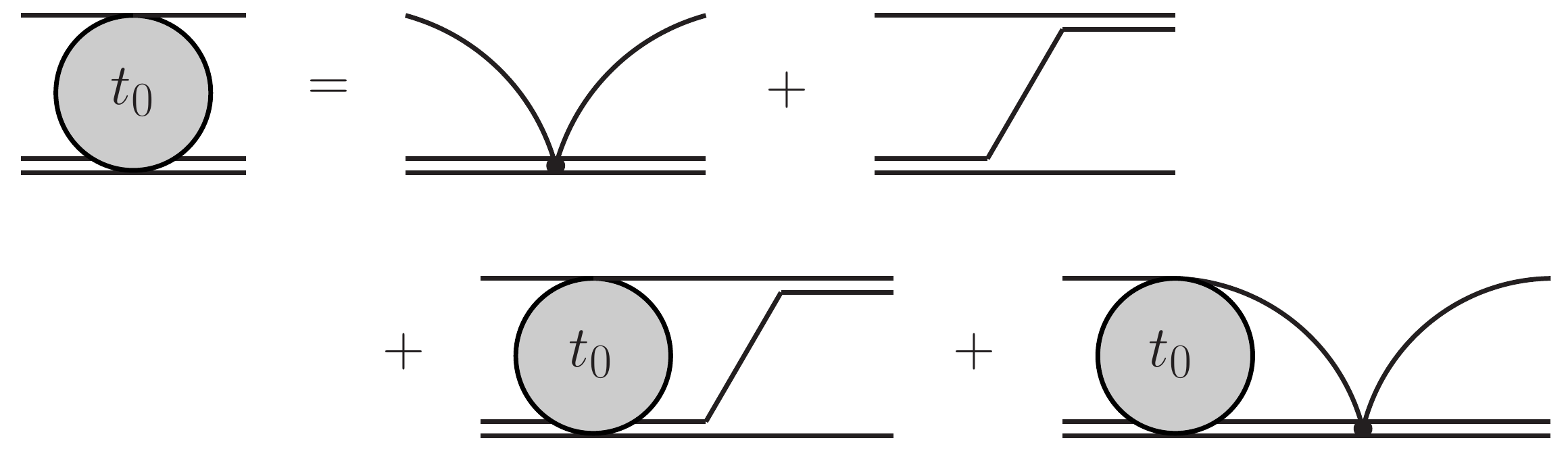}}
\caption{Leading-order atom-dimer amplitude, resulting from the iteration of two- and three-body interactions.}
\label{pic:stm}
\end{figure}

Three-body observables may be computed if we can solve for the atom-dimer scattering t-matrix, $t$. As illustrated in Fig.~\ref{pic:stm}, interactions contributing to $t(q,p)$ include the exchange of an atom between dimers, the three-body counterterm, and the iteration of these two diagrams to arbitrary order. This results in the solution to the non-relativistic Faddeev equation including both two- and three-body contact interactions, which is called the modified Skorniakov-Ter-Martirosian (STM) equation~\cite{Bedaque:1998kg,Bedaque:1998km}. The coupling constant for the three-body conterterm, $h$, is introduced to ensure that the resulting three-body observables are cutoff
independent.

In our perturbative analysis, we expand the quantities involved in the modified STM equation in powers of $\gamma r_0$ and $k r_0$ up to N$^2$LO:
\begin{eqnarray}
\mathcal{D}(p_0,\mathbf{p})&=&\mathcal{D}^{(0)}(p_0,\mathbf{p})+\mathcal{D}^{(1)}(p_0,\mathbf{p})+\mathcal{D}^{(2)}(p_0,\mathbf{p})+\cdots
\nn
t(q,p;E)&=&t_{0}(q,p;E)+t_{1}(q,p;E)+t_{2}(q,p;E)+\cdots
\nn
\mathcal{H}&=&H_0(\Lambda)+H_1(\Lambda)+H_2(E,\Lambda)+\cdots,
\end{eqnarray}
where $\mathcal{H} = \Lambda^2 h/2mg^2$. The on-shell t-matrix is related to the S-wave scattering amplitude $T(k)$ through:
\begin{equation}
\label{eq:T-t}
T(k) = \mathcal{Z} \sum_{n=0}^\infty t_{n}(k,k;E),
\end{equation}
where magnitudes of incoming momentum $\mathbf{p}$ and outgoing momentum $\mathbf{q}$ equal
to the on-shell value $k$, which satisfies $mE=3k^2/4-\gamma^2$. The renormalization factor $\mathcal Z$ can also be expanded in powers of $\gamma r_0$:
\begin{equation}
\label{eq:norm-Z}
\mathcal{Z} = \sum_{n=0}^\infty \mathcal{Z}_n, \hspace{10mm} \mathcal{Z}_n= \frac{8\pi\gamma}{m^2g^2} \left(\gamma r_0\right)^{n}.
\end{equation}

We now review established results at LO and NLO which we use to calculate
$t_2$, and, from it, three-body observables at N$^2$LO.

\subsection{The Leading-Order Three-Body Amplitude}
The leading-order (LO) $t_0$ is calculated via the modified STM equation, which, projected to the S-wave, is:
\begin{equation}
\label{eq:t0}
\tilde{t}_0(q,p;E)=M(q,p;E)+\frac{2}{\pi}\int_0^\Lambda dq'\
\frac{q'^2}{-\gamma+\sqrt{3q'^3/4-mE-i\epsilon}} M(q,q';E) \tilde{t}_0(q',p;E),
\end{equation}
where
\begin{equation}
\label{eq:kernel-M}
 M(q,p;E)=\frac{1}{qp} \log\left(
  \frac{q^2+p^2+qp-mE}{q^2+p^2-qp-mE}\right)
+\frac{2H_0(\Lambda)}{\Lambda^2}. 
\end{equation}
The t-matrix in Eq.~\eqref{eq:t0} is rescaled as $t_0(q,p;E) = mg^2 \tilde{t}_0(q,p;E)$ to absorb the unphysical constant, $g$.

In the bound-state region ($E<-\gamma^2/m$ if $\gamma > 0$ and $E < 0$ if $\gamma < 0$), the $i\epsilon$ in Eq.~\eqref{eq:t0} can be eliminated, and $\tilde{t}_0$ is therefore real. In the scattering case, $\tilde{t}_0$ is complex. To simplify our calculation, we introduce a real K-matrix that satisfies a modified STM equation with the $i\epsilon$ in Eq.~\eqref{eq:t0} replaced by a principal-value integration (denoted by $\mathcal{P}$):
\begin{equation}
\label{eq:K0}
\tilde{K}_0(q,p;E) = M(q,p;E) +\frac{2}{\pi} \mathcal{P}\int_0^\Lambda dq' \frac{q'^2}{-\gamma+\sqrt{3q'^2/4-mE}}M(q,q';E) \tilde{K}_0(q',p;E).
\end{equation}
The relation between t- and K-matrix at LO is
\begin{equation}
\tilde{t}_0(k,p;E) = \frac{\tilde{K}_0(k,p;E)}{1-i\frac{8\gamma k}{3} \tilde{K}_0(k,k;E)}.
\end{equation}
The fully-off-shell t-matrix at LO is related to the K-matrix (see, e.g., Ref.~\cite{Newton}) by
\begin{equation}
\label{eq:t0-K0}
\tilde{t}_0(q,p;E) = \tilde{K}_0(q,p;E) + i\frac{8\gamma k}{3}\frac{\tilde{K}_0(k,q;E) \tilde{K}_0(k,p;E)}{1-i\frac{8\gamma k}{3} \tilde{K}_0(k,k;E)}~.
\end{equation}

The three-body force parameter $H_0$ is tuned to fit one three-body
observable to ensure a cutoff independent result. $H_0$ has also been studied
at an analytic level in Refs.~\cite{Bedaque:1998km, Braaten:2011sz}, where the result
\begin{equation}
  \label{eq:h0}
  H_0(\Lambda)=c \frac{\sin(s_0 \ln(\Lambda/\bar{\Lambda})+\arctan(s_0))}
{\sin(s_0\ln(\Lambda/\bar{\Lambda})-\arctan(s_0))},
\end{equation}
valid up to $\mathcal{O}(1/\Lambda)$ corrections, was found. In Eq.~(\ref{eq:h0})
$s_0 = 1.00624$, and $\bar{\Lambda}$ is the parameter determined by the LO renormalization condition. For example, the relation between $\bar{\Lambda}$ and the atom-dimer scattering length $a_{ad}$ is given by~\cite{Braaten:2004rn}
\begin{equation}
\label{eq:a3-Lbar}
a_{ad} = \left[1.46 + 2.15 \cot\left(s_0\ln(a\bar{\Lambda})+0.09\right)\right] a.
\end{equation}  

Ref.~\cite{Bedaque:1998km} deduced $c=1$ in Eq.~(\ref{eq:h0}), based on an analytic calculation; while, in Ref.~\cite{Braaten:2011sz}, Braaten et al. found $c=0.879$ yielded much better agreement ($\sim 10^{-3}$) with the numerical value of $H_0$. This $\approx 10$\% effect originates from the details of the regularization. For example, when a hard-cutoff regularization with a finite $\Lambda$ is performed in the numerical calculation, the asymptotic behavior of $K_0$ at $q, p \sim \Lambda$ is distorted from the analytic result, which assumes $\Lambda \rightarrow \infty$ 
(see Sec.~\ref{sec:result-lo-halfon}). 
In order to ensure that such distortion does not affect the running of sub-leading three-body forces in our perturbative calculation, we first calculate the LO $\tilde{K}_0$ at a large cutoff $\Lambda_{\infty}$ and insert $\tilde{K}_0$ in perturbative integrations with a new cutoff at a lower value, $\Lambda < \Lambda_\infty$. By doing so, the numerically calculated $\tilde{K}_0$ agrees with the analytic expression up to $q,p \sim \Lambda$. The discrepancy induced by details of regularization is thus limited to $\mathcal{O}(\Lambda/\Lambda_\infty)$. However, the $H_0$ needed to renormalize $\tilde{K}_0$ in these calculations is still sensitive to the details of the regularization at $\Lambda_\infty$ and is described by Eq.~(\ref{eq:h0}) with $c=0.879$, in agreement with Ref.~\cite{Braaten:2011sz}.

\subsection{The Next-To-Leading-Order Three-Body Amplitude}
\begin{figure}[tbpic:t1]
\centerline{\includegraphics[width=16cm,angle=0,clip=true]{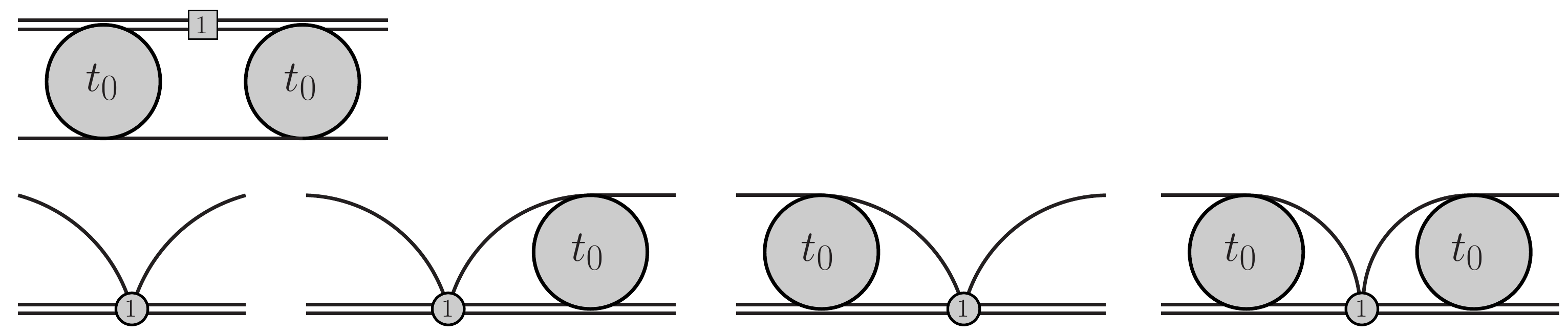}}
\caption{The diagrams for the atom-dimer t-matrix at NLO. The NLO piece of the dimer propagator is denoted by the square labeled ``1'' and NLO corrections to the atom-dimer contact interaction are indicated by the circles labeled ``1''.}
\label{pic:t1}
\end{figure}

The next-to-leading-order (NLO) calculation of effective-range corrections, $\sim r_0$, to the three-body amplitude was first done perturbatively for fixed $\gamma$ by Hammer and Mehen in Ref.~\cite{Hammer:2001gh}, and recalculated for variable $\gamma$ by Ji {\it et al.}~\cite{Ji:2010su,Ji:2011qg}. Contributions to range corrections at NLO are illustrated in Fig.~\ref{pic:t1}. These graphs lead to the $\mathcal{O}(r_0)$ piece of the atom-dimer t-matrix, which is calculated in perturbation theory using the LO t-matrix obtained from Eq.~(\ref{eq:t0}). The NLO on-shell t-matrix, projected to the S-wave, is then
\begin{eqnarray}
\label{eq:t1}
\tilde{t}_1(k,k;E)
&=&\frac{1}{\pi}\int_0^\Lambda dq'
q'^2\frac{\gamma+\sqrt{3q'^2/4-mE}}{-\gamma+\sqrt{3q'^2/4-mE-i\epsilon}}\tilde{t}_0^2(k,
q';E)\nn
&&+\frac{2 
\tilde{H}_1(\Lambda)}{\Lambda^2}\left[1+\frac{2}{\pi}\int_0^\Lambda
dq'\frac{q'^2}{-\gamma+\sqrt{3q'^2/4-mE-i\epsilon}}\tilde{t}_0(k,q';E)\right]^2,
\end{eqnarray}
where $\tilde{t}_1$ and $\tilde{H}_1$ are rescaled as $t_1 \equiv r_0
mg^2 \tilde{t}_1$, and $ H_1 \equiv r_0 \tilde{H}_1$. Here we only show
the elastic scattering case $p=q=k$, which is what we are interested for the computation of, e.g. the phase shift.

Using a similar approach to the LO case, the complex t-matrix at NLO, $\tilde{t}_1$, can be related to a real K-matrix, which satisfies an expression with principal-value integrations~\cite{Ji:2011qg}:
\begin{eqnarray}
\label{eq:K1}
\tilde{K}_1(k,k;E) &=& \frac{1}{\pi} \mathcal{P}\int_0^\Lambda dq\, \frac{q^2 (\gamma+\sqrt{3q^2/4-mE}\, )}{-\gamma+\sqrt{3q^2/4-mE}} \tilde{K}_0^2(k,q;E)
\nn
&&\hspace{2cm} +\frac{2\tilde{H}_1(\Lambda)}{\Lambda^2} \left[1+\frac{2}{\pi} \mathcal{P} \int_0^\Lambda dq\, \frac{q^2 \tilde{K}_0(k,q;E)}{-\gamma+\sqrt{3q^2/4-mE}} \right]^2.
\end{eqnarray}
The NLO pieces of the on-shell t-matrix and K-matrix are related by
\begin{equation}
\label{eq:t1-K1}
\tilde{t}_1(k,k;E) = \left[1-i\frac{8\gamma k}{3} \tilde{K}_0(k,k;E)\right]^{-2}
\left[\tilde{K}_1(k,k;E) + i\frac{8\gamma^2 k}{3} \tilde{K}_0(k,k;E)\right].
\end{equation}
The introduction of a real $\tilde{K}_1$ not only simplifies the calculation but also straightforwardly leads to the NLO effective-range correction to the phase shift.

Refs.~\cite{Ji:2010su,Ji:2011qg} pointed out that the three-body force $\tilde{H}_1(\Lambda)$ will in general be $\gamma$-dependent:
\begin{equation}
\label{eq:h1}
\tilde{H}_1(\Lambda) = \Lambda h_{10}(\Lambda) + \gamma h_{11}(\Lambda),
\end{equation}
where both $h_{10}$ and $h_{11}$ are functions of $\Lambda$, whose analytic expressions were derived in Ref.~\cite{Ji:2011qg}. $h_{10}$ is of $\mathcal{O}(\Lambda^0)$ and is log-periodic in $\Lambda$
\begin{equation}
\label{eq:h10}
h_{10}(\Lambda)=-\frac{3\pi (1+s_0^2)}{64
  \sqrt{1+4s_0^2}}\frac{\sqrt{1+4s_0^2} 
- \cos \left(2 s_0 \ln (\Lambda/\bar{\Lambda}) - \arctan 2s_0\right)} {\sin^2
\left( s_0 \ln (\Lambda/\bar{\Lambda}) - \arctan s_0\right)}~.
\end{equation}
The $\gamma h_{11}$ piece of $\tilde{H}_1$ is $\sim \gamma \ln \Lambda$ (see Ref.~\cite{Ji:2011qg} for details). If we are only interested in the fixed-$\gamma$ case, as is relevant for three-nucleon systems or the case of $^4$He trimers, then it is appropriate to think of $\tilde{H}_1$ as one overall number, i.e. the $\gamma h_{11}$ part of Eq.~\eqref{eq:h1} can be combined with $h_{10}$. The two pieces can, however, be disentangled by experiments in systems with a variable scattering length, such as the recombination of ultracold $^7$Li or $^{133}$Cs atoms~\cite{Kraemer:2006,Gross:2009}.

\section{The Three-Body Scattering Amplitude At N$^2$LO}
\label{sec:n2lo-amplitude}
We now calculate the next-to-next-to-leading-order (N$^2$LO) effective-range corrections to the atom-dimer amplitude perturbatively. We focus on the elastic-scattering
situation $p=q=k$ and study N$^2$LO corrections to three-body bound-state energies and the
S-wave elastic-scattering phase shift in three-body systems with fixed
scattering length (i.e. fixed $\gamma$) and effective range. 

Since three-body energy is conserved in an elastic channel, we omit the third argument, $E$, in expressions of t-matrices in this section to avoid cluttering notation (i.e. $\tilde{t}_0(q,p) \equiv \tilde{t}_0(q,p;E)$). We will restore the third argument in subsequent sections.

\subsection{Diagrams and t-matrix}
\label{sec:n2lo-amp-subA}
\begin{figure}[tbpic:t2a]
\centerline{\includegraphics[width=5.5cm,angle=0,clip=true]{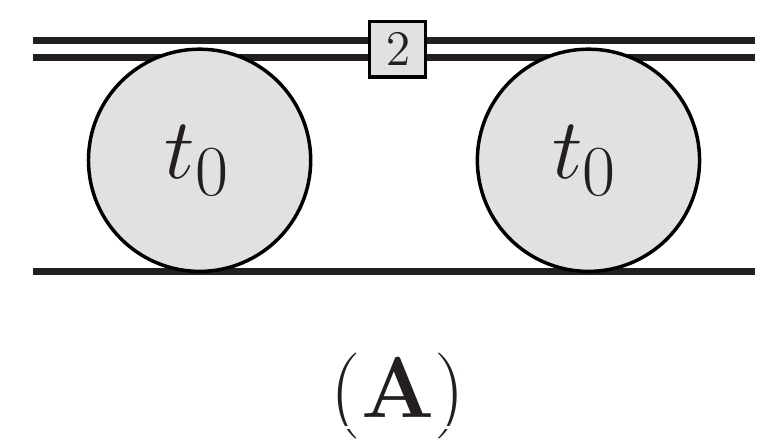}}
\caption{N$^2$LO t-matrix diagram part A: Insertion of N$^2$LO dimer propagator (square labeled ``2'').}
\label{pic:t2a}
\end{figure}

\begin{figure}[tbpic:t2b]
\centerline{\includegraphics[width=12cm,angle=0,clip=true]{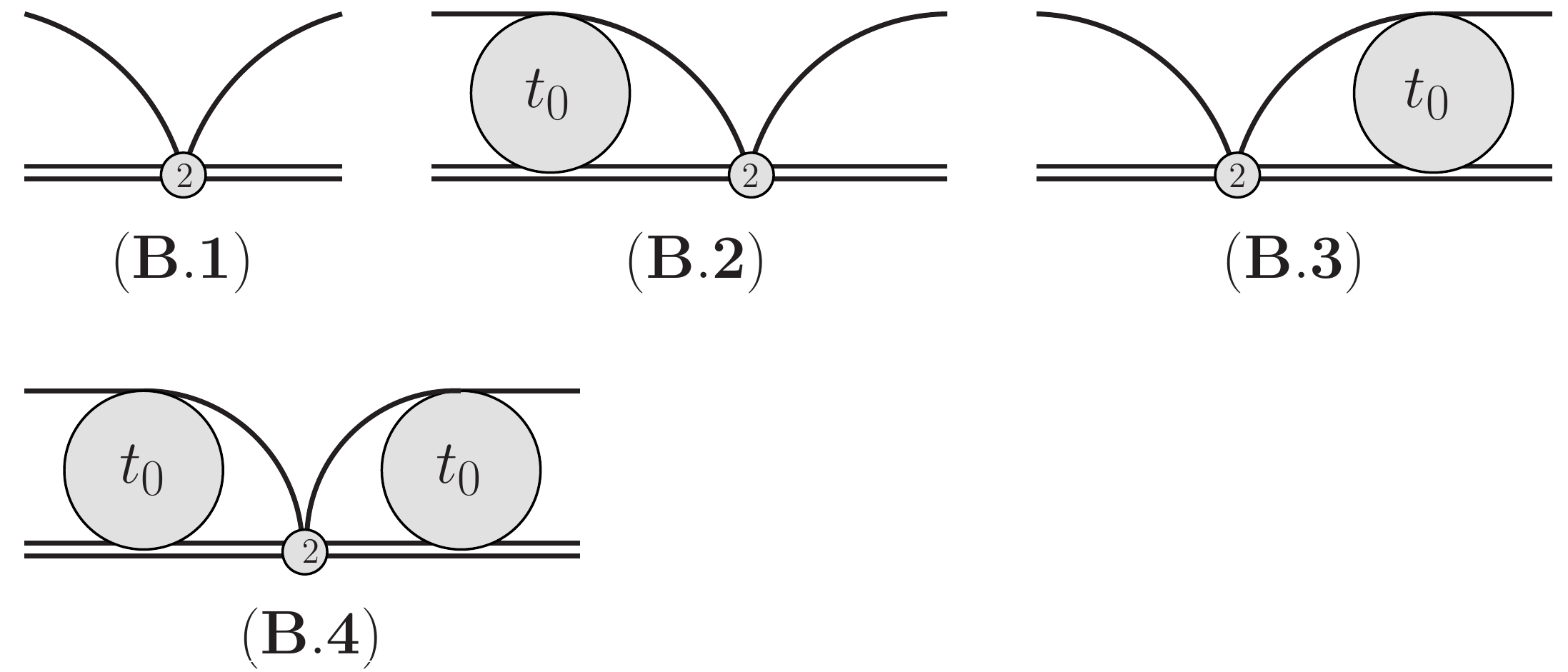}}
\caption{N$^2$LO t-matrix diagram part B: Insertion of N$^2$LO 3-body contact interaction (circle labeled ``2'').}
\label{pic:t2b}
\end{figure}

\begin{figure}[tbpic:t2c]
\centerline{\includegraphics[width=8cm,angle=0,clip=true]{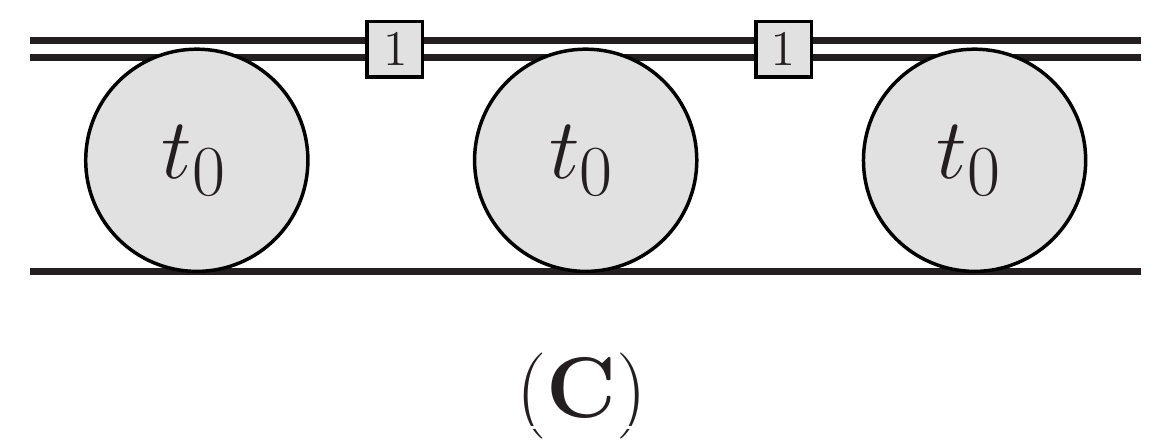}}
\caption{N$^2$LO t-matrix diagram part C: Double insertion of NLO dimer progators (squares labeled ``1'').}
\label{pic:t2c}
\end{figure}

\begin{figure}[tbpic:t2d]
\centerline{\includegraphics[width=15cm,angle=0,clip=true]{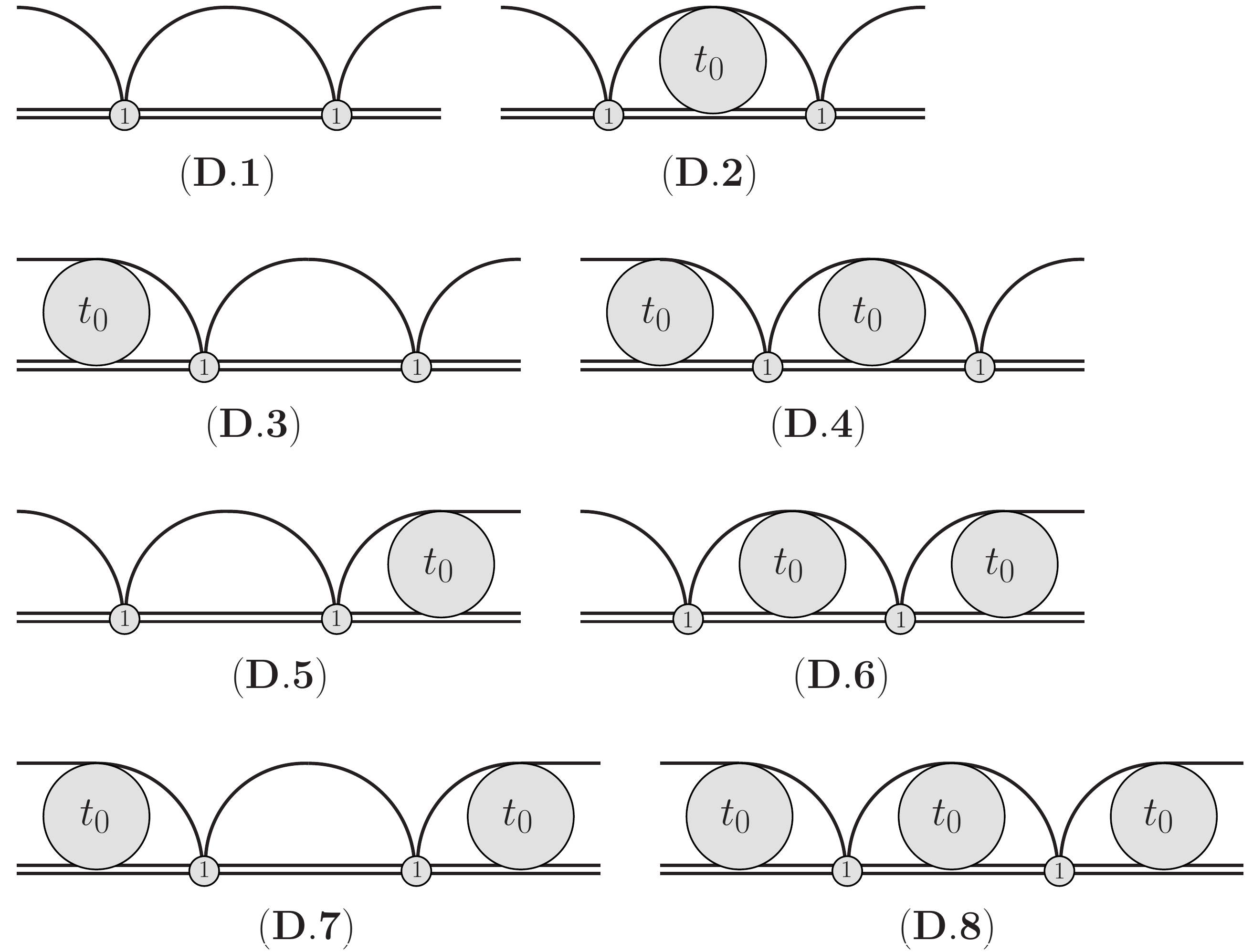}}
\caption{N$^2$LO t-matrix diagram part D: double insertion of NLO 3-body contact interactions (circles labeled ``1'').}
\label{pic:t2d}
\end{figure}

\begin{figure}[tbpic:t2e]
\centerline{\includegraphics[width=15cm,angle=0,clip=true]{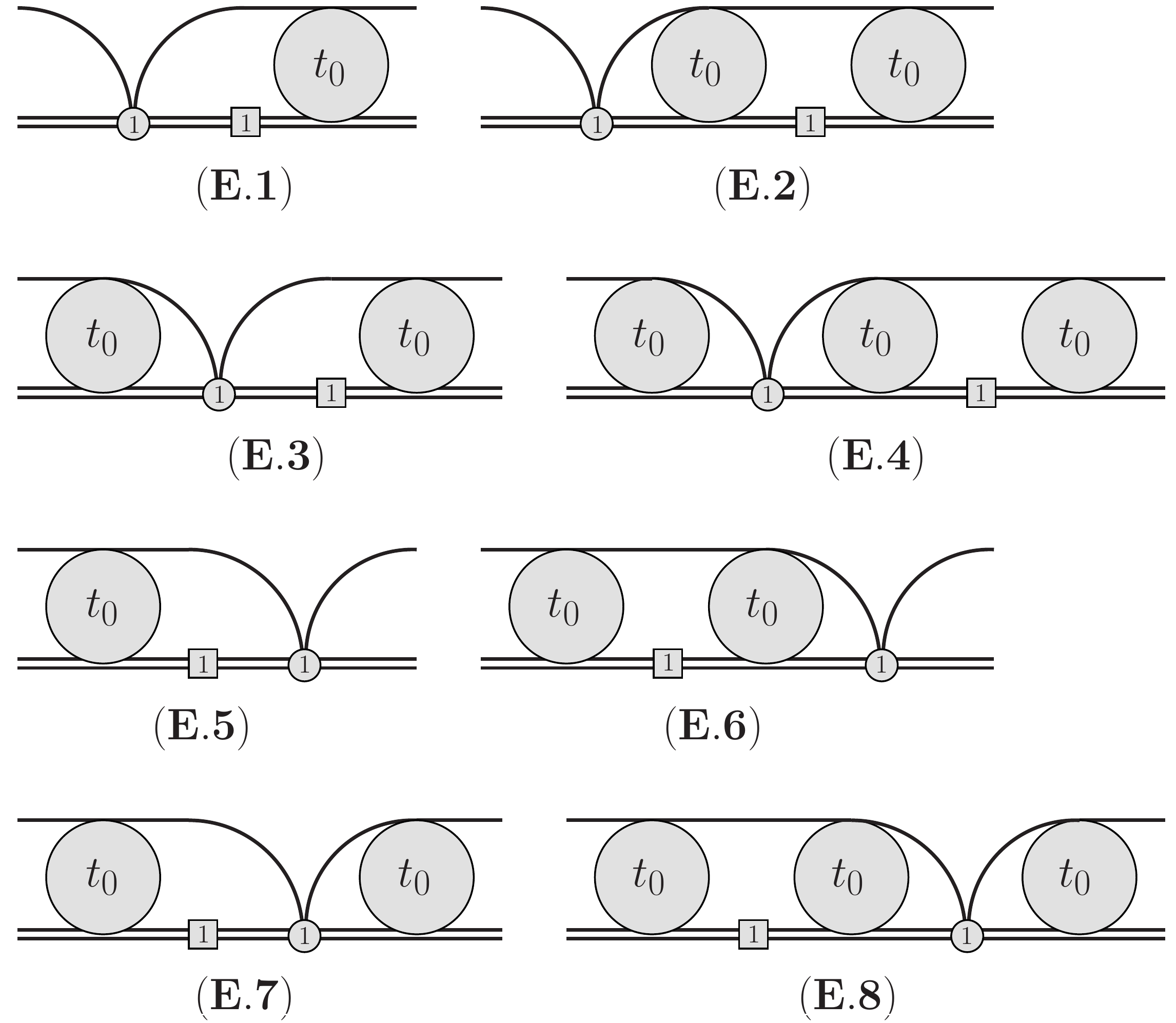}}
\caption{N$^2$LO t-matrix diagram part E: These arise from the insertion of one NLO dimer propagator (squares labeled ``1'') and one NLO three-body contact interaction (circles labeled ``1'').}
\label{pic:t2e}
\end{figure}

In our perturbative analysis at N$^2$LO, we insert effective-range corrections
to the dimer propagator and the three-body counterterm between LO t-matrices, which are obtained from the modified STM equation~\eqref{eq:t0}. Compared to the calculation of the t-matrix at NLO, here we insert both NLO and N$^2$LO terms and consider their contributions to the $\mathcal{O}(r_0^2)$ piece of the atom-dimer t-matrix. The resulting diagrams are categorized in classes from A to E, that are illustrated in Figs.~\ref{pic:t2a}, \ref{pic:t2b}, \ref{pic:t2c}, \ref{pic:t2d}, \ref{pic:t2e}.

Fig.~\ref{pic:t2a} shows the effect of the N$^2$LO correction to the dimer propagator. This constitutes diagram A, whose contribution to the N$^2$LO t-matrix is
\begin{eqnarray}
\label{eq:t2a-hf}
it_2^A(\mathbf{k},\mathbf{p})  &=& \int \frac{d^4 q}{(2\pi)^4}\, i\mathcal{D}^{(2)}(q_0,\mathbf{q})\, iS(E-q_0,-\mathbf{q})
\,
it_0(\mathbf{k},\mathbf{q})\, it_0(\mathbf{q},\mathbf{p}),
\end{eqnarray}
where we only consider the on-shell case that $|\mathbf{k}|=|\mathbf{p}|=k$ and
$mE=3k^2/4-\gamma^2$. By projecting $t_2^A(\mathbf{k},\mathbf{p})$ onto the S-wave we obtain
\begin{eqnarray}
t_2^A(k,k)
&=& \frac{r_0^2}{2\pi mg^2}\int_0^\Lambda dq\, q^2 \frac{(\gamma+\sqrt{3q^2/4-mE}\
)^2}{-\gamma+\sqrt{3q^2/4-mE-i\epsilon}}\, t_0^2(k,q).
\end{eqnarray}

At N$^2$LO three-body ($\equiv$ atom-dimer) contact interactions will be required to renormalize the amplitude. We denote these as $H_2$. Based on the power-counting arguments given in Ref.~\cite{Bedaque:1998km} we anticipate that, in addition to being dependent on $\Lambda$, $H_2$ will also depend on the three-body energy $E$, i.e. we have $H_2(E,\Lambda)$. 
Fig.~\ref{pic:t2b} defines class B, the contribution of such contact interactions to the atom-dimer scattering amplitude.
From the Feynman rules, it's formulated as
\begin{eqnarray}
\label{eq:t2B-vec}
it_2^B(\mathbf{k},\mathbf{p})&=&i\frac{2mg^2H_2(E,\Lambda)}{\Lambda^2}
\left[1+\int \frac{d^4 q}{(2\pi)^4}\, i\mathcal{D}^{(0)}(q_0,\mathbf{q})\, iS(E-q_0,-\mathbf{q})\, it_0(\mathbf{k},\mathbf{q})\right]
\nn
&&\times\left[1+\int \frac{d^4 q}{(2\pi)^4}\, i\mathcal{D}^{(0)}(q_0,\mathbf{q})\, iS(E-q_0,-\mathbf{q})\, it_0(\mathbf{p},\mathbf{q})\right]~.
\end{eqnarray}
Here we define the part in the square bracket in Eq.~\eqref{eq:t2B-vec} as a function $\mathcal{L}(\mathbf{p})$:
\begin{equation}
\mathcal{L}(\mathbf{p}) \equiv 1+\int_0^\Lambda \frac{d^4 q}{(2\pi)^4}\, i\mathcal{D}^{(0)}(q_0,\mathbf{q})\, iS(E-q_0,-\mathbf{q})\, it_0(\mathbf{p},\mathbf{q}),
\end{equation}
whose S-wave projection is
\begin{equation}
\label{eq:math-L}
\mathcal{L}(k) = 1+\frac{2}{\pi mg^2}\int_0^\Lambda dq\, \frac{q^2}{-\gamma+\sqrt{3q^2/4-mE-i\epsilon}}t_0(k,q).
\end{equation}
The function $\mathcal{L}$ will also appear as part of the N$^2$LO contributions from diagrams in classes D and E. The S-wave projection of $t_2^B$ in the on-shell case is then
\begin{eqnarray}
t_2^B(k,k) 
&=&\frac{2mg^2 H_2(E,\Lambda)}{\Lambda^2}\, \mathcal{L}^2(k).
\end{eqnarray}

This accounts for contributions directly from the N$^2$LO dimer propagator and N$^2$LO three-body counterterms. We now compute contributions arising from two insertions of NLO terms. These constitute classes C--E.

Fig.~\ref{pic:t2c} represents double insertions of the NLO part of the dimer propagator: class C. The resulting contribution to the N$^2$LO t-matrix is 
\begin{eqnarray}
it_2^C(\mathbf{k},\mathbf{p}) &=& \int\frac{d^4 q}{(2\pi)^4} \int\frac{d^4 q'}{(2\pi)^4} iS(E-q_0,-\mathbf{q}) i\mathcal{D}^{(1)}(q_0,\mathbf{q}) iS(E-q_0',-\mathbf{q'}) i\mathcal{D}^{(1)}(q_0',\mathbf{q'})
\nn
&&\times it_0(\mathbf{k},\mathbf{q})\, it_0(\mathbf{q},\mathbf{q'})\, it_0(\mathbf{q'},\mathbf{p}),
\end{eqnarray}
whose on-shell S-wave projection is
\begin{eqnarray}
t_2^C (k,k) 
&=& \frac{r_0^2}{\pi^2 m^2 g^4}\int\limits_0^\Lambda \int\limits_0^\Lambda dq\, dq'\, \frac{q^2(\gamma+\sqrt{3q^2/4-mE}\, ) }{-\gamma+\sqrt{3q^2/4-mE-i\epsilon}}\, \frac{q'^2(\gamma+\sqrt{3q'^2/4-mE}\, ) }{-\gamma+\sqrt{3q'^2/4-mE-i\epsilon}}
\nn
&&
\times\, t_0(k,q)\, t_0(q,q')\, t_0(k,q').
\end{eqnarray}

\begin{figure}[tbpic:t2d-eq]
\centerline{\includegraphics[width=15cm,angle=0,clip=true]{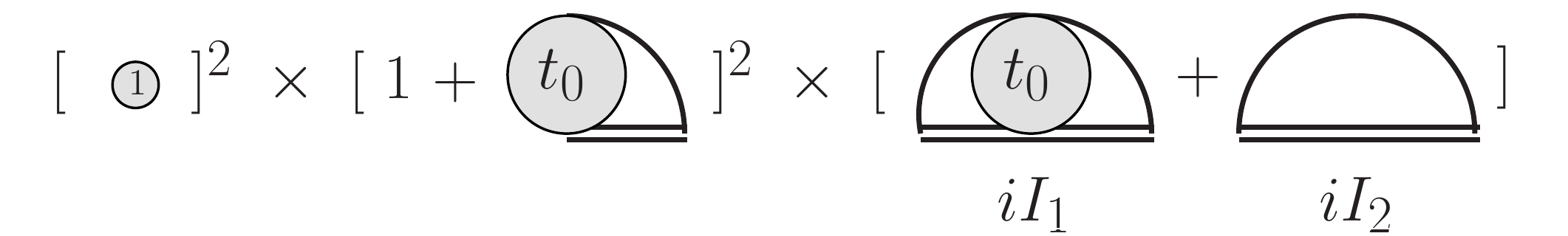}}
\caption{An alternative way of writing part D of the N$^2$LO t-matrix diagrammatically.}
\label{pic:t2d-eq}
\end{figure}

Fig.~\ref{pic:t2d} represents class D: graphs with two insertions of the NLO three-body counterterm. The
on-shell contribution can be re-expressed in a form that factorizes counterterms and closed loops (see Fig.~\ref{pic:t2d-eq} for a diagrammatic formulation). The loops $\mathcal{I}_1$ and $\mathcal{I}_2$ defined in Fig.~\ref{pic:t2d-eq} are
\begin{subequations}
\begin{eqnarray}
i\mathcal{I}_1 &=& \int\frac{d^4 q}{(2\pi)^4} \int\frac{d^4 q'}{(2\pi)^4}\, i\mathcal{D}^{(0)}(q_0,\mathbf{q})\, iS(E-q_0,-\mathbf{q})\, 
it_0(\mathbf{q},\mathbf{q'})
\nn
&&\times i\mathcal{D}^{(0)}(q_0',\mathbf{q'})\, iS(E-q_0',-\mathbf{q'})~,
\end{eqnarray}
\begin{eqnarray}
i\mathcal{I}_2 &=& \int \frac{d^4 q}{(2\pi)^4}\, i\mathcal{D}^{(0)}(q_0,\mathbf{q})\, iS(E-q_0,-\mathbf{q})~,
\end{eqnarray}
\end{subequations}
which are both only functions of the three-body energy. 
In terms of these the on-shell S-wave projection of class D is
\begin{eqnarray}
t_2^D(k,k) &=& \left[\frac{2mg^2 H_1(\Lambda)}{\Lambda^2}\right]^2
\mathcal{L}^2(k)\,
\left[\mathcal{I}_1 +\mathcal{I}_2\right]
\nn
&=&\frac{8mg^2H_1^2(\Lambda)}{\pi\Lambda^4} \mathcal{L}^2(k)
 \left[\int_0^\Lambda dq\, \frac{q^2}{-\gamma+\sqrt{3q^2/4-mE-i\epsilon}}
\right.
\nn
&&\hspace{-5mm}
+\left.
\frac{2}{\pi mg^2} \int\limits_0^\Lambda \int\limits_0^\Lambda dq dq'\, \frac{q^2}{-\gamma+\sqrt{3q^2/4-mE-i\epsilon}} \frac{q'^2}{-\gamma+\sqrt{3q'^2/4-mE-i\epsilon}} t_0(q,q')\right].
\nn
\end{eqnarray}

\begin{figure}[tbpic:t2e-eq]
\centerline{\includegraphics[width=15cm,angle=0,clip=true]{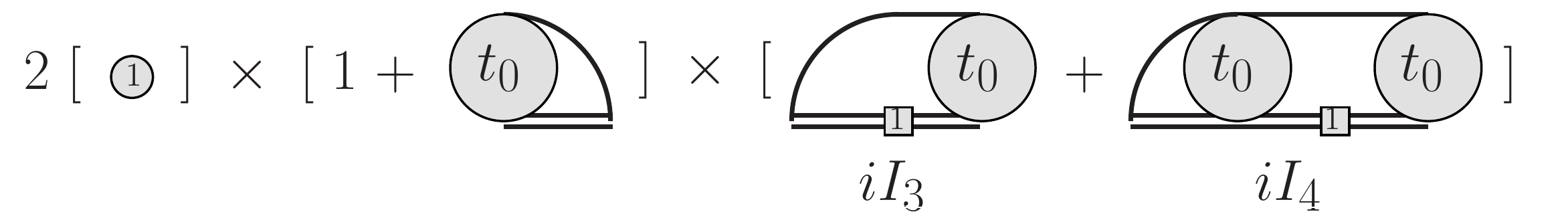}}
\caption{An alternative way of writing part E of the N$^2$LO t-matrix diagrammatically.}
\label{pic:t2e-eq}
\end{figure}

Fig.~\ref{pic:t2e} includes terms with one insertion of the NLO dimer propagator and
one of the NLO three-body counterterm. These make up class E. Their factorization is shown diagrammatically in Fig.~\ref{pic:t2e-eq}, where the loop integrations $I_3$ and $I_4$ are, respectively, defined as
\begin{subequations}
\begin{eqnarray}
\mathcal{I}_3 &=& \int_0^\Lambda \frac{d^4 q}{(2\pi)^4}\, iS(E-q_0,-\mathbf{q})\, i\mathcal{D}^{(1)}(q_0,\mathbf{q})\, it_0(\mathbf{q},\mathbf{k}),
\end{eqnarray}
\begin{eqnarray}
\mathcal{I}_4 &=& \int\frac{d^4 q}{(2\pi)^4} \int\frac{d^4 q'}{(2\pi)^4}\, i\mathcal{D}^{(0)}(q_0,\mathbf{q})\, iS(E-q_0,-\mathbf{q})\, i\mathcal{D}^{(1)}(q_0',\mathbf{q'})\, iS(E-q_0',-\mathbf{q'})
\nn
&&\times it_0(\mathbf{q},\mathbf{q'})\, it_0(\mathbf{q'},\mathbf{k}).
\end{eqnarray}	
\end{subequations}

Therefore, the contribution of class E to the S-wave on-shell N$^2$LO atom-dimer t-matrix is
\begin{eqnarray}
&&t_2^E(k,k)
= \frac{4 mg^2 H_1(\Lambda)}{\Lambda^2}
\mathcal{L}(k)
\left[\mathcal{I}_3+\mathcal{I}_4\right]
\nn
&& =\frac{4 r_0 H_1(\Lambda)}{\pi \Lambda^2} \mathcal{L}(k)
\left[ \int_0^\Lambda dq\, \frac{q^2 (\gamma+\sqrt{3q^2/4-mE}\, )}{-\gamma+\sqrt{3q^2/4-mE-i\epsilon}} t_0(k,q) \right.
\nn
&& \left.+\frac{2}{\pi mg^2} \int\limits_0^\Lambda \int\limits_0^\Lambda dq dq'\, \frac{q^2}{-\gamma+\sqrt{3q^2/4-mE-i\epsilon}}
 \frac{q'^2 (\gamma+\sqrt{3q'^2/4-mE}\, )}{-\gamma+\sqrt{3q'^2/4-mE-i\epsilon}} t_0(k,q') t_0(q,q') \right].\nn
\end{eqnarray}

\subsection{The atom-dimer K-matrix at N$^2$LO}
The integrals in the previous subsection include terms $i\epsilon$ in the denominator, and so, in the scattering case, the N$^2$LO t-matrix $t_2$ is complex.
Each term in the dimer propagator's $r_0$ expansion~\eqref{eq:dimer-expan}, has the same denominator, and so they all have a pole at the on-shell momentum for atom-dimer scattering. We therefore separate the
propagator into a Cauchy principal-value and imaginary part:
\begin{equation}
\label{eq:cauchy}
\frac{1}{-\gamma+\sqrt{3q^2/4-mE-i\epsilon}}=\ensuremath{\mathcal{P}}\frac{1}{-\gamma+\sqrt{3q^2/4-mE}}+i\frac{4\pi\gamma}{3k}\delta(q-k).
\end{equation}
In the following exposition, we do not include the principal-value symbol. Except where especially noted, all the integrals with a real-number singular pole should be considered as principal-value integrals.

We now isolate the delta-function part in Eq.~\eqref{eq:cauchy} for each piece of the N$^2$LO contribution to the t-matrix. In a bound-state problem with a negative three-body energy $E$, these imaginary parts need to be omitted. It is therefore useful to relate $t_2$ to an N$^2$LO K-matrix, in which only real integrals appear. 

After using Eq.~(\ref{eq:cauchy}) in the expressions for $t_2^A$--$t_2^E$, and employing Eqs.~(\ref{eq:t0-K0}) and (\ref{eq:K1}), 
we find that the full expression for the N$^2$LO piece of the t-matrix takes the form
\begin{equation}
\label{eq:t2-K2}
\tilde{t}_2(k,k)=\frac{\tilde{K}_2(k,k)}{\left(1-i\frac{8\gamma k}{3}\tilde{K}_0(k,k)\right)^2} + i\frac{8\gamma k}{3}\frac{ \left(\gamma\tilde{K}_0(k,k) +\tilde{K}_1(k,k)\right)^2}{\left(1-i\frac{8\gamma k}{3}\tilde{K}_0(k,k)\right)^3},
\end{equation}
where we have rescaled $t_2$ according to $t_2 = r_0^2mg^2\tilde{t}_2$. Here:
\begin{eqnarray}
\label{eq:K2}
\tilde{K}_2(k,k) &=& \frac{1}{2\pi} \int_0^\Lambda dq\, \frac{q^2 (\gamma+\sqrt{3q^2/4-mE}\, )^2}{-\gamma+\sqrt{3q^2/4-mE}} \tilde{K}_0^2(k,q)
\nn
&&+\frac{2\tilde{H}_2(E,\Lambda)}{\Lambda^2} \left[1+\frac{2}{\pi} \int_0^\Lambda dq\, \frac{q^2 \tilde{K}_0(k,q)}{-\gamma+\sqrt{3q^2/4-mE}} \right]^2
\nn
&&+\frac{1}{\pi^2} \int\limits_0^\Lambda \int\limits_0^\Lambda dq dq'\, \frac{q^2 (\gamma+\sqrt{3q^2/4-mE}\, )}{-\gamma+\sqrt{3q^2/4-mE}} \frac{q'^2 (\gamma+\sqrt{3q'^2/4-mE}\, )}{-\gamma+\sqrt{3q'^2/4-mE}}
\nn
&&\hspace{2.5cm}\times\, \tilde{K}_0(k,q) \tilde{K}_0(k,q') \tilde{K}_0(q,q')
\nn
&&\hspace{-16mm}+\frac{8\tilde{H}_1^2(\Lambda)}{\pi\Lambda^4} \left[1+\frac{2}{\pi} \int_0^\Lambda dq\, \frac{q^2 \tilde{K}_0(k,q)}{-\gamma+\sqrt{3q^2/4-mE}} \right]^2
\nn
&&\hspace{-12mm}\times \left[ \int_0^\Lambda dq\, \frac{q^2}{-\gamma+\sqrt{3q^2/4-mE}}\left(1+\frac{2}{\pi}\int_0^\Lambda dq' \frac{q'^2 \tilde{K}_0(q,q')}{-\gamma+\sqrt{3q'^2/4-mE}}\right)\right]
\nn
&&\hspace{-16mm}+\frac{4\tilde{H}_1(\Lambda)}{\pi\Lambda^2}\left[1+\frac{2}{\pi} \int_0^\Lambda dq\, \frac{q^2 \tilde{K}_0(k,q)}{-\gamma+\sqrt{3q^2/4-mE}} \right]
\nn
&&\hspace{-12mm}\times \left[ \int_0^\Lambda dq\, \frac{q^2 (\gamma+\sqrt{3q^2/4-mE}\, )}{-\gamma+\sqrt{3q^2/4-mE}} \tilde{K}_0(k,q)\left(1+\frac{2}{\pi}\int_0^\Lambda dq' \frac{q'^2 \tilde{K}_0(q,q')}{-\gamma+\sqrt{3q'^2/4-mE}}\right)\right]
\nn
&\equiv&\tilde{K}_2^A + \tilde{K}_2^B + \tilde{K}_2^C + \tilde{K}_2^D + \tilde{K}_2^E,
\end{eqnarray}
with the N$^2$LO three-body counterterm parameter rescaled as $H_2 = r_0^2 \tilde{H}_2$.
Each of the integrations summed up in Eq.~\eqref{eq:K2} is identified with $\tilde{K}_2^{n}$, $n=A$--$E$, thus associating each contribution to the N$^2$LO K-matrix with a particular category of diagrams defined in Subsection~\ref{sec:n2lo-amp-subA}. 
Since Eq.~(\ref{eq:t2-K2}) can also be derived by considering the general relation between t- and K-matrices and expanding to second order in $r_0$ the fact that we obtain $\tilde{t}_2$ with this structure is a non-trivial check that we have perturbative unitarity, i.e. our t-matrix satisfies unitarity up to corrections of $\mathcal{O}(r_0^3)$.

\section{Three-body observables at N$^2$LO}
\label{sec:n2lo-observables}
Here we discuss how to calculate three-body observables at N$^2$LO within our perturbative approach. In an atom-dimer scattering state, we consider the elastic-scattering phase shift in the S-wave; for trimer bound states, we focus on the binding energy. These quantities are expanded in a series of $\gamma r_0$ and/or $k r_0$, with the leading order representing universal physics. We derive NLO and N$^2$LO shifts in these quantities as corrections beyond universality.

\subsection{S-Wave Phase Shifts}
The S-wave amplitude $T(k)$ of an atom-dimer scattering state is related to the phase shift by
\begin{equation}
T(k) = \frac{3\pi}{m} \frac{1}{k\cot\delta(k) -ik}.
\end{equation}
$k \cot \delta$ can be expanded in powers of $r_0$
\begin{equation}
k\cot\delta = k\cot\delta_0 +r_0{[k\cot\delta]}_1 +r_0^2{[k\cot\delta]}_2+\cdots,
\end{equation}
where ${[\, ]}_1$ and ${[\, ]}_2$ indicate respectively the NLO and N$^2$LO part of $k\cot \delta$. Therefore, the scattering amplitude $T(k)$ is expanded correspondingly as
\begin{eqnarray}
\label{eq:ampT}
T(k) &=& \frac{3\pi}{m}  
\left[
\frac{1}{k\cot\delta_0-ik}
-r_0\frac{{[k\cot\delta]}_1}{\left(k\cot\delta_0-ik\right)^2}
\right.
\nn 
&&\hspace{8mm}\left.+ r_0^2\left(
\frac{\left({[k\cot\delta]}_1\right)^2}{\left(k\cot\delta_0-ik\right)^3} 
- \frac{{[k\cot\delta]}_2}{\left(k\cot\delta_0-ik\right)^2}
\right)
+\cdots
\right]
\nn
&\equiv& T_0(k) + T_1(k) + T_2(k) +\cdots.
\end{eqnarray}

The LO amplitude is related to the K-matrix by Eqs.~\eqref{eq:T-t}, \eqref{eq:norm-Z}, and \eqref{eq:t0-K0}:
\begin{eqnarray}
T_0(k) &=& \mathcal{Z}_0 t_0(k,k;E)
\nn
&=& \frac{8\pi\gamma}{m} \tilde{t}_0(k,k;E) 
\nn
&=& \frac{3\pi}{m} \frac{1}{\frac{3}{8\gamma}\tilde{K}_0^{-1}-ik},
\end{eqnarray}
which means that we recover the LO relation between phase shift and K-matrix
\begin{equation}
\label{eq:kcot0}
k \cot\delta_0 = \frac{3}{8\gamma}\tilde{K}_0^{-1}(k,k;E).
\end{equation}

Similarly, the NLO amplitude is related to K-matrices as:
\begin{eqnarray}
T_1(k) &=& \mathcal{Z}_1 t_0(k,k;E) +\mathcal{Z}_0 t_1(k,k;E)
\nn
&=& r_0 \frac{8\pi\gamma}{m} \left[\gamma  \tilde{t}_0(k,k;E) + \tilde{t}_1(k,k;E)\right]~,
\end{eqnarray}
which leads to the relation for the NLO part of $k\cot\delta$:
\begin{equation}
\label{eq:kcot1}
{[k\cot\delta]}_1 = -\frac{3}{8\gamma} \tilde{K}_0^{-2}(k,k;E) \left(\tilde{K}_1(k,k;E) + \gamma \tilde{K}_0(k,k;E)\right)~.
\end{equation}

Proceeding in the same way at N$^2$LO, we derive the N$^2$LO relation between the scattering amplitude and K-matrices as:
\begin{eqnarray}
\label{eq:T2-series}
T_2(k) &=& \mathcal{Z}_2 t_0(k,k;E) + \mathcal{Z}_1 t_1(k,k;E) + \mathcal{Z}_0 t_2(k,k;E)
\nn
&=& r_0^2 \frac{3\pi}{m} \left\lbrace
\frac{\left[\frac{3}{8\gamma} \tilde{K}_0^{-2} \left(\tilde{K}_1 +\gamma\tilde{K}_0\right)\right]^2}
{\left(\frac{3}{8\gamma}\tilde{K}_0^{-1} -ik\right)^3}
+\frac{\frac{3}{8\gamma} \tilde{K}_0^{-2} \left(\tilde{K}_2 - \gamma\tilde{K}_1 -\tilde{K}_1^2/\tilde{K}_0\right)}
{\left(\frac{3}{8\gamma}\tilde{K}_0^{-1} -ik\right)^2}
\right\rbrace~,
\end{eqnarray}
where we substitute the expressions of $\tilde{t}_1$~\eqref{eq:t1-K1} and $\tilde{t}_2$~\eqref{eq:t2-K2} into Eq.~\eqref{eq:T2-series}. After recognizing that part of the expression \eqref{eq:T2-series} can be written as $k\cot\delta_0$ and $[k\cot\delta]_1$ we compare it with the N$^2$LO piece of Eq.~\eqref{eq:ampT}, and find
\begin{equation}
{[k\cot\delta]}_2 = -\frac{3}{8\gamma} \tilde{K}_0^{-2}(k,k;E) \left(
\tilde{K}_2(k,k;E) - \gamma\tilde{K}_1(k,k;E) -\frac{\tilde{K}_1^2(k,k;E)}{\tilde{K}_0(k,k;E)}
\right)~.
\end{equation}

\subsection{Three-Body Bound States}
When the three-body energy is below the atom-dimer threshold ($E<0$ for $\gamma<0$ and $E<-\gamma^2/m$ when $\gamma>0$) all integrals become purely real and the principal-value prescription can be dropped, since no singularity appears in the integrand. Thus $\tilde{t}=\tilde{K}$ in this case.

A three-body bound state exists with the binding energy $B$ when the three-body K-matrix has a pole at $E=-B$. The K-matrix can be  expanded around the position of its singularity
\begin{equation}
\label{eq:K-ZR}
\tilde{K}(q,p;E) = \frac{\tilde{Z}(q,p)}{E+B} +\mathcal{R}(q,p;E),
\end{equation}
where functions $\tilde{Z}$ and $\mathcal{R}$ are defined as the residue and regular part of this expansion. 

In our perturbative approach, the binding energy $B$ is expanded in powers of $r_0$ as
\begin{equation}
B=B_0+r_0 B_1+r_0^2 B_2~.
\end{equation}
We can also expand each term in Eq.~\eqref{eq:K-ZR} in powers of $r_0$:
\begin{eqnarray}
\tilde{K}_0 + r_0\tilde{K}_1 +r_0^2\tilde{K}_2 &=& \frac{\tilde{Z}_0 + r_0\tilde{Z}_1 + r_0^2\tilde{Z}_2}{E+B_0+r_0 B_1+r_0^2 B_2} +\mathcal{R}_0 +r_0\mathcal{R}_1 +r_0^2\mathcal{R}_2
\end{eqnarray}

The next-to-leading-order energy shift is then
\begin{equation}
\label{eq:B1}
B_1 = -\frac{\lim_{E\to -B_0} (E+B_0)^2 \tilde{K}_1(q,p;E)}{\tilde{Z}_0(q,p)},
\end{equation}
which is independent of incoming and outgoing momenta, $q$ and $p$, so we can take $q=p=k$ and calculate $\tilde{K}_1(k,k;E)$ from Eq.~\eqref{eq:K1} just for convenience. In fact, the residue function $\tilde{Z}_0(q,p)$ takes a separable form~\cite{Ji:2011qg}:
\begin{equation}
\label{eq:Z0-Ga}
\tilde{Z}_0(q,p) = \Gamma(q)\Gamma(p)~,
\end{equation}
with $\Gamma(q)$ satisfying a homogeneous integral equation
\begin{equation}
\label{eq:Gamma}
\Gamma(q) = \frac{2}{\pi} \int_0^\Lambda dq' M(q,q';-B_0) \frac{q'^2}{-\gamma+\sqrt{3q'^2/4+mB_0}} \Gamma(q').
\end{equation}
Therefore, $B_1$ can, instead, be calculated from
\begin{eqnarray}
\label{eq:B1-Ga}
B_1 &=& -\frac{1}{\pi} \int_0^\Lambda dq\, q^2\frac{ \gamma+\sqrt{3q^2/4+mB_0}}{-\gamma+\sqrt{3q^2/4+mB_0}} \Gamma^2(q)
-\frac{8\tilde{H}_1(\Lambda)}{(\pi\Lambda)^2} \left[\int_0^\Lambda dq \frac{q^2\, \Gamma(q)}{-\gamma+\sqrt{3q^2/4+mB_0}} \right]^2.
\nn
\end{eqnarray}

Similarly, the N$^2$LO shift of the binding energy is related to residues of the double pole and triple pole in the expansion of Eq.~\eqref{eq:K-ZR}:
\begin{equation}
\label{eq:B2}
B_2 = -\frac{\lim_{E\to -B_0} \left[ (E+B_0)^2 \tilde{K}_2(q,p;E) +(E+B_0)B_1 \tilde{K}_1(q,p;E)\right]}{\tilde{Z}_0(q,p)}.
\end{equation}
In Eq.~(\ref{eq:B2}) $(E+B_0)^2 \tilde{K}_2$ has a pole at $E=-B_0$ which is canceled by the corresponding pole of $(E+B_0)B_1 \tilde{K}_1$. This cancellation is seen explicitly if we derive an expression for $B_2$ analogous to Eq.~\eqref{eq:B1-Ga} for $B_1$.

First, we insert Eq.~\eqref{eq:Z0-Ga} into Eqs.~\eqref{eq:K1}, and expand $(E+B_0)\tilde{K}_1$ as
\begin{eqnarray}
\label{eq:K1-pole}
(E+B_0)\tilde{K}_1(k,k;E) 
&=& 
\frac{\Gamma^2(k)}{\pi(E+B_0)}\left[\int_0^\Lambda dq\, q^2 \frac{\gamma+\sqrt{3q^2/4+mB_0}}{-\gamma+\sqrt{3q^2/4+mB_0}} \Gamma^2(q)\right] 
\nn
&&\hspace{-35mm}
+\frac{\Gamma^2(k)}{E+B_0}\frac{8\tilde{H}_1(\Lambda)}{(\pi\Lambda)^2} \left[\int_0^\Lambda dq\, \frac{q^2\, \Gamma(q)}{-\gamma+\sqrt{3q^2/4+mB_0}} \right]^2
\nn
&&\hspace{-35mm}
+\Gamma(k)\frac{8\tilde{H}_1(\Lambda)}{\pi\Lambda^2}
\left[\int_0^\Lambda dq\, \frac{q^2\, \Gamma(q)}{-\gamma+\sqrt{3q^2/4+mB_0}} \right]
\left[ 1+\frac{2}{\pi}\int_0^\Lambda dq\, \frac{q^2 \mathcal{R}_0(k,q;-B_0)}{-\gamma+\sqrt{3q^2/4+mB_0}}\right]
\nn
&&\hspace{-35mm}
+\frac{2\Gamma(k)}{\pi} \left[\int_0^\Lambda dq\, q^2  \frac{\gamma+\sqrt{3q^2/4+mB_0}}{-\gamma+\sqrt{3q^2/4+mB_0}}\Gamma(q)\mathcal{R}_0(k,q;-B_0)\right]
+\cdots,
\end{eqnarray}
where ellipses indicate terms of $\mathcal{O}\left((E+B_0)^1\right)$, that vanish in the limit $E\rightarrow -B_0$, and the pole at $E=-B_0$ is now explicit.

$(E+B_0)^2\tilde{K}_2$ can be computed using Eqs.~(\ref{eq:K1}), (\ref{eq:K2}), and (\ref{eq:Z0-Ga}) in a similar way, and expanded
 around $E=-B_0$. It too has a first-order pole at $E=-B_0$. Inserting $(E+B_0)^2\tilde{K}_2$'s and $(E+B_0)\tilde{K}_1$'s pole expansions into Eq.~\eqref{eq:B2} and replacing $B_1$ by the expression \eqref{eq:B1-Ga}, we find that the terms singular at $E=-B_0$ cancel, leaving:
\begin{eqnarray}
\label{eq:B2-Ga}
B_2 
&=& -\frac{1}{2\pi} \int_0^\Lambda dq\, q^2\frac{(\gamma+\sqrt{3q^2/4+mB_0}\, )^2}{-\gamma+\sqrt{3q^2/4+mB_0}} \Gamma^2(q)
\nn
&& -\frac{8\tilde{H}_2(-B_0,\Lambda)}{(\pi\Lambda)^2} \left[\int_0^\Lambda dq\, \frac{q^2\, \Gamma(q)}{-\gamma+\sqrt{3q^2/4+mB_0}} \right]^2
\nn
&&\hspace{-10mm}
-\frac{1}{\pi^2} \int\limits_0^\Lambda \int\limits_0^\Lambda dq dq'\, q^2 {q'}^2 
\frac{\gamma+\sqrt{3q^2/4+mB_0}}{-\gamma+\sqrt{3q^2/4+mB_0}}
\frac{\gamma+\sqrt{3q'^2/4+mB_0}}{-\gamma+\sqrt{3q'^2/4+mB_0}}
\Gamma(q) \Gamma(q') \mathcal{R}_0(q,q';-B_0)
\nn
&& -\frac{32\tilde{H}_1^2(\Lambda)}{\pi^3\Lambda^4} 
\left[\int_0^\Lambda dq\, \frac{q^2\, \Gamma(q)}{-\gamma+\sqrt{3q^2/4+mB_0}} \right]^2
\nn
&&\hspace{15mm}
\times \left[ \int_0^\Lambda dq\, \frac{q^2}{-\gamma+\sqrt{3q^2/4+mB_0}}
\left(1+\frac{2}{\pi}\int_0^\Lambda dq' \frac{q'^2 \mathcal{R}_0(q,q';-B_0)}{-\gamma+\sqrt{3q'^2/4+mB_0}}\right)\right]
\nn
&& -\frac{8\tilde{H}_1(\Lambda)}{(\pi\Lambda)^2}
\left[\int_0^\Lambda dq\, \frac{q^2\, \Gamma(q)}{-\gamma+\sqrt{3q^2/4+mB_0}} \right]
\nn
&&\hspace{5mm}
\times \left[ \int_0^\Lambda dq\, q^2 \frac{\gamma+\sqrt{3q^2/4+mB_0}}{-\gamma+\sqrt{3q^2/4+mB_0}} \Gamma(q)\left(1+\frac{2}{\pi}\int_0^\Lambda dq' \frac{q'^2 \mathcal{R}_0(q,q';-B_0)}{-\gamma+\sqrt{3q'^2/4+mB_0}}\right)\right]~,
\nn
\end{eqnarray}
which is finite at $E=-B_0$. This form for $B_2$ shows that, after renormalization in which any significant cutoff dependence in $B_2$ will be cancelled, 
$B_2$ is only a function of the LO binding energy $B_0$. It is independent of incoming and outgoing momenta.

\section{Asymptotic Behavior of LO K-Matrix}
\label{sec:result-lo}

We list several important asymptotic features of the LO K-matrix, $\tilde{K}_0$, in this section, for both half- and fully-off-shell cases.
These properties are essential in the calculation of $H_2(E,\Lambda)$ when the N$^2$LO renormalization is performed. A detailed derivation of these asymptotics is given in Appendix~\ref{app:asymptotic-K0}. 

General features of $\tilde{K}_0$'s asymptotic behavior are renormalization-condition independent. However, the normalization factors of $\tilde{K}_0$, in both half- and fully-off-shell case, are determined by a specific renormalization condition at leading order. To compare analytic forms of $\tilde{K}_0$'s asymptotics with numerical values, we employed a specific physical condition in this section: the atom-dimer scattering length is fixed to $\gamma a_{ad} = 1.5$ at leading order. Fitting the normalization factors and a few other parameters to their numerical values in this physical choice, we can study general features of $\tilde{K}_0$'s asymptotics with high accuracy.

\subsection{Asymptotics of half-shell K-matrix $\tilde{K}_0(k,p;E)$}
\label{sec:result-lo-halfon}
The asymptotic behavior of the half-on-shell $\tilde{K}_0(k,p;E)$ has an expansion in powers of $\gamma/p$ and $mE/p^2$ at large $p$:
\begin{equation}
\label{eq:K0-asymptotic0}
\tilde{K}_0(k,p;E) \propto p^{is_0-1}\left(1 +D_1 \frac{\gamma}{p} + D_2\frac{\gamma^2}{p^2} + C_1 \frac{mE}{p^2} +\cdots \right),
\end{equation}
where ellipses indicate higher-order terms of $\mathcal{O}(\gamma^3/p^3)$ or $\mathcal{O}(m^2E^2/p^2)$. The constants $D_n$ and $C_1$ are computed to be (see Appendix of Ref.~\cite{Bedaque:2002yg} where we have corrected their result for $D_1$ by a factor of 1/3)
\begin{subequations}
\begin{eqnarray}
D_n &=& \left(\frac{2}{\sqrt{3}}\right)^n \frac{I(is_0-n)}{\prod^n_{k=1}
[1-I(is_0-k)]}~,\\
C_1 &=& \frac{\frac{2}{3} I(is_0-2) +L(is_0)}{1-I(is_0-2)}~,
\end{eqnarray}
\end{subequations}
where $I(s)$ and $L(s)$ are functions calculated from Mellin transforms (see Ref.~\cite{Ji:2012} for derivation):
\begin{subequations}
\label{eq:mellin-IL}
\begin{eqnarray}
I(s) &=& \frac{8}{\sqrt{3}s} \frac{ \sin \frac{\pi s}{6}}{\cos \frac{\pi s}{2}},
\\
L(s) &=& -\frac{8}{3}\, \frac{\sin \frac{\pi(s-1)}{6}}{\cos \frac{\pi s}{2}}.
\end{eqnarray}
\end{subequations}

After fixing a renormalization condition at LO, the K-matrix is a real number associated with observables. Therefore, we can rewrite its expression as 
\begin{equation}
\label{eq:K0-asymptotic1}
\tilde{K}_0(k,p;E) = a_{\gamma}(k)\, \left[\frac{1}{p}\phi_0(p)
+\frac{\gamma}{p^2}\phi_1(p) +\frac{\gamma^2}{p^3}\phi_2(p)
+\frac{mE}{p^3}\psi_1(p)+\cdots\right] ,
\end{equation}
with
\begin{subequations}
\label{eq:phi-psi}
\begin{eqnarray}
\phi_n(p) &=& |D_n|\sin\left(s_0 \ln\frac{p}{\bar{\Lambda}} +\arg D_n\right)~,
\\
\psi_1(p) &=& |C_1|\sin\left(s_0 \ln\frac{p}{\bar{\Lambda}} +\arg C_1\right)~.
\end{eqnarray}
\end{subequations}
The normalization factor $a_{\gamma}(k)$ in Eq.~\eqref{eq:K0-asymptotic1} is generally a function of both $\gamma$ and $k$. Since in this paper we only consider a system with a fixed scattering length, $\gamma$ is also fixed and so is treated as a constant parameter. With this constraint in our analysis, $a_{\gamma}$ only varies with the on-shell momentum $k$. 

Furthermore, at the zero-energy threshold, $E=0$, the asymptotics of $\tilde{K}_0(k,p;E)$ can be expanded just in powers of $\gamma/p$, as
\begin{equation}
\label{eq:K0-asymptotic2}
\tilde{K}_0(k,p;E) = a_{\gamma}(k)\,
\sum^\infty_{n=0}\frac{\gamma^n}{p^n}|D_n|\sin\left(s_0
\ln\frac{p}{\bar{\Lambda}} +\arg D_n\right),
\end{equation}
where $k=\frac{2\gamma}{\sqrt{3}}$ at $E=0$. 

\subsection{Asymptotics of fully-off-shell K-matrix $\tilde{K}_0(q,p;E)$}
\label{sec:result-lo-fulloff}
When the incoming and outgoing momenta are both large, i.e. $q,p\gg k,\gamma$, the fully-off-shell K-matrix $\tilde{K}_0(q,p;E)$'s asymptotic behavior is dominated by the leading piece, $\sim 1/(qp)$, in the expansion at large $p$ and $q$. Here, we focus on the behavior of this leading term and do not consider sub-leading terms in the expansion.

Using power-counting analysis and numerical fitting (see also Ref.~\cite{Griesshammer:2005}) we find that $\tilde{K}_0(q,p;E)$ at large momenta $p,q \gg k,\gamma$ is dominated by
\begin{equation}
\label{eq:K0-as-qp0}
\tilde{K}_0(q,p;E) = \frac{1}{qp}\left[\mathcal{F}_0(q,p) + \mathcal{G}_0(q,p;E)\right]~.
\end{equation}
The energy-independent part of $\tilde{K}_0$ is represented by the function $\mathcal{F}_0$, which obeys
\begin{equation}
\label{eq:F0-as}
\mathcal{F}_0(q,p)= 
\begin{cases}
b_{\gamma}\, \phi_0(q)\, \bar{\phi}_0(p) + \rho(q,p), &
\mbox{if }(q> p) \gg k,\gamma\\
b_{\gamma}\, \bar{\phi}_0(q)\, \phi_0(p) + \rho(p,q), &
\mbox{if }(p> q) \gg k,\gamma
\end{cases}~,
\end{equation}
where $b_{\gamma}$ is a constant since $\gamma$ is fixed  (its value is determined by the LO renormalization condition), $\phi_0$ is defined in Eq.~\eqref{eq:phi-psi}, and $\bar{\phi}_0$ is defined as
\begin{equation}
\label{eq:phi-bar}
\bar{\phi_0}(p) = \cos\left(s_0\ln\frac{p}{\bar{\Lambda}}\right)~.
\end{equation}
The energy-dependent part of $\tilde{K}_0$ is written as a function $\mathcal{G}_0$:
\begin{equation}
\label{eq:G0-as}
\mathcal{G}_0(q,p;E) = c_{\gamma}(E)\, \phi_0(q) \phi_0(p)~,
\end{equation}
where the factor $c_{\gamma}(E)$ only depends on $E$ at fixed $\gamma$, and is also determined by the LO renormalization condition.

The product $\phi_0 \bar{\phi}_0$ in function $\mathcal{F}_0$, together with $\phi_0^2$ in function $\mathcal{G}_0$, dominate off-diagonal (i.e. $p\gg q$ or $p\ll q$) elements of the LO K-matrix:  analysis of the modified STM equation \eqref{eq:t0} shows that $\rho$ tends to zero if $q\gg p$ or $p\gg q$. The function $\rho$
thus represents a remainder, which plays an important role in the near-diagonal (i.e. $p\sim q$) K-matrix elements. The remainder function depends on the renormalization condition and we have only determined how to numerically fit it to a chosen parametrization.

\begin{figure}[tbpic:Kq0p]
\centerline{\includegraphics[width=12cm,angle=0,clip=true]{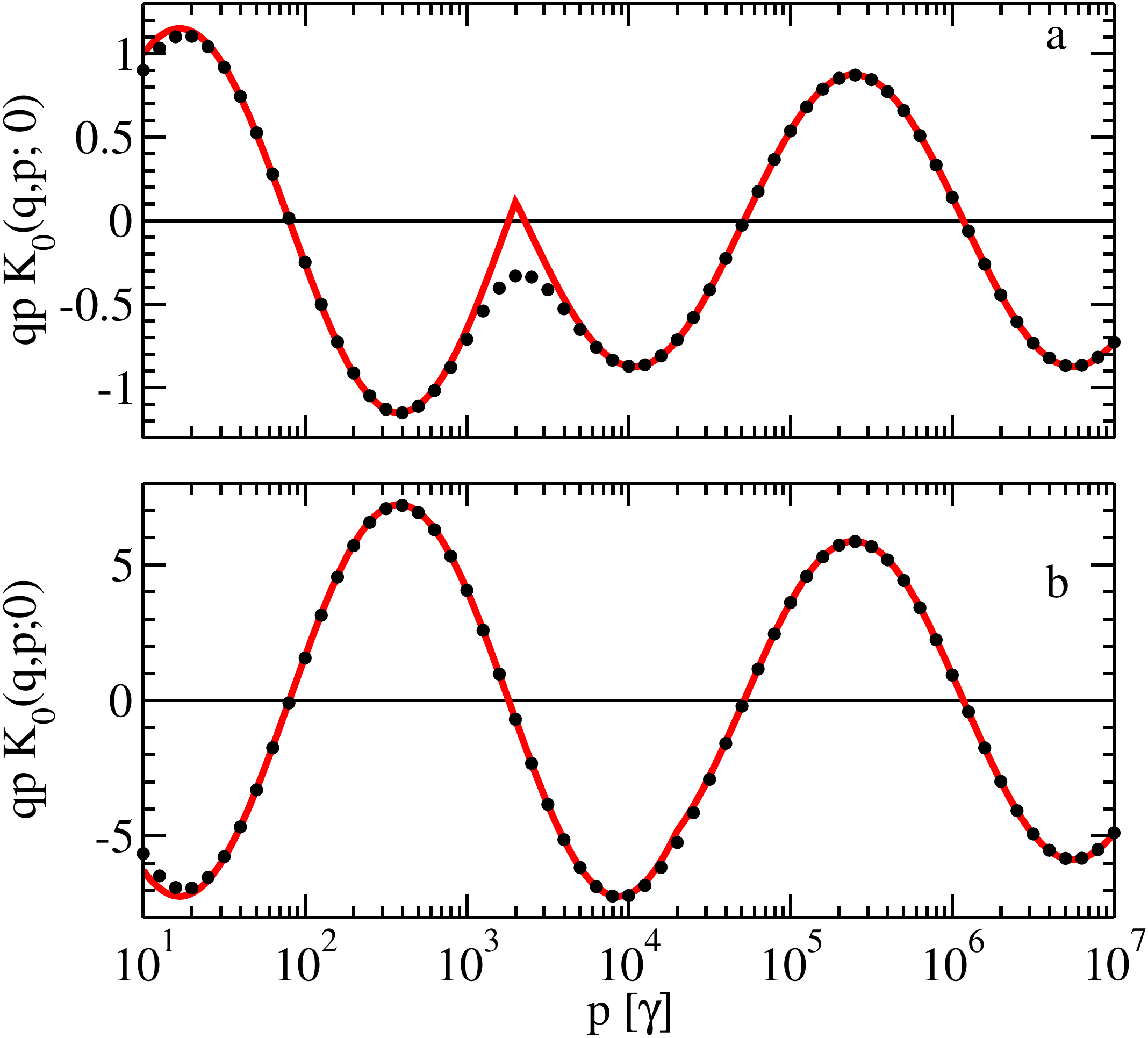}}
\caption{Amplitudes $q p\tilde{K}_0(q,p;0)$ as functions of the momentum $p$ at fixed values of $q$: $q= 2\times10^3 \gamma$ (upper panel, a) and $q=2\times10^4 \gamma$ (lower panel, b). Dots
are the numerical results. Solid lines (red) represent the analytic function defined by the first term of Eq.~\eqref{eq:K0-as-E0}. In each case, $\tilde{H}_0(\Lambda)$ is chosen to reproduce $\gamma a_{ad}=1.5$.}
\label{pic:Kq0p}
\end{figure}

Here we will verify each piece of the approximate form of the fully-off-shell $\tilde{K}_0$ in Eqs.~(\ref{eq:K0-as-qp0}-\ref{eq:G0-as}) by comparing these results with numerical calculations. In these calculations, we first fix the LO atom-dimer scattering length  to be $\gamma a_{ad} =1.5$, and then calculate $\tilde{K}_0$ at $E=0$. 

Firstly, at $E=0$, we can combine functions $\mathcal{F}_0$ and $\mathcal{G}_0$ and express $\tilde{K}_0$ as
\begin{equation}
\label{eq:K0-as-E0}
\tilde{K}_0(q,p;0)= 
\begin{cases}
\phi_0(q)\, \chi_0(p) + \rho(q,p), &
\mbox{if }(q> p) \gg \gamma\\
\chi_0(q)\, \phi_0(p) + \rho(p,q), &
\mbox{if }(p> q) \gg \gamma
\end{cases}~,
\end{equation}
where $\chi_0$ obeys
\begin{equation}
\chi_0(p) = \sqrt{b_{\gamma}^2 + c_{\gamma}^2(0)}\, \sin\left(s_0\ln\frac{p}{\bar{\Lambda}} + \arctan \frac{b_{\gamma}}{c_{\gamma}(0)}\right)~.
\end{equation}

We show in Fig.~\ref{pic:Kq0p} that the first term, proportional to $\phi_0(q) \chi_0(p)$ or $\chi_0(q) \phi_0(p)$, in our approximation~\eqref{eq:K0-as-E0} agrees well
with the numerical results when $q$ and $p$ are not close to each other. Here, in order to compare with Eq.~\eqref{eq:K0-as-E0} for $\tilde{K}_0$, we performed a best fit to the amplitude and phase of the function $\chi_0$ within an accuracy of $10^{-3}$, which determines $b_{\gamma} = -2.012$ and $c_{\gamma}(0)=-8.612$. Meanwhile, $\bar{\Lambda}$ is parametrized in Eq.~\eqref{eq:a3-Lbar}, and, when fitted, turns out to be $\bar{\Lambda}=4.421 \gamma$ (a $0.15\gamma$ correction to the result from Ref.~\cite{Braaten:2003yc}, to obtain better accuracy).

\begin{figure}[tbpic:Rqp-num]
\centerline{\includegraphics[width=12cm,angle=0,clip=true]{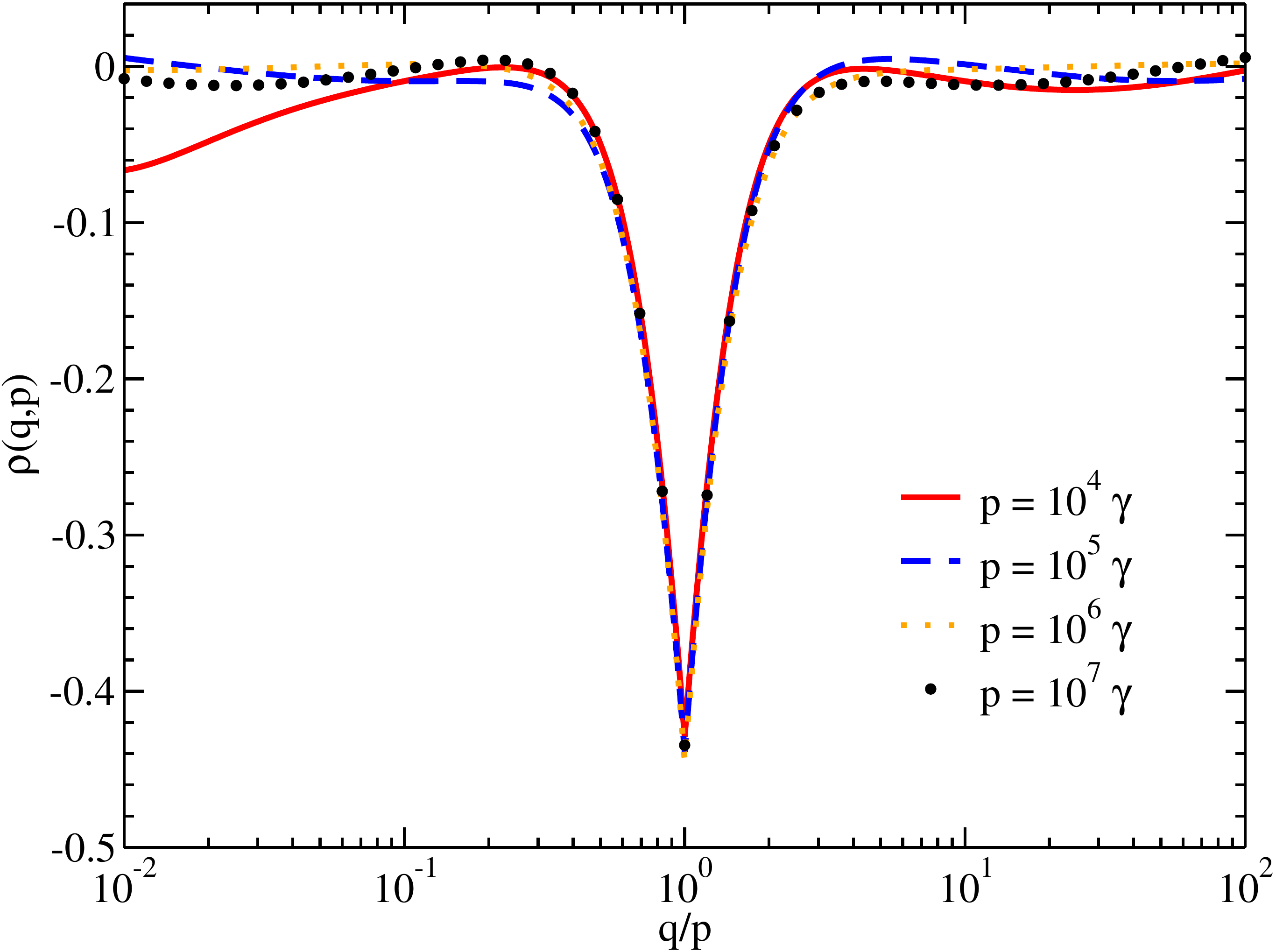}}
\caption{Remainder $\rho(q,p)$ as a function of the ratio $q/p$, with each curve
representing a different value of $p$. As in Fig.~\ref{pic:Kq0p}, the results are for a renormalization condition $\gamma a_{ad}=1.5$. } 
\label{pic:Rqp-num}
\end{figure}

The remainder function $\rho(q,p)$ describes discrepancies between the numerical
values of $\tilde{K}_0(q,p;0)$ and this off-diagonal part (i.e. $\phi_0(q) \chi_0(p)$ or $\chi_0(q) \phi_0(p)$). Since the final expression for $\tilde{K}_0(q,p;0)$ displays discrete-scale invariance we choose a form for $\rho(q,p)$ which depends only on the ratio of $q/p$.  And indeed, we show in
Fig.~\ref{pic:Rqp-num} that $\rho(q,p)$ depends only explicitly on the ratio $q/p$ if $q/p\sim1$ (the same parameters used in Fig.~\ref{pic:Kq0p} were chosen in producing Fig.~\ref{pic:Rqp-num}).
We  must also respect the symmetry under the exchange of incoming and outgoing momenta ($\tilde{K}_0(q,p;E) = \tilde{K}_0(p,q;E)$). Hence we fit $\rho$
to the approximate form: \begin{equation}
\label{eq:R-x}
\rho(x) = \alpha x e^{-\beta/ x}; \hspace{1cm}
x =
\begin{cases}
q/p, & \mbox{if } q<p\\
p/q, & \mbox{if } q>p
\end{cases}
\end{equation}
where $\alpha$ and $\beta$ are renormalization-scheme dependent. The particular function employed in Eq.~(\ref{eq:R-x}) has no theoretical justification and is chosen solely for its ability to give a good fit. 
Once the same parameters as in Fig.~\ref{pic:Kq0p} are chosen, we find $\alpha=-1.665$ and $\beta=1.322$, which are fitted at the region $x\sim 1$ within an accuracy of $10^{-3}$. 

\begin{figure}[tbpic:Rqp-anl]
\centerline{\includegraphics[width=12cm,angle=0,clip=true]{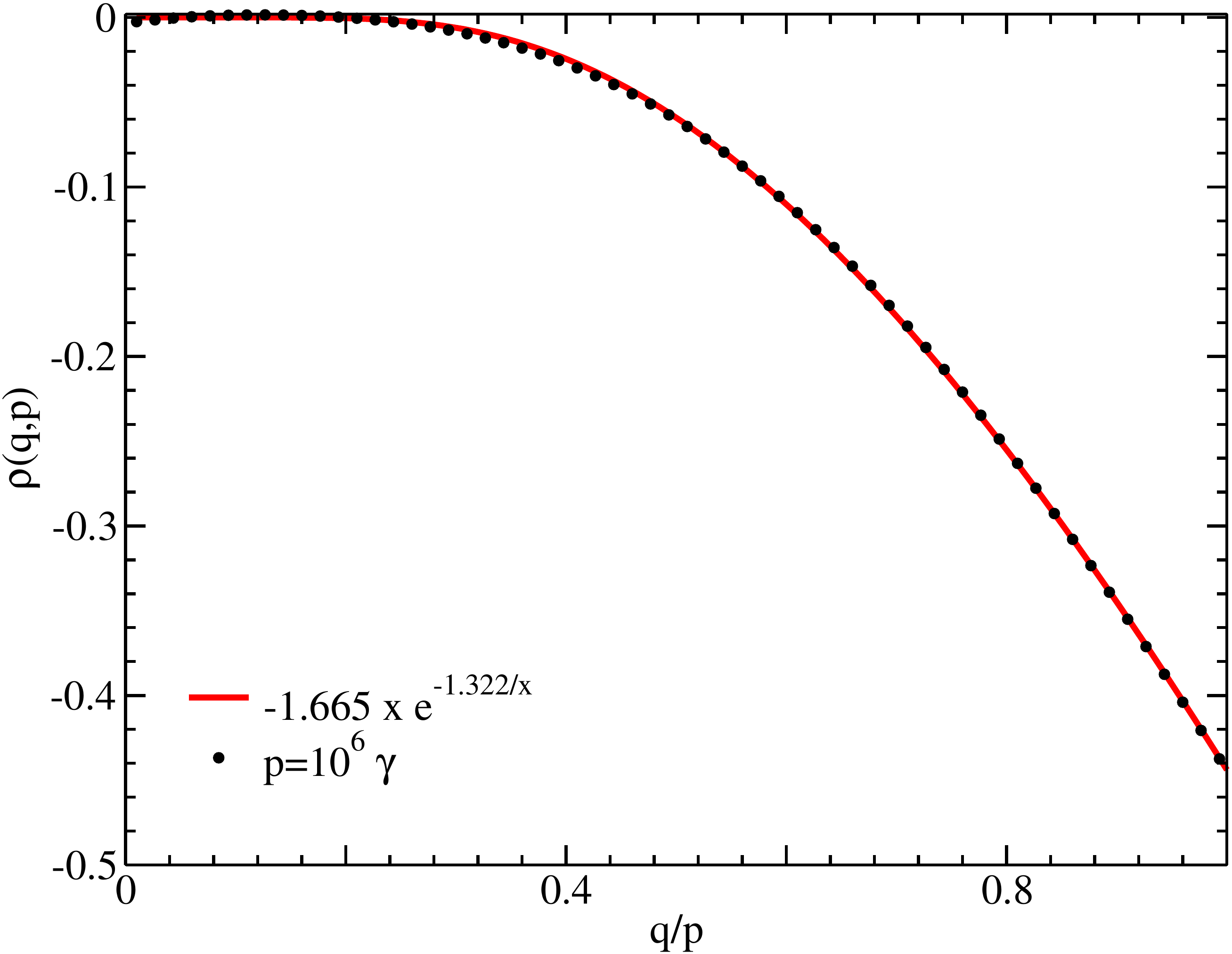}}
\caption{Remainder $\rho(q,p)$ as a function of the ratio $q/p$, where $p$ is fixed to $10^{6}\gamma$. The numerical result (dots) is renormalized to reproduce $\gamma a_{ad}=1.5$. The analytic expression (red solid line) is a fit~\eqref{eq:R-x} to the numerical result.}
\label{pic:Rqp-anl}
\end{figure}

In Fig.~\ref{pic:Rqp-anl}, the numerical result of $\rho(q,p)$ is plotted as a function of $q/p$ at a fixed value of $p$, and it agrees well with our analytic formula with the parameters given above. As suggested in Fig.~\ref{pic:Rqp-num}, such an approximation loses accuracy if $q/p$ is not close to $1$. We show in Sec.~\ref{sec:renorm-h20} that the effect of this on the determination of the N$^2$LO three-body-force parameter, $h_{20}$, is small.

\begin{figure}[tbpic:c-gamma-E]
\centerline{\includegraphics[width=12cm,angle=0,clip=true]{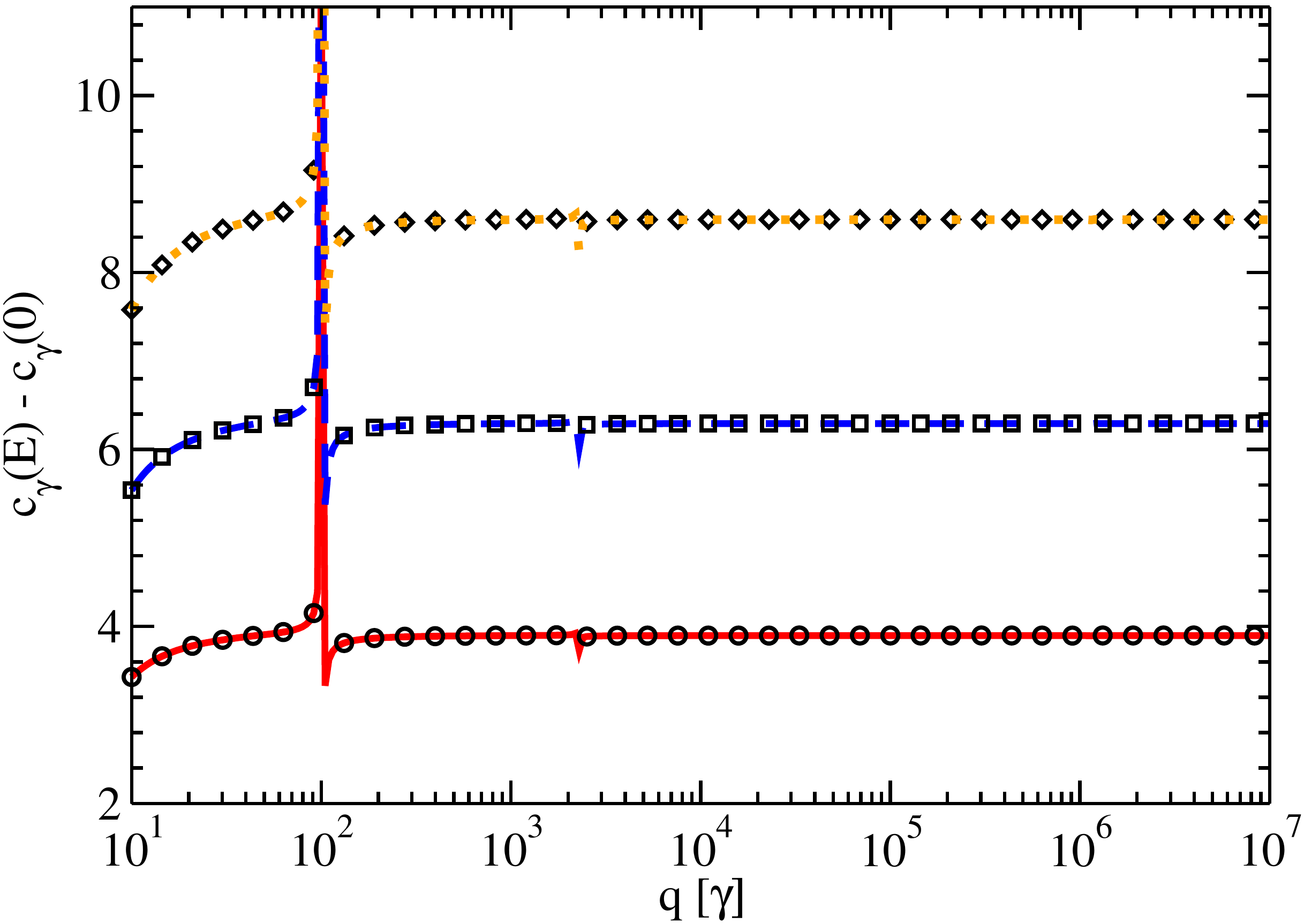}}
\caption{$c_{\gamma}(E)-c_{\gamma}(0)$ as a function of $q$ at different fixed values of $p$ and $E$. When $p$ is fixed to $10^{3}\gamma$, three lines respectively represent the cases for $mE=-\gamma^2/3$ (red solid line), $mE=-2\gamma^2/3$ (blue dashed line) and $mE=-\gamma^2$ (orange dotted line). When $p$ is fixed to $10^4\gamma$, dots represent the cases for $mE=-\gamma^2/3$ (circles), $mE=-2\gamma^2/3$ (squares) and $mE=-\gamma^2$ (diamonds). 
Results are renormalized to reproduce $\gamma a_{ad} =1.5$.}
\label{pic:c-gamma-E}
\end{figure}

Furthermore, we can also obtain $c_{\gamma}(E)$'s value with simple algebra based on Eq.~\eqref{eq:K0-as-qp0}:
\begin{equation}
\label{eq:c-gamma}
c_{\gamma}(E)=c_{\gamma}(0) + \frac{\tilde{K}_0(q,p;E) - \tilde{K}_0(q,p;0)} {\phi_0(q)\phi_0(p)}. 
\end{equation}
Assuming our approximation is correct, we should observe that $c_{\gamma}$ is $q$- or $p$-independent when we insert numerical results for $\tilde{K}_0$ into Eq.~\eqref{eq:c-gamma}. The outcome of this test is shown in Fig.~\ref{pic:c-gamma-E}, where we plotted the second term of Eq.~\eqref{eq:c-gamma} at different values of $q$, $p$ and $E$. It shows that Eq.~\eqref{eq:c-gamma} is a function of $E$ alone: it does not change when $q$ or $p$ varies, provided that $q~\simge~10^2 \gamma$. The divergence below this value is due to higher-order corrections to $\tilde{K}_0$ (i.e. $\mathcal{O}(\gamma/q,\gamma/p)$), which shift the numerator in Eq.~\eqref{eq:c-gamma} from zero at $\phi_0(q)=0$ or $\phi_0(p)=0$. 

We also expect that the near-diagonal part of $\tilde{K}_0$, $\rho(q,p)$, is energy independent when only the dominant part of $\tilde{K}_0$ is considered. Our expectation is supported by our findings in Fig.~\ref{pic:c-gamma-E}, since no bump is observed in the curves at $q=p=10^3\gamma$ or $q=p=10^4\gamma$.

In this way, we describe the asymptotic behavior of the dominant part ($\sim 1/(qp)$) of the fully-off-shell $\tilde{K}_0(q,p;E)$.

\section{Cutoff dependence and renormalization}
\label{sec:renormalization}
In the absence of additional renormalization, $\tilde{K}_2$ has cutoff dependence $\sim\Lambda^2$ at N$^2$LO, which renders unrenormalized results from N$^2$LO
calculations unphysical. A sub-leading short-range piece of the three-body/atom-dimer interaction, here denoted $\tilde{H}_2$, is thus needed at N$^2$LO to
absorb this cutoff dependence. Since $\tilde{H}_2$ contains powers up to $\mathcal{O}(\Lambda^2)$, we write the N$^2$LO
three-body force parameter $\tilde{H}_2$ as
\begin{equation}
\label{eq:H2}
\tilde{H}_2 = \Lambda^2 h_{20}(\Lambda) + mE  h_{22}(\Lambda),
\end{equation}
In this section, we study both the leading piece of $\tilde{H}_2$, $h_{20}$, and its
subleading piece $h_{22}$.
$\gamma$-dependent pieces of this three-body force are absorbed in $h_{20}$ as a
fine-tuning modification in problems with a fixed-but-finite $\gamma$. Renormalization must be carried out with such terms made explicit if we want to study systems with varying $\gamma$~\cite{Ji:2010su,Ji:2011qg}.

\subsection{Relevant integrals}
\label{sec:renorm-int}
The principal-value part of the integration in $\mathcal{L}(k)$, with $t_0$ replaced by $\tilde{K}_0$, appears repeatedly in the LO, NLO and N$^2$LO perturbative calculation, i.e. in Eqs. (\ref{eq:K0}, \ref{eq:K1}, \ref{eq:K2}). Here we denote the integral itself (with the infrared regularization dropped) by $\xi$: 
\begin{equation}
\xi \equiv \frac{2}{\pi} \int^\Lambda dq\, \frac{q^2 \tilde{K}_0(k,q)}{-\gamma + \sqrt{3 q^2/4 - mE}},
\end{equation}
where, as usual, a principal-value prescription is to be understood as the means of dealing with any pole that is present in the integrand.
Upon inserting the asymptotic form of $\tilde{K}_0(k,q)$,~\eqref{eq:K0-asymptotic2}, up to terms $\sim 1/q^3$, we then have:
\begin{eqnarray}
\label{eq:int1}
\xi &=&
\frac{4}{\sqrt{3}\pi}a_{\gamma}(k)\,\int^{\Lambda} dq
\left(1+\frac{2}{\sqrt{3}}\frac{\gamma}{q}
+\frac{4}{3}\frac{\gamma^2}{q^2} +\frac{2}{3}\frac{mE}{q^2}\right)
\left(\phi_0
+\frac{\gamma}{q}\phi_1 +\frac{\gamma^2}{q^2}\phi_2
+\frac{mE}{q^2}\psi_1\right).
\end{eqnarray}

In order to simplify the notation, we define integrations that contain functions $\phi_n$ as
\begin{equation}
\varPhi^{(n)}_{m}(\Lambda) = \frac{1}{\pi}\int^\Lambda dq\, q^{m-1}\phi_n.
\end{equation}
We can also define
\begin{equation}
\varPsi_{m}(\Lambda) = \frac{1}{\pi}\int^\Lambda dq\, q^{m-1} \psi_1.
\end{equation}

Since $\phi_n$ and $\psi_n$ are $\mathcal{O}(1)$, the subscript $m$ of the defined functions $\varPhi^{(n)}_{m}$ and $\varPsi_{m}$ indicates they have cutoff dependence, $\sim\Lambda^m$, except that $m=0$ indicates the possible presence of both terms $\sim\ln \Lambda$ and $\mathcal{O}(1)$. Eq.~\eqref{eq:int1} can then be written as an expansion in $\gamma / \Lambda$ and $mE/\Lambda^2$
\begin{eqnarray}
\label{eq:xi-serie}
\xi &=& \frac{4a_{\gamma}(k)\,}{\sqrt{3}} 
\left[\varPhi^{(0)}_1 + \gamma \left(\frac{2}{\sqrt{3}}\varPhi^{(0)}_0 +\varPhi^{(1)}_0\right)
+\gamma^2\left(\frac{4}{3}\varPhi^{(0)}_{-1}+\frac{2}{\sqrt{3}}\varPhi^{(1)}_{-1}+\varPhi^{(2)}_{-1}\right)\right.
\nn
&&\left.+mE\left(\frac{2}{3}\varPhi^{(0)}_{-1} +\varPsi_{-1}\right)\right].
\end{eqnarray}
The infrared regularization of some integrals (e.g. $\varPhi^{(0)}_1$, $\varPhi^{(0)}_0$ and $\varPhi^{(1)}_0$) in~\eqref{eq:xi-serie} could affect the $\mathcal{O}(1)$ pieces of $\mathcal{L}$. However, we have numerically verified that, when computing ${\mathcal L}(k)$, the combination of infrared parts of these integrals is canceled by the ``$1$" term in~\eqref{eq:math-L}. This cancelation yields $\mathcal{L}(k)=\xi$.

When $\gamma$ is fixed, the $\gamma$-dependent terms in Eq.~\eqref{eq:xi-serie} can be excluded in the analysis of cutoff dependence, since they can be absorbed by fine tuning of $\tilde{H}_0$, $h_{10}$ and $h_{20}$ respectively in the LO, NLO and N$^2$LO renormalization. In the rest of this section, we do not consider such $\gamma$-dependent terms. Therefore, of all the functions $\phi_n$, only $\phi_0$ is needed for our purposes here. We can thus further simplify our notation for $\varPhi^{(n)}_{m}$ by dropping the superscript $(n)$:
\begin{equation}
\varPhi_{m}(\Lambda) = \frac{1}{\pi}\int^\Lambda dq\, q^{m-1}\phi_0.
\end{equation}
$\xi$'s expression is then reduced to
\begin{equation}
\label{eq:xi-fixg}
\xi = \frac{4a_{\gamma}(k)\,}{\sqrt{3}} 
\left[\varPhi_1
+mE\left(\frac{2}{3}\varPhi_{-1} +\varPsi_{-1}\right)\right],
\end{equation}
where each integral in Eq.~\eqref{eq:xi-fixg} is calculated in Appendix~\ref{app:integrals}. $\varPhi_1$ is used for calculating $h_{20}$, and the addition of terms $\sim mE$ is needed for the calculation of $h_{22}$.

For later convenience, here we also define some other integrals that are needed in the rest of this section. Integrals containing $\bar{\phi}_0$ are defined similarly to $\Phi_n$:
\begin{equation}
\bar{\varPhi}_n(\Lambda) = \frac{1}{\pi}\int^\Lambda dq\, q^{n-1} \bar{\phi}_0(q)~.
\end{equation}
Analytic expressions for the $\varPhi_n$ and $\bar{\varPhi}_n$ are given in Appendix~\ref{app:integrals}.

Integrals that contain a product of two functions, e.g. $\phi_0$ and $\bar{\phi}_0$, are defined by
\begin{equation}
\mathcal{U}_{\phi \bar{\phi}, n}(\Lambda) = \frac{1}{\pi}
\int^\Lambda dq\, q^{n-1} \phi_0(q) \bar{\phi}_0(q),
\end{equation}
where the subscript $n$ also indicates $\mathcal{U}$'s power-law dependence on $\Lambda$. Analytic expressions for $\mathcal{U}_{\phi \bar{\phi}, n}$ and $\mathcal{U}_{\phi \phi, n}$ are calculated in Appendix~\ref{app:integrals}.

The double integrals are defined in a similar way, for example:
\begin{subequations}
\begin{eqnarray}
\mathcal{W}_{\phi\phi(\bar{\phi}), m+n} (\Lambda)
&=& \frac{1}{\pi} \int^\Lambda dq\, q^{m-1}  \phi_0^2(q) \bar{\varPhi}_n(q)
\nn
&=& 
\frac{1}{\pi^2}\int^\Lambda dq\, q^{m-1}  \phi_0^2(q) \int^q dp\, p^{n-1} \bar{\phi}_0(p),
\\
\mathcal{W}_{\phi(\phi\bar{\phi}), m+n} (\Lambda)
&=& \frac{1}{\pi}\int^\Lambda dq\, q^{m-1}  \phi_0(q) \mathcal{U}_{\phi \bar{\phi}, n}(q)
\nn
&=& 
\frac{1}{\pi^2}\int^\Lambda dq\, q^{m-1}  \phi_0(q) \int^q dp\, p^{n-1} \phi_0(p) \bar{\phi}_0(p),
\end{eqnarray}
\end{subequations}
where $m+n$ denotes the integrals' overall powers of $\Lambda$. In Appendix~\ref{app:integrals}, we write the expressions for $\mathcal{W}_{\phi(\bar{\phi}), 1+1}$, $\mathcal{W}_{\phi\phi(\bar{\phi}), 1+1}$, $\mathcal{W}_{\phi(\phi\bar{\phi}), 1+1}$ and $\mathcal{W}_{\phi\phi(\bar{\phi}\bar{\phi}), 1+1}$, which are all shown to be $\sim\Lambda^2$.

Another type of double integral, that contains the near-diagonal function $\rho(q,q')$, is defined by
\begin{equation}
\Omega_{\phi\rho\phi}(\Lambda) =\frac{1}{\pi^2}\int\limits^\Lambda \int\limits^\Lambda dq dq' \, \phi(q)\rho(q,q')\phi(q'),
\label{eq:Omegadef}
\end{equation}
where the $q$-dependent ($q'$-dependent) function appears on the left (right) side of $\rho$ in the subscript. These integrals will be calculated directly in the next subsection. We also will use the notation $\Omega_{\rho \phi}$, $\Omega_{\phi \rho}$ and $\Omega_{\rho}$. If $\phi$ does not appear in the designated location in a $\Omega$ subscript then it is to be replaced by ``1" in the integrand in Eq.~(\ref{eq:Omegadef}). 

\subsection{The dominant cutoff dependence and $h_{20}$} 
\label{sec:renorm-h20}

We now calculate the dominant cutoff dependence in each of the terms in Eq.~\eqref{eq:K2} categorized from A to E. Inserting the leading piece of Eq.~\eqref{eq:K0-asymptotic1}, we see that the first term, (A), has dominant cutoff dependence:
\begin{equation}
\tilde{K}_2^A
= a_{\gamma}^2(k)\, \frac{\sqrt{3}}{4\pi} \int^\Lambda dq\, q \phi_0^2(q)
= a_{\gamma}^2(k)\, \frac{\sqrt{3}}{4} \mathcal{U}_{\phi\phi,2}(\Lambda).
\end{equation}

After inserting the dominant pieces of $\xi$ from Eq.~\eqref{eq:xi-fixg} and the dominant piece of $\tilde{H}_2(E,\Lambda)$ from Eq.~(\ref{eq:H2}) we find that the second term in $\tilde{K}_2$, $\tilde{K}_2^B$ (the part of $\tilde{K}_2$ that results from the insertion of the N$^2$LO 3-body force) behaves as
\begin{equation}
\tilde{K}_2^B 
= \frac{2\Lambda^2 h_{20}(\Lambda)}{\Lambda^2} \xi^2
=a_{\gamma}^2(k)\,\frac{32}{3} h_{20}(\Lambda) \varPhi_1^2(\Lambda).
\end{equation}

By inserting the off-shell $\tilde{K}_0$'s asymptotic form~\eqref{eq:K0-as-qp0}, we find that the  third term from two insertions of the NLO piece of the dimer propagator (C) is calculated as
\begin{eqnarray}
\label{eq:1d1d}
\tilde{K}_2^C 
&=& \frac{a_{\gamma}^2(k)\,}{\pi^2} \int\limits^\Lambda \int\limits^\Lambda dq dq'\, \phi_0(q) \phi_0(q') \left[\mathcal{F}_0(q,q') +\mathcal{G}_0(q,q')\right]
\nn
&=& a_{\gamma}^2(k)\,c_{\gamma}(E)\, \left[\frac{1}{\pi}\int^\Lambda dq \phi_0^2(q)\right]^2
+
\frac{a_{\gamma}^2(k)\,}{\pi^2} \int\limits^\Lambda \int\limits^\Lambda dq dq' \phi_0(q) \phi_0(q') \rho(q,q')
\nn
&&+ \frac{a_{\gamma}^2(k)\,b_{\gamma}}{\pi^2} \left( \int^\Lambda dq \phi_0^2(q)\int^q dq' \phi_0(q')\bar{\phi}_0(q') + \int^\Lambda dq \phi_0(q) \bar{\phi}_0(q) \int^\Lambda_q dq' \phi_0^2(q') \right)
\nn
&=& a_{\gamma}^2(k)\,\left[c_{\gamma}(E)\, \mathcal{U}_{\phi \phi,1}^2(\Lambda)
+2b_{\gamma} \mathcal{W}_{\phi\phi(\phi\bar{\phi}),1+1}(\Lambda) + \Omega_{\phi\rho\phi}(\Lambda)\right].
\end{eqnarray}
where we used the fact that $\int^\Lambda dq \int^\Lambda_q dq' = \int^\Lambda dq' \int^{q'} dq$ in the last step.

The fourth term (D), from double insertions of the NLO piece of the 3-body force, is
\begin{eqnarray}
\label{eq:1t1t}
\tilde{K}_2^D
&=&\frac{8\Lambda^2 h_{10}^2(\Lambda)}{\Lambda^4} \xi^2
\frac{2}{\sqrt{3}\pi}\int^\Lambda dq\, \left(q+\frac{4}{\sqrt{3}\pi}\int^\Lambda dq' \left[ \mathcal{F}_0(q,q')+\mathcal{G}_0(q,q')\right]\right)
\nn
&=& \left[\frac{32 h_{10}(\Lambda)}{3\Lambda}\right]^2 
a_{\gamma}^2(k)\,\varPhi_1^2(\Lambda) 
\left\lbrace\frac{\sqrt{3}}{4\pi}\int^\Lambda q dq
+ c_{\gamma}(E)\, \left[\frac{1}{\pi}\int^\Lambda dq \phi_0^2(q)\right]^2\right.
\nn
&&\left. 
+ \frac{2b_{\gamma}}{\pi^2} \int^\Lambda dq \phi_0(q) \int^q dq' \bar{\phi}_0(q') 
+ \frac{1}{\pi^2} \int\limits^\Lambda \int\limits^\Lambda dq dq' \rho(q,q')\right\rbrace
\nn
&=&\left[\frac{32 h_{10}(\Lambda)}{3\Lambda}\right]^2 
a_{\gamma}^2(k)\,\varPhi_1^2(\Lambda) 
\nn
&&\times\left[
c_{\gamma}(E)\, \varPhi_1^2(\Lambda) +\frac{\sqrt{3}\Lambda^2}{8\pi}
+ 2b_{\gamma}\mathcal{W}_{\phi(\bar{\phi}),1+1}(\Lambda)
+\Omega_{\rho}(\Lambda)\right],
\end{eqnarray}
where the leading piece of $\tilde{H}_1$, $\Lambda h_{10}(\Lambda)$, is inserted in the calculation. 

Finally, the last term (E), from the insertion of one NLO dimer and one NLO 3-body force, is
\begin{eqnarray}
\label{eq:1d1t}
\tilde{K}_2^E
&=& \frac{4\Lambda h_{10}(\Lambda)}{\Lambda^2} \xi
\frac{a_{\gamma}(k)\,}{\pi}\int^\Lambda dq\, \phi_0(q) \left(q+\frac{4}{\sqrt{3}\pi}\int^\Lambda dq' \left[ \mathcal{F}_0(q,q')+\mathcal{G}_0(q,q')\right]\right)
\nn
&=& \frac{64 h_{10}(\Lambda)}{3\Lambda} a_{\gamma}^2(k)\, \varPhi_1(\Lambda)
\left\lbrace
\frac{\sqrt{3}}{4\pi}\int^\Lambda dq\, q \phi_0(q)
+\frac{c_{\gamma}(E)}{\pi^2} \left[\int^\Lambda dq \phi_0^2(q)\right]
\left[\int^\Lambda dq\, \phi_0(q)\right]
\right.
\nn
&&+\frac{1}{\pi^2}\int^\Lambda dq \phi_0(q) \int^\Lambda dq' \rho(q,q')
\nn
&&\left.
+\frac{b_{\gamma}}{\pi^2} \left[\int^\Lambda dq \phi_0^2(q) \int^q dq' \bar{\phi}_0(q')
+\int^\Lambda dq \phi_0(q) \int^q dq' \phi_0(q') \bar{\phi}_0(q')\right]
\right\rbrace
\nn
&=&\frac{64 h_{10}(\Lambda)}{3\Lambda} a_{\gamma}^2(k)\, \varPhi_1(\Lambda)
\left\lbrace c_{\gamma}(E)\, \varPhi_1(\Lambda)\mathcal{U}_{\phi\phi,1}(\Lambda)
+\frac{\sqrt{3}}{4}\varPhi_2(\Lambda)
\right. 
\nn
&&\left.
+b_{\gamma}[\mathcal{W}_{\phi\phi(\bar{\phi}),1+1}(\Lambda)+\mathcal{W}_{\phi(\phi\bar{\phi}),1+1}(\Lambda)]
+\Omega_{\phi\rho}(\Lambda) \frac{}{} \right\rbrace.
\end{eqnarray}

The infrared regularization of each integral in $\tilde{K}_2^A$-$\tilde{K}_2^E$ can in fact contribute to the $\mathcal{O}(1)$ part of $\tilde{K}_2$; however, their contributions are two orders smaller in powers of $\Lambda$ compared to the dominant cutoff-dependent pieces of $\tilde{K}_2$. We therefore drop the infrared regularization in each integral for our convenience. By doing so, we treat expressions for these indefinite integrals (calculated in Appendix~\ref{app:integrals}) as exact results, with unsolved infrared pieces absorbed in the renormalized $\tilde{K}_2$. 

In order to simplify the calculation, we regroup integrations
\begin{equation}
\tilde{K}_2^A +\tilde{K}_2^C + \tilde{K}_2^D + \tilde{K}_2^E 
= 
a_{\gamma}^2(k)\, ( \Sigma_{E} + \Sigma_{1D} + \Sigma_{2D} +\Sigma_{\rho}),
\end{equation}
where $\Sigma_{E}$ indicates the summation of terms proportional to $c_{\gamma}(E)$, $\Sigma_{1D}$ sums up terms that contain a $1$D integral, $\Sigma_{2D}$ denotes the summation of $2$D integrals $\mathcal{W}$, and $\Sigma_{\rho}$ refers to the summation of integrals including $\rho(q,q')$.

We first sum up the three terms $\propto c_{\gamma}(E)$ (in $\tilde{K}_2^C$, $\tilde{K}_2^D$ and $\tilde{K}_2^E$) by
\begin{equation}
\Sigma_{E} = c_{\gamma}(E)\,  \left( \mathcal{U}_{\phi \phi,1}(\Lambda)
+ \frac{32 h_{10}(\Lambda)}{3\Lambda}\varPhi_1^2(\Lambda)\right)^2~,
\end{equation}
The $\Lambda$-dependence here is canceled in the NLO renormalization, because $h_{10}$ is tuned at NLO to make the $\sim\Lambda$ pieces in the square bracket cancel with each other: 
\begin{eqnarray}
\label{eq:h10-uphi}
\frac{32 h_{10}(\Lambda)}{3\Lambda} &=& -\frac{\mathcal{U}_{\phi \phi,1}(\Lambda)}{\varPhi_1^2(\Lambda)}
\nn
&=& -\frac{\pi}{2\Lambda}\, 
\frac{(1+s_0^2)}{\sqrt{1+4s_0^2}}\frac{\sqrt{1+4s_0^2} 
- \cos \left(2 s_0 \ln (\Lambda/\bar{\Lambda}) - \arctan 2s_0\right)} {\sin^2
\left( s_0 \ln (\Lambda/\bar{\Lambda}) - \arctan s_0\right)}~,
\end{eqnarray}
which agrees with $h_{10}(\Lambda)$'s expression~\eqref{eq:h10} that is obtained in \cite{Ji:2011qg}.

By substituting Eq.~\eqref{eq:h10-uphi} for $h_{10}$, we write the summation of $1$D integrals (in $\tilde{K}_2^A$, $\tilde{K}_2^D$ and $\tilde{K}_2^E$) as
\begin{equation}
\Sigma_{1D} = \frac{\sqrt{3}}{4}\left[ \mathcal{U}_{\phi\phi,2}(\Lambda)
-2\frac{\mathcal{U}_{\phi \phi,1}(\Lambda)}{\varPhi_1(\Lambda)}\varPhi_2(\Lambda)
+ \frac{\mathcal{U}_{\phi \phi,1}^2(\Lambda)}{\varPhi_1^2(\Lambda)} \frac{\Lambda^2}{2\pi}\right]~.
\end{equation}
Since the subscript of all integrals defined above displays the order of their cutoff dependence, readers can easily verify that each term in $\Sigma_{1D}$ $\sim \Lambda^2$.

We then sum up the three $2$D integrations of the off-diagonal amplitude (in C-E), and obtain
\begin{equation}
\label{eq:sigma-d}
\frac{\Sigma_{2D}}{2 b_\gamma}=
\mathcal{W}_{\phi\phi(\phi\bar{\phi}),1+1}(\Lambda)
+\frac{\mathcal{U}_{\phi \phi,1}^2(\Lambda)}{\varPhi_1^2(\Lambda)}
\mathcal{W}_{\phi(\bar{\phi}),1+1}(\Lambda)
-\frac{\mathcal{U}_{\phi \phi,1}(\Lambda)}{\varPhi_1(\Lambda)}
\left[\,
\mathcal{W}_{\phi\phi(\bar{\phi}),1+1}(\Lambda)
+
\mathcal{W}_{\phi(\phi\bar{\phi}),1+1}(\Lambda)
\,\right].
\end{equation}
Analytic expressions for the four $\mathcal{W}$'s in Eq.~\eqref{eq:sigma-d} are given in Appendix~\ref{app:integrals}. Each term in $\Sigma_{2D}$ $\sim \Lambda^2$.

The summation of integrals containing the near-diagonal remainder function $\rho$ results in:
\begin{equation}
\Sigma_\rho = \Omega_{\phi\rho\phi}(\Lambda) 
-2\frac{\mathcal{U}_{\phi \phi,1}(\Lambda)}{\varPhi_1(\Lambda)}
\Omega_{\phi\rho}(\Lambda)
+\frac{\mathcal{U}_{\phi \phi,1}^2(\Lambda)}{\varPhi_1^2(\Lambda)}
\Omega_{\rho}(\Lambda),
\end{equation}
where
\begin{subequations}
\begin{eqnarray}
\Omega_{\phi\rho\phi}(\Lambda) &=&\frac{2}{\pi^2}\int^\Lambda dq\, q\phi_0(q) \int^1_0 dx\, \phi_0(xq)\rho(x)
\nn
&=& \frac{2}{\pi} \left[ \mathcal{U}_{\phi\phi,2}(\Lambda) \int^1_0 dx \cos(s_0\ln x)\, \rho(x)
+\mathcal{U}_{\phi\bar{\phi},2}(\Lambda)
\int^1_0 dx \sin(s_0\ln x)\, \rho(x)\right], 
\nn
\end{eqnarray}
and, similarly,
\begin{eqnarray}
\Omega_{\phi\rho}(\Lambda) 
&=& 
\frac{\varPhi_2(\Lambda)}{\pi}
\left[\int^1_0 dx \rho(x) + \int^1_0 dx \cos(s_0\ln x) \rho(x)\right]
+\frac{\bar{\varPhi}_2(\Lambda)}{\pi}
\int^1_0 dx \sin(s_0\ln x) \rho(x),
\nn
\end{eqnarray}
\begin{equation}
\Omega_{\rho}(\Lambda) 
\equiv \frac{1}{\pi^2}\int\limits^\Lambda \int\limits^\Lambda dq dq' \rho(q,q')
= \frac{\Lambda^2}{\pi^2} \int^1_0 dx\, \rho(x),
\end{equation}
\end{subequations}
Here we have used the identity
\begin{equation}
\phi_0(xq) = \phi_0(q) \cos(s_0\ln x) + \bar{\phi}_0(q) \sin(s_0\ln x).
\end{equation}

We define those integrals including $\rho(x)$ as
\begin{subequations}
\label{eq:theta}
\begin{eqnarray}
\theta_1 &=& \frac{1}{\pi}\int^1_0 dx \rho(x) 
= \frac{\alpha}{\pi} \int^1_0 dx\, x e^{-\beta/x};
\\
\theta_2 &=& \frac{1}{\pi}\int^1_0 dx \cos(s_0 \ln x) \rho(x) 
= \frac{\alpha}{\pi} \int^1_0 dx\, \cos (s_0 \ln x) x e^{-\beta/x};
\\
\theta_3 &=& \frac{1}{\pi}\int^1_0 dx \sin(s_0 \ln x) \rho(x) 
= \frac{\alpha}{\pi} \int^1_0 dx\, \sin(s_0 \ln x) x e^{-\beta/x}.
\end{eqnarray}	
\end{subequations}
The $\theta_i$'s are constant numbers for a given renormalization condition, and are cutoff independent. Once the parameters $\alpha$ and $\beta$ are numerically fitted, we can obtain the $\theta_i$'s from Eq.~\eqref{eq:theta}.

Therefore, we now rewrite $\Sigma_\rho$ as
\begin{eqnarray}
\Sigma_\rho
&=&
2\theta_2 \mathcal{U}_{\phi\phi,2}(\Lambda) 
+2\theta_3 \mathcal{U}_{\phi\bar{\phi},2}(\Lambda) 
-2\frac{\mathcal{U}_{\phi \phi,1}(\Lambda)}{\varPhi_1(\Lambda)}
\left[(\theta_1+\theta_2)\varPhi_2(\Lambda)
+\theta_3\bar{\varPhi}_2(\Lambda)\right]
+\frac{\mathcal{U}_{\phi \phi,1}^2(\Lambda)}{\varPhi_1^2(\Lambda)}
\frac{\theta_1 \Lambda^2}{\pi}~,
\nn
\end{eqnarray}
and see immediately that $\Sigma_\rho \sim \Lambda^2$. As illustrated in Subsection~\ref{sec:result-lo-fulloff}, $\rho(q,p)$'s analytic expression~\eqref{eq:R-x} can differ from its numerical values if $q/p\ll1$. One might think that this could affect our obtained values of $\theta_i$'s since the integrals~\eqref{eq:theta} are from $q/p=0$ to $q/p=1$. But $\rho(q,p)$ is dominated by the region $q/p\sim 1$, so integrations in the region $q/p \ll 1$ do not affect the results significantly. We have numerically verified that the uncertainty in determining $\theta_i$ is $< 1$\%. 

As the defined summations, $\Sigma_{1D}$, $\Sigma_{2D}$ and $\Sigma_\rho$, are all $\sim \Lambda^2$, we must add them all together to cancel the dominant cutoff dependence $\sim \Lambda^2$. Therefore, $h_{20}$ must satisfy 
\begin{equation}
\label{eq:h20}
\Sigma_{1D} + \Sigma_{2D} + \Sigma_\rho + \frac{32}{3}h_{20}(\Lambda)  \varPhi_1^2(\Lambda)= 0,
\end{equation}
which yields the result that the leading term of $\tilde{H}_{2}$, $\Lambda^2 h_{20}$, is---as expected---of order $\Lambda^2$. The analytic expression for $h_{20}$ obtained from Eq. (100) is determined solely by the LO renormalization condition, and is thus independent of the additional input required at N$^2$LO.

\begin{figure}[tbpic:H20]
\centerline{\includegraphics[width=12cm,angle=0,clip=true]{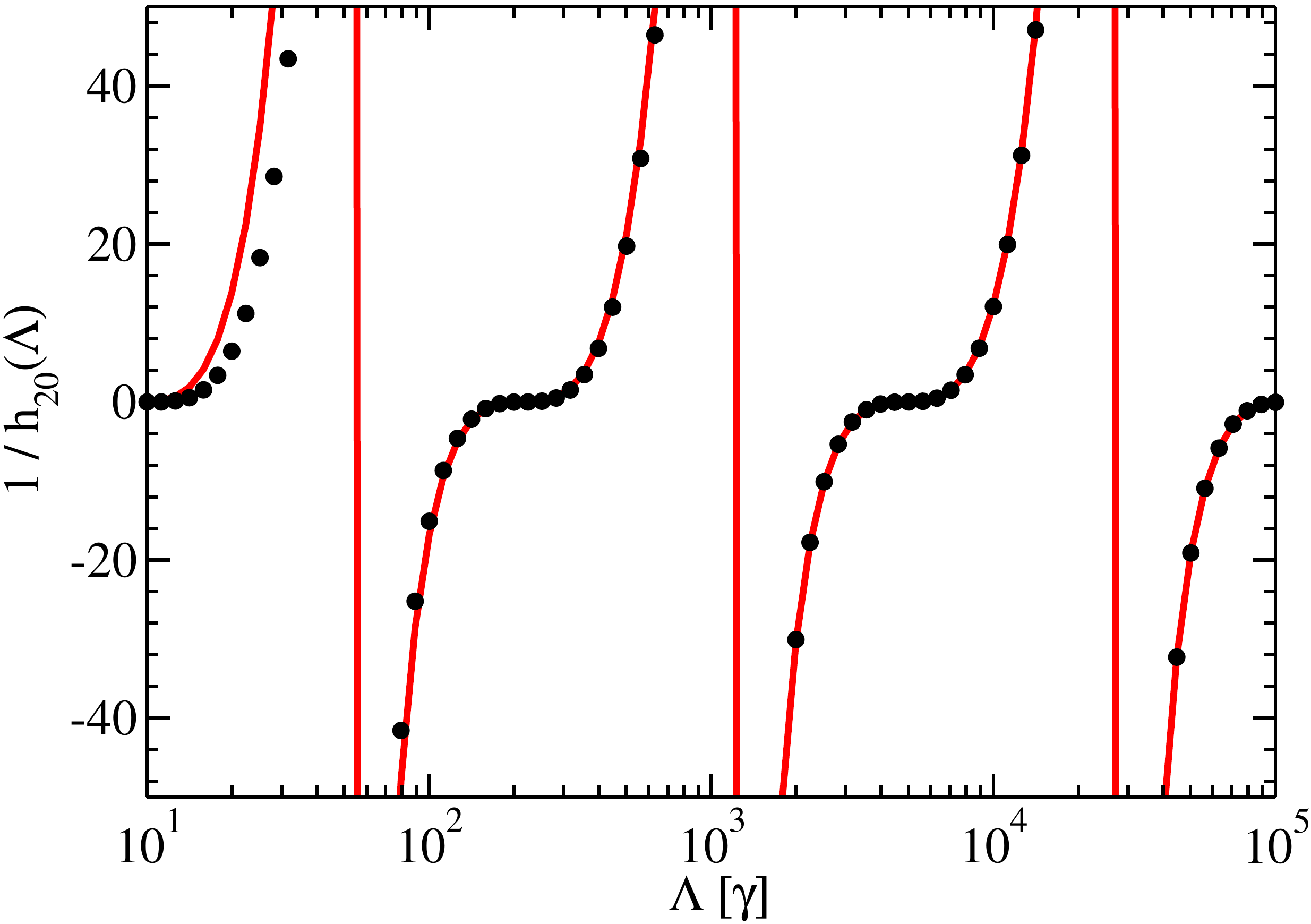}}
\caption{Leading term of the N$^2$LO three-body force, $h_{20}(\Lambda)$: The solid (red) line is the analytic expression, and dots are the numerical result. Results are renormalized to reproduce $\gamma a_{ad}=1.5$.}
\label{pic:H20}
\end{figure}

We compare this prediction with a numerical calculation of $h_{20}$ in Fig.~\ref{pic:H20}. After reproducing the atom-dimer scattering length $\gamma a_{ad}=1.5$ at leading order, we have LO parameters $b_{\gamma}=-2.012$, $\alpha=-1.665$ and $\beta=1.322$, as obtained in Sec.~\ref{sec:result-lo-fulloff} for this renormalization condition. This yields values of the $\theta_i$'s, $\theta_1=-0.038$, $\theta_2=-0.036$, and $\theta_3=0.0088$. Since the $\theta_i$'s are small, $\Sigma_\rho$ is generally two orders of magnitude smaller than other terms in Eq.~\eqref{eq:h20}. Therefore, the $1\%$ error in the determination of $\theta_i$'s discussed above results in only a $10^{-4}$ error in the determination of $h_{20}$. In the numerical calculation of $h_{20}$,  we also fix the NLO and N$^2$LO $a_{ad}=1.5/\gamma$ and the N$^2$LO trimer excited-state binding energy shift $B_2=0$,
although the result for $h_{20}$ is not affected by the 
N$^2$LO renormalization condition. 

The numerical values of $h_{20}$ shown in Fig.~\ref{pic:H20} agree with the analytic calculation from Eq.~\eqref{eq:h20} within an accuracy of $1\%$ for $\Lambda>10^3\gamma$ (the accuracy improves to $0.1\%$ in the regime $3\times10^4\gamma<\Lambda<10^5\gamma$). This proves that our expression for $h_{20}$ is an accurate prediction. Fig.~\ref{pic:H20} also shows discrepancies between the analytic and numerical values of $h_{20}$ at small $\Lambda$. These are due to higher-order $\gamma/\Lambda$ corrections.

\subsection{The sub-leading cutoff dependence and $h_{22}$}
\label{sec:renorm-h22}

With a fixed $\gamma$, $h_{22}$ must then cancel the sub-leading cutoff dependence, which is expected to be $\sim\ln\Lambda$ in Eq.~\eqref{eq:K2}. Since this sub-leading cutoff dependence is two orders lower in powers of $\Lambda$ compared to the leading one, to extract it from a calculation requires high accuracy. In order to obtain a complete analytic expression for $h_{22}$, the asymptotic behavior of the K-matrix, $\tilde{K}_0(q,p;E)$, must be expanded up to N$^2$LO in all regions ($q\gg p$, $q\sim p$, {\it etc.}). We are only able to obtain an accurate result for the half-on-shell K-matrix, $\tilde{K}_0(k,q;E)$;  the near-diagonal ($q\sim p$) part of the fully-off-shell $\tilde{K}_0(q,p;E)$ was only obtained in an approximate form. This approximation likely lacks the high accuracy required for a complete calculation of $h_{22}$.

Therefore in this subsection, we first demonstrate numerically that $h_{22}$ is necessary for renormalization. We then perform a partial analytic calculation of $h_{22}$: we drop the insertions of NLO terms in Eq.~\eqref{eq:K2}, and consider only contributions from the N$^2$LO dimer and the N$^2$LO three-body force.
This incomplete picture should not influence our understanding of $\tilde{H}_2$'s sub-leading structure on a power-counting level. In other words, if the sub-leading piece of $\tilde{H}_2$ is required to be $\sim mE$ so as to cancel divergences from the insertion of the N$^2$LO dimer, we would not expect this feature to be changed by the addition of contributions in classes C--E to the calculation.

\begin{figure}[tbpic:B31]
\centerline{\includegraphics[width=12cm,angle=0,clip=true]{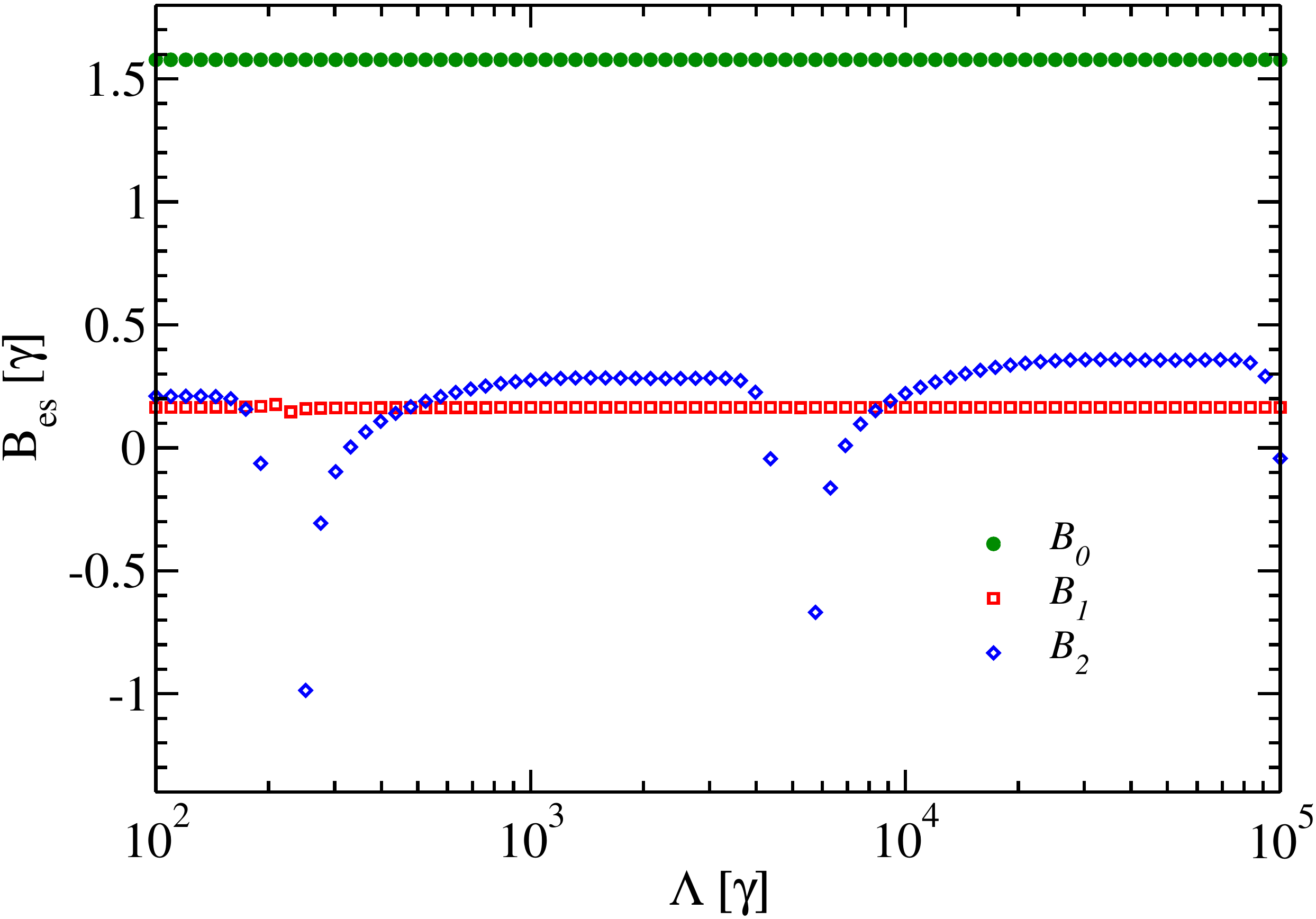}}
\caption{The trimer excited-state binding energy as a function of the cutoff $\Lambda$: The dots (dark green) are the LO value, the squares (red) are the NLO shifts. The diamonds (blue) are the N$^2$LO shifts obtained from using an energy-independent $H_2$. $\gamma a_{ad}=1.5$ is reproduced at all orders.}
\label{pic:N2LOrenorm}
\end{figure}

Turning first to the numerical calculation, we employ the same parameters as in the previous subsection: the atom-dimer scattering length $\gamma a_{ad}=1.5$ is fixed at leading order, and we demand that this value is not altered at NLO or N$^2$LO. In Fig.~\ref{pic:N2LOrenorm} the LO value of the trimer excited-state binding energy, $B_0^{(1)}$, together with the NLO and N$^2$LO shifts $B_1^{(1)}$ and $B_2^{(1)}$, are shown as a function of the cutoff $\Lambda$. While $B_0^{(1)}$ and $B_1^{(1)}$ are cutoff independent, $B_2^{(1)}$ has a noticeable cutoff dependence, if only one renormalization condition is used, i.e. if $H_2$ is assumed to be independent of energy. This demonstrates the necessity for the inclusion of an additional counterterm at N$^2$LO. The absence of such cutoff dependence in $B_1^{(1)}$ shows that no energy-dependent counterterm is needed at NLO. 
 
\begin{figure}[tbpic:B30]
\centerline{\includegraphics[width=12cm,angle=0,clip=true]{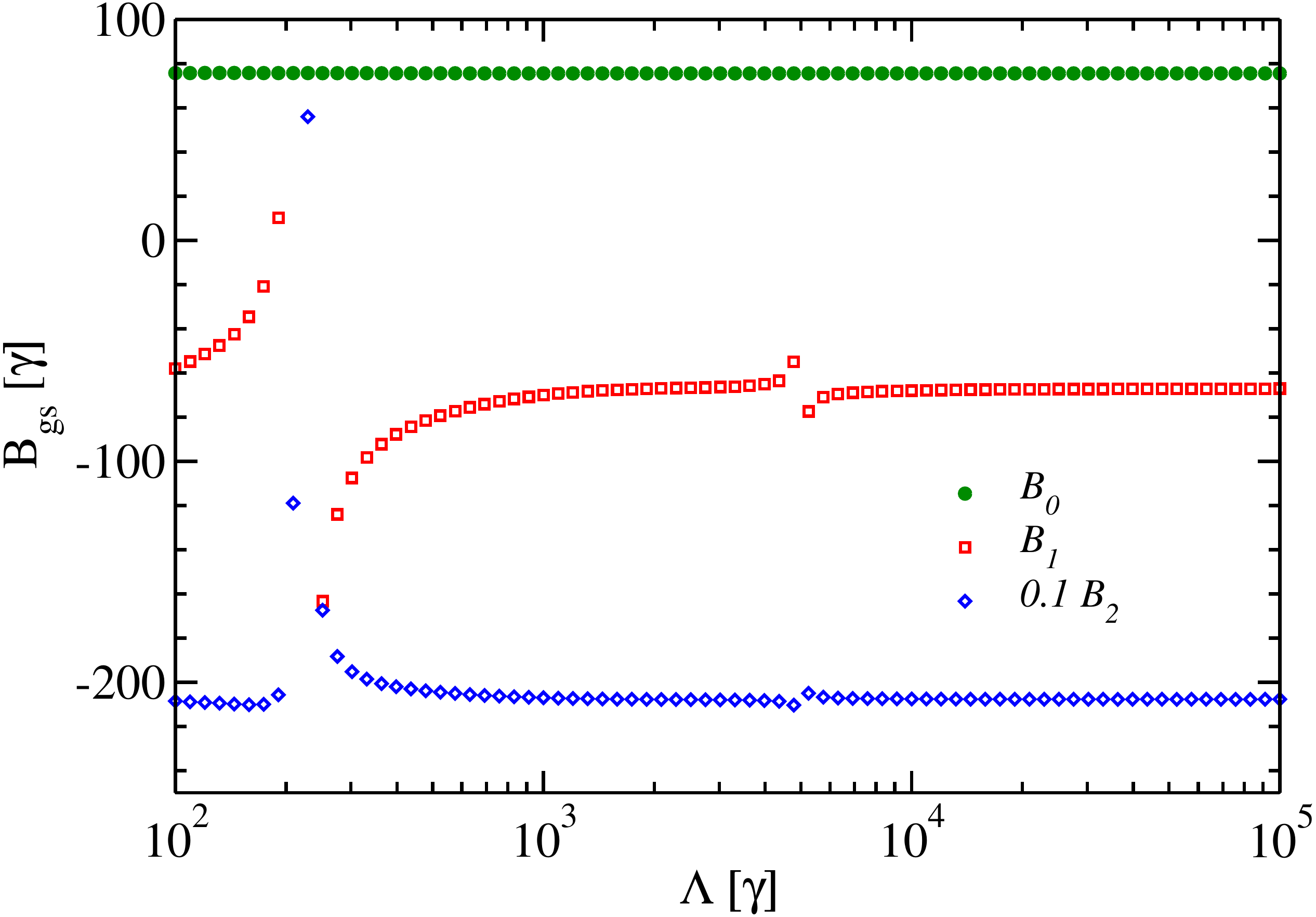}}
\caption{The trimer ground-state binding energy as a function of the cutoff $\Lambda$: The dots (dark green) are the LO value, the squares (red) are the NLO shifts. The diamonds (blue) are the N$^2$LO shifts obtained from using an energy-dependent $H_2$. $\gamma a_{ad}=1.5$ is reproduced at all orders, the trimer excited-state energy's N$^2$LO shift is fixed to zero.}
\label{pic:N2LOrenorm-2}
\end{figure}

With the inclusion of the linear-in-energy piece of $H_2$~\eqref{eq:H2} in the numerical calculation, we need an additional parameter at N$^2$LO. Here we fix that parameter by requiring a zero N$^2$LO shift of the trimer excited-state binding energy. We then calculate the LO value of the trimer ground-state energy, and its NLO and N$^2$LO shifts. These results are plotted as a function of the cutoff $\Lambda$ in Fig.~\ref{pic:N2LOrenorm-2}. At cutoffs $\Lambda < 500 \gamma$ cutoff dependence is seen in both the NLO and N$^2$LO shift. This cutoff dependence is present because, while such a $\Lambda$ obeys $\Lambda \gg \gamma$, $\Lambda$ is not well above the typical bound state momentum $k$. The fact that we are examining the deepest bound state means that higher cutoffs must be considered, and, indeed, results at all orders converge to a fixed result provided that $\Lambda \gg \gamma, k$. We note in passing that the N$^2$LO shift is very large in the ``natural units" for this problem, $\gamma^2 B_d$. 
Presumably this is because the presence of the counterterm $\sim k^2$ means that the N$^2$LO shift is really proportional to $k^2$, not $\gamma^2$, and is concomitantly larger for this state where $k^2$ is considerably bigger than $\gamma^2$. We shall return to this issue in the next section. 

To try and better understand the need for a linear-in-energy piece of $H_2$ 
we define a partially summed N$^2$LO K-matrix as
\begin{eqnarray}
\label{eq:K2-H22}
\tilde{K}_2^{\rm part}(k,k;E) &=& \frac{1}{2\pi} \int^\Lambda_0 dq\, \frac{q^2 (\gamma+\sqrt{3q^2/4 -mE}\, )^2}{-\gamma+\sqrt{3q^2/4 -mE}} \tilde{K}_0^2(k,q;E)
+\frac{2\tilde{H}_2}{\Lambda^2} \mathcal{L}^2(k).
\end{eqnarray}

As the $\gamma\Lambda$ dependence is absorbed in the fine tuning of $h_{20}$, we consider only the $\mathcal{O}(mE/q^3)$ term in $\tilde{K}_0(k,q;E)$'s asymptotic expansion \eqref{eq:K0-asymptotic1}. Inserting this into Eq.~\eqref{eq:K2-H22}, we study the cutoff dependence of the partially-summed $\tilde{K}_2$ in an expansion in powers of $\Lambda$, and find:
\begin{eqnarray}
\zeta  &=& a_{\gamma}^2(k)\,\left(
\frac{\sqrt{3}}{4}\mathcal{U}_{\phi\phi,2} + \frac{32 h_{20}}{3} \varPhi_1^2
\right)
\nn
&&+ a_{\gamma}^2(k)\, mE\left[
\frac{1}{2\sqrt{3}}
\left(3\mathcal{U}_{\phi\psi,0}-\mathcal{U}_{\phi\phi,0}\right)
+\frac{128 h_{20}}{9} \varPhi_1 \left(\varPhi_{-1} +\frac{3}{2}\varPsi_{-1}\right)
+\frac{32 h_{22}}{3\Lambda^2} \varPhi_1^2
\right].
\end{eqnarray}

In order to cancel the leading cutoff dependence ($\sim \Lambda^2$) of this partial summation, $h_{20}$ in $\tilde{K}_2^{\rm part}$ is required to satisfy:
\begin{equation}
\label{eq:h20-partial}
\frac{\sqrt{3}}{4}\mathcal{U}_{\phi\phi,2} + \frac{32 h_{20}}{3} \varPhi_1^2 = 0.
\end{equation}
Similarly, $h_{22}$ is required to cancel the sub-leading cutoff dependence ($\sim \ln \Lambda$ and $\mathcal{O}(1)$). Combining with Eq.~\eqref{eq:h20-partial}, we obtain
\begin{equation}
\label{eq:h22-partial}
\frac{1}{2\sqrt{3}}
\left(3\mathcal{U}_{\phi\psi,0}-\mathcal{U}_{\phi\phi,0}\right)
-\frac{1}{\sqrt{3}} \frac{\mathcal{U}_{\phi\phi,2}}{\varPhi_1}
\left(\varPhi_{-1} +\frac{3}{2}\varPsi_{-1}\right)
+\frac{32 h_{22}}{3\Lambda^2} \varPhi_1^2 =0.
\end{equation}
Therefore, the analytic formula for $h_{22}$ in this calculation of the partially-summed $\tilde{K}_2$ is:
\begin{eqnarray}
\label{eq:h22-partial-analytic}
h_{22}(\Lambda) &=& \frac{\sqrt{3}\pi}{128} (1+s_0^2) \frac{1}{\sin^2\left[s_0\ln(\Lambda/\bar{\Lambda})-\tan^{-1}s_0\right]}
\nn
&&\times \left\lbrace  \left[1-3|C_1| \cos(\arg C_1)\right] \ln (\Lambda/\lambda_*)\right.
\nn
&&\hspace{0.5cm} -\frac{1}{2s_0} \left(\sin\left[2s_0\ln(\Lambda/\bar{\Lambda})\right] - 3|C_1| \sin\left[2s_0 \ln(\Lambda/\bar{\Lambda})+\arg C_1\right]\right)
\nn
&&\hspace{0.5cm}- \frac{1- \frac{1}{\sqrt{1+s_0^2}} \cos \left[2 s_0\ln(\Lambda/\bar{\Lambda})-\tan^{-1}s_0\right]}{ \sin\left[s_0\ln(\Lambda/\bar{\Lambda})-\tan^{-1}s_0\right] }
\nn
&&\hspace{0.5cm}\left.\times \left( \sin\left[s_0\ln(\Lambda/\bar{\Lambda})+\tan^{-1}s_0\right] + \frac{3|C_1|}{2} \sin\left[s_0\ln(\Lambda/\bar{\Lambda})+\arg C_1 +\tan^{-1}s_0\right]\right)\right\rbrace,
\nn
\end{eqnarray}
where $\lambda_*$ determines the fine tuning of $h_{22}(\Lambda)$, that is introduced to represent the infrared regularization of $\mathcal{U}_{\phi\psi,0}$ and $\mathcal{U}_{\phi\phi,0}$. $\lambda_*$ is independent of the cutoff, but dependent on the renormalization condition at N$^2$LO.

\begin{figure}[tbpic:H22]
\centerline{\includegraphics[width=12cm,angle=0,clip=true]{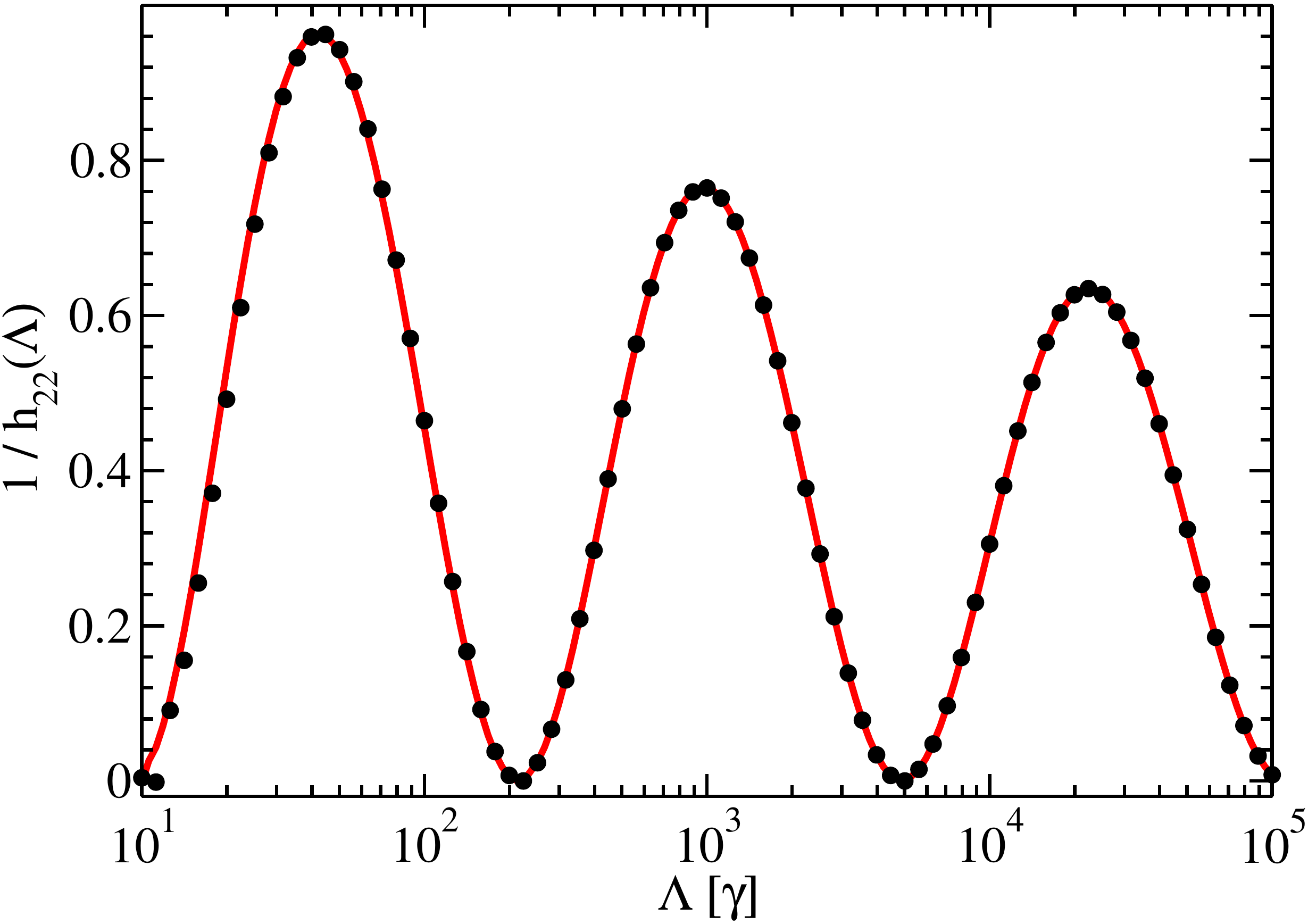}}
\caption{The subleading term of the N$^2$LO three-body force $h_{22}(\Lambda)$: The solid (red) line is the analytic expression, the dots are the numerical result. $\gamma a_{ad}=1.5$ is reproduced at LO, and Eq.~\eqref{eq:n2lo-cond} determines the N$^2$LO renormalization condition.}
\label{pic:H22}
\end{figure}

We compare the analytic expression \eqref{eq:h22-partial-analytic} with the numerical result of $h_{22}$ obtained from Eq.~\eqref{eq:K2-H22} in Fig.~\ref{pic:H22}. We again renormalize to $\gamma a_{ad}=1.5$ at LO. This determines the value $\bar{\Lambda}$ in Eq.~\eqref{eq:h22-partial-analytic}. Since the partially summed $\tilde{K}_2$ excludes two insertions of NLO terms, it cannot rigorously represent the N$^2$LO corrections to any physical observable. We therefore choose two artificial renormalization conditions to determine the unknown constants $h_{20}$ and $h_{22}$ which appear in Eq.~\eqref{eq:K2-H22} for $\tilde{K}_2^{\rm part}$:
\begin{equation}
\label{eq:n2lo-cond}
\tilde{K}_2^{\rm part}\left(\frac{2\gamma}{\sqrt{3}},\frac{2\gamma}{\sqrt{3}};0\right) = \tilde{K}_2^{\rm part}\left(0,0;-B_d\right)=0. 
\end{equation}
Choosing different N$^2$LO renormalization conditions would yield a different value of the parameter $\lambda_*$ in Eq.~\eqref{eq:h22-partial-analytic}. Using numerical results obtained under the N$^2$LO renormalization condition (\ref{eq:n2lo-cond}) we find that the best fit to $h_{22}(\Lambda)$ is obtained with $\lambda_*=1.752\times10^{-4}\gamma$. The remaining behavior of $h_{22}$ is a prediction---at least within the context of this subsection's approximation---and that prediction agrees with numerical results for $h_{22}$ obtained within the same truncated calculation of the N$^2$LO atom-dimer amplitude, $\tilde{K}_2^{\rm part}$, to an accuracy of $10^{-3}$.

In this section we have demonstrated the need to include an energy-dependent piece in $\tilde{H}_2$ when performing N$^2$LO calculations in the three-body system in SREFT. This was done numerically, in a full N$^2$LO calculation, and analytically, in an N$^2$LO calculation that only included diagrams of classes A and B. This means that only after reproducing {\it two} three-body observables in SREFT can we predict other observables at N$^2$LO accuracy. This finding agrees with the results of Bedaque {\it et al.} in~\cite{Bedaque:2002yg} and Barford and Birse in Ref.~\cite{BB05}. However, it contradicts the conclusion reached by Platter and Phillips in Ref.~\cite{Platter:2006ev}. The problem with the latter analysis is that the N$^2$LO piece of the dimer propagator was included in the modified STM equation, and thereby iterated to all orders. As discussed in Ref.~\cite{Platter:2006ev}, this softens the ultra-violet behavior of the atom-dimer K-matrix, i.e. the $K(0,q;-B_d)$ found in Ref.~\cite{Platter:2006ev} 
falls off more steeply at large 
$q$ than does $\tilde{K}_0(0,q;-B_d)$. The integrals computed there thus all converge in the limit $\Lambda \rightarrow \infty$. A definite $\Lambda \rightarrow \infty$ result is found, and no additional renormalization appears necessary at N$^2$LO. 

However, as emphasized in Ref.~\cite{EG09}, the existence of a definite $\Lambda \rightarrow \infty$ limit in an EFT calculation does not guarantee that the results found in that limit are rigorous consequences of the EFT. In particular, in contrast to what was seen in the work of Platter and Phillips, the perturbative treatment of $\sim r_0$ corrections performed here shows that there is {\it no} modification of the fall-off at large momenta of the atom-dimer amplitude as that amplitude is computed to higher orders in SREFT. The fall off seen at LO ($\sim 1/q$) then determines the behavior of integrands in loop integrals to all orders in perturbation theory. We conclude that the approach taken by Platter and Phillips amounts to using the physics at scale $r_0$ to regulate the loop integrals that constitute the perturbations to the LO result. While the results of Ref.~\cite{Platter:2006ev} suggest that this is phenomenologically efficacious, it does not constitute an EFT calculation if $r_0 \sim l$, since in 
that case other short-distance effects can enter the computation of atom-dimer scattering at the same scale. Platter and Phillips' non-perturbative treatment of the $r_0$ corrections could perhaps be justified from an EFT perspective if $r_0 \gg \ell$. 

\section{Helium Trimers}
\label{sec:he-trimer}
To test whether this disagreement  between Platter and Phillips'~\cite{Platter:2006ev} result
and our finding regarding the necessity of $\tilde{H}_2$ at N$^2$LO causes differences in the prediction of three-body observables at N$^2$LO, we here apply our full N$^2$LO analysis to observables in trimers of Helium-4 atoms. We compare our results with the partially resummed SREFT calculations by Platter and Phillips~\cite{Platter:2006ev} and calculations from realistic potentials by Roudnev and Yakovlev~\cite{Roudnev2000,Roudnev2003} and Kolganova {\it et al.}~\cite{Kolganova:2004}. 

 It was first observed by Luo {\it et al.}~\cite{luo1993} that two $^4$He atoms can form a shallow bound state, a dimer. The $^4$He dimer's scattering length, $a = 104^{+8}_{-18} \rm$ \AA, and binding energy, $B_d = 1.1^{+0.3}_{-0.2}$ mK, has been evaluated from measurements of the dimer bond length by Grisenti {\it et al.}~\cite{Grisenti:2000zz}. The effective-range expansion in Eq.~\eqref{eq:a-gamma} can then be used to infer an effective range of the order of $10$ \AA.

However, experimental data for systems of three $^4$He atoms are still limited. Although a three-$^4$He-atom bound state, i.e. a trimer, has been observed~\cite{Schollkopf:1994}, the trimer binding energies and the atom-dimer scattering length have not yet been measured. Meanwhile, many independent calculations based on realistic potential models~\cite{Roudnev2000, Roudnev2003, Motovilov:1999iz, Barletta:2001za, Kolganova:2004} have studied $^4$He trimers. Fairly good agreement has been achieved for trimer binding energies and the atom-dimer scattering length. For a review of these calculations, see Ref.~\cite{Kolganova:2011uc}.
These realistic potentials differ at short distances, but all mimic a long-range van der Waals potential $\sim C_6/r^6$ for $^4$He dimers. In Sec.~\ref{sec:pionfree-eft} we used the form of the effective-range expansion derived by Gao~\cite{Gao1998} for such a potential to show that SREFT at N$^2$LO can be used to describe systems interacting via such a potential, and 
only parameters $a$ and $r_0$ are needed if observables are to be calculated up to an accuracy $\mathcal{O}(r_0^2/a^2)$ or $\mathcal{O}(k^2 r_0^2)$. This provides an opportunity for a benchmark comparison between calculations of atom-dimer scattering using realistic potentials and SREFT. Such benchmarks have proven very instructive in the study of three-nucleon scattering~\cite{Friar:1990zza,Friar:1995zz,Kievsky:1998gt}. Therefore,
in this section we compare our EFT results for $^4$He trimers and the atom-dimer scattering length at $\mathcal{O}(r_0^2/a^2)$ with calculations from the TTY potential~\cite{Roudnev2000, Roudnev2003}. 

The two-body parameters $a$ and $r_0$, which are inputs in the SREFT analysis, must then be obtained from this potential model. Motovilov {\it et al.} found that the TTY potential~\cite{Tang:1995} predicts a $^4$He-$^4$He scattering length of $a=100.01$ \AA~and a $^4$He dimer binding energy   $B_d=1.30962$ mK~\cite{Motovilov:1999iz}. From the effective-range expansion, we obtain the atom-atom effective range as $r_0=7.50(5)$ \AA, which provides our EFT expansion parameter $\gamma r_0=0.0781$. We therefore estimate that typical NLO corrections will be $\sim 8\%$ and N$^2$LO corrections $< 1$\%. 

The SREFT calculations also require one (at LO and NLO) or two (at N$^2$LO) three-body inputs. Three-body
calculations by Roudnev and Yakovlev from the TTY potential~\cite{Roudnev2000, Roudnev2003} showed $^4$He trimer binding energies in the ground and first-excited state of $B_t^{(0)} = 96.33 B_d$ and $B_t^{(1)} = 1.738 B_d$ respectively. The
atom-dimer scattering length was obtained as $a_{ad}= 1.205 \gamma^{-1}$. After some initial disagreement, revised numbers for $B_t^{(1)}$ and $a_{ad}$ from Kolganova {\it et al.}~\cite{Kolganova:2004} agree with these values~\cite{Kolganova:2011uc}. In what follows we take the results of Ref.~\cite{Roudnev2000,Roudnev2003}, to the precision quoted here, as ``the TTY results". 

Upon inserting the N$^2$LO dimer propagator in the modified STM equation, Platter and Phillips obtained $B_t^{(1)} = 1.7375(5) B_d$ and $a_{ad} = 1.204(1) \gamma^{-1}$, where errors in the brackets indicate differences from two renormalization schemes. In their calculation, where only one three-body input is needed, they can reproduce either $B_t^{(1)}$ or $a_{ad}$ at LO, NLO and N$^2$LO, and predict the other at the corresponding order. Their results are consistent with the TTY results with a remaining error $\sim 0.1\%$, consistent with the corrections above N$^2$LO. Platter and Phillips also predicted the $^4$He trimer ground-state binding energy as $B_t^{(0)} = 89.45(7) B_d$, which differs from TTY number by $8\%$. Since $k$ for this state is large this discrepancy can be explained by corrections beyond $\mathcal{O}(k^2 r_0^2)$.

In contrast to these results, we argue that a correct treatment of $\mathcal{O}(r_0^2)$ effects as perturbations means that two input pieces of three-body data are needed for proper renormalization. In Table.~\ref{tb:he4}, we show our results for $^4$He trimer observables in two renormalization schemes: we fix either $B_t^{(1)}$ or $a_{ad}$ to the TTY results at LO and NLO, and use the other as the additional parameter at N$^2$LO. If we reproduce $a_{ad}$ at all orders, and, in addition, $B_t^{(1)}$ at N$^2$LO, we predict the effective range of atom-dimer scattering to be:
\begin{equation}
r_{ad} \gamma=0.835 + 0.070 + 0.008=0.913(1),
\label{eq:radpred}
\end{equation}
where we have separated the LO, NLO, and N$^2$LO contributions to $r_{ad}$. This shows the excellent convergence pattern for this observable, which is very consistent with the predicted expansion parameter $r_0/a \sim 0.1$. Based on this convergence pattern, we conservatively predict an N$^3$LO-and-beyond shift in $r_{ad}$ of at most $0.001 \gamma^{-1}$. That this is a conservative estimate is supported by the difference between the two results for $r_{ad}$ obtained in the two different renormalization schemes  (fix $B_t^{(1)}$/$a_{ad}$ at LO and NLO), which is an order of magnitude smaller: the difference between predictions for $r_{ad}$ is $5\%$ at LO, decreases to $0.9\%$ at NLO, and to $0.02\%$ at N$^2$LO. 

A different situation is presented by the $^4$He trimer's ground-state energy. We find (in the same renormalization scheme as was used to obtain (\ref{eq:radpred})):
\begin{equation}
B_t^{(0)}/B_d=97.1 - 7.40 + 27.2=116(11).
\label{eq:Bt0pred}
\end{equation}
Since corrections of the form $k^2 r_0^2$, which are the largest effects for $B_t^{(0)}$, only enter at even orders, this may explain the increase in the size of the N$^2$LO correction, compared to the NLO one. Here again, the quoted uncertainty on our final result for $B_t^{(0)}$ comes from examining the convergence pattern; while the difference between renormalization schemes does not decrease from NLO to N$^2$LO it is markedly smaller than the $\sim 8$\% uncertainty we quote here. This uncertainty encompasses the $B_t^{(0)}$ result from the full TTY calculation only at the 2$\sigma$ level. Platter and Phillips made a similar argument regarding the accuracy of their N$^2$LO result, $B_t^{(0)}=90 B_2$. In their case, the convergence pattern is less peculiar, and the final result obtained reproduces the full TTY calculation within the expected accuracy.

\begin{table}
\label{tb:he4}
\begin{tabular*}{0.75\textwidth}{@{\extracolsep{\fill}} l l c c c c}
\hline
\multicolumn{2}{c}{Input} & $B_t^{(1)}$ [$B_d$] & $B_t^{(0)}$ [$B_d$] & $a_{ad}$ [$\gamma^{-1}$] & $r_{ad}$ [$\gamma^{-1}$]\\
\hline
\hline
$a_{ad}$ & LO & 1.723 & 97.12 & \textcolor{red}{1.205} & 0.8352 \\
$a_{ad}$ & NLO & 1.736 & 89.72 & \textcolor{red}{1.205} & 0.9049 \\
$a_{ad}$ , $B_t^{(1)}$ & N$^2$LO & \textcolor{red}{1.738} & 116.9 & \textcolor{red}{1.205} & 0.9132 \\
\hline
$B_t^{(1)}$ & LO & \textcolor{red}{1.738} & 99.37 & 1.178 & 0.8752 \\
$B_t^{(1)}$ & NLO & \textcolor{red}{1.738} & 89.77 & 1.201 & 0.9130 \\
$B_t^{(1)}$ , $a_{ad}$ & N$^2$LO & \textcolor{red}{1.738} & 115.9 & \textcolor{red}{1.205} & 0.9135 \\
\hline
\hline
\multicolumn{2}{c}{TTY~\cite{Roudnev2000,Roudnev2003}}  & 1.738 & 96.33 & 1.205 & \\
\hline
\end{tabular*}
\caption{EFT predictions for the $^4$He trimer binding energies and atom-dimer scattering length $a_{ad}$ and effective range $r_{ad}$ up to N$^2$LO. Energies and lengths are in units of the dimer binding energy and binding momentum respectively. Text in red denotes renormalization conditions (inputs) at each order. Results are compared with calculations from the TTY potential~\cite{Roudnev2000,Roudnev2003}.}
\end{table}

\begin{figure}[tbpic:cot1]
\centerline{
\includegraphics[width=12cm,angle=0,clip=true]{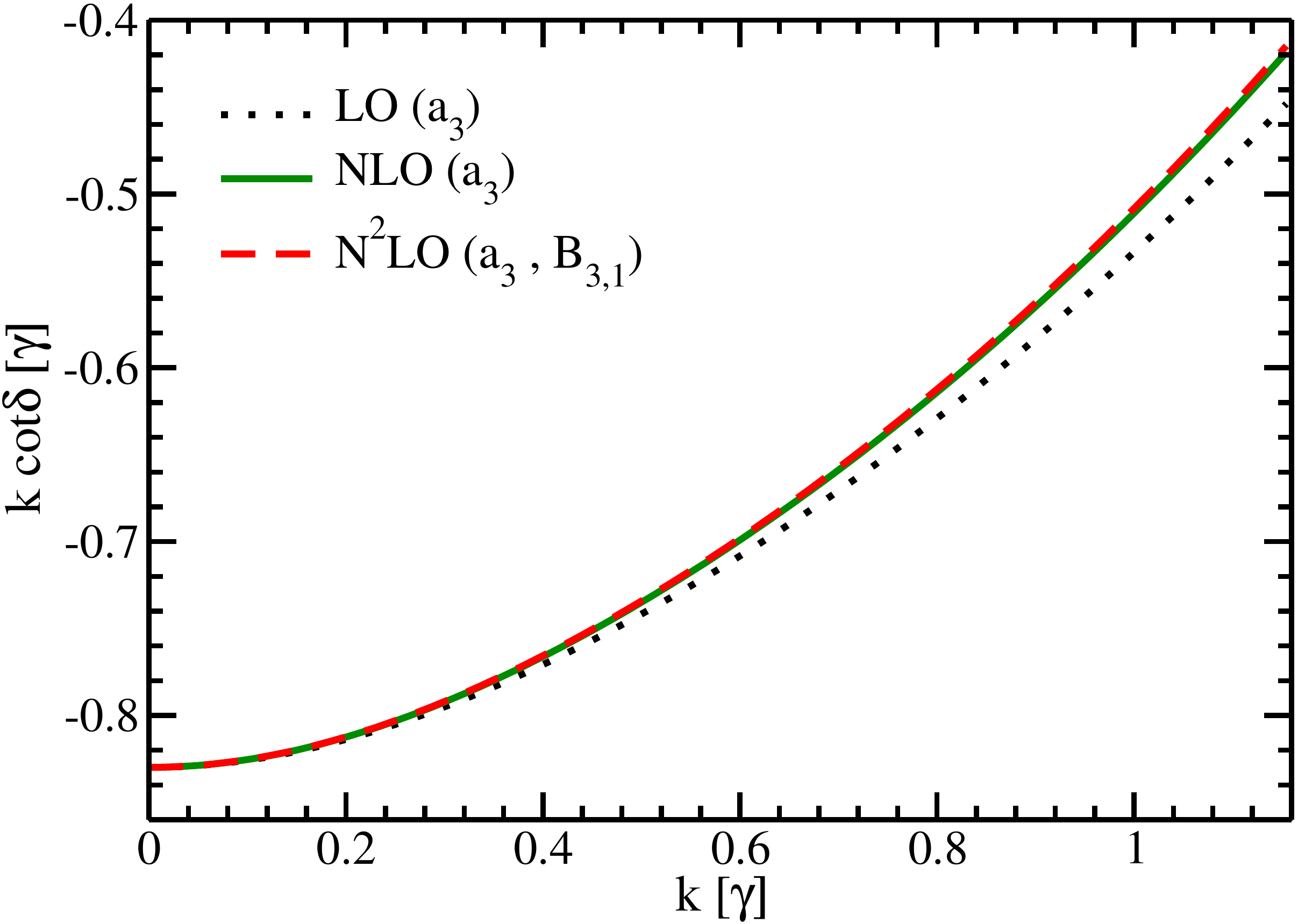}
}
\caption{$^4$He atom-dimer scattering phase shifts: $a_{ad}=1.205\gamma^{-1}$ is fixed at LO (dotted line), and also NLO (solid line). Both $a_{ad}$ and $B_t^{(1)}=1.738 B_d$ are fixed at N$^2$LO (dashed line). $k\cot\delta$ is in units of the $^4$He dimer binding momentum.}
\label{pic:cot1}
\end{figure}

\begin{figure}[tbpic:cot2]
\centerline{
\includegraphics[width=12cm,angle=0,clip=true]{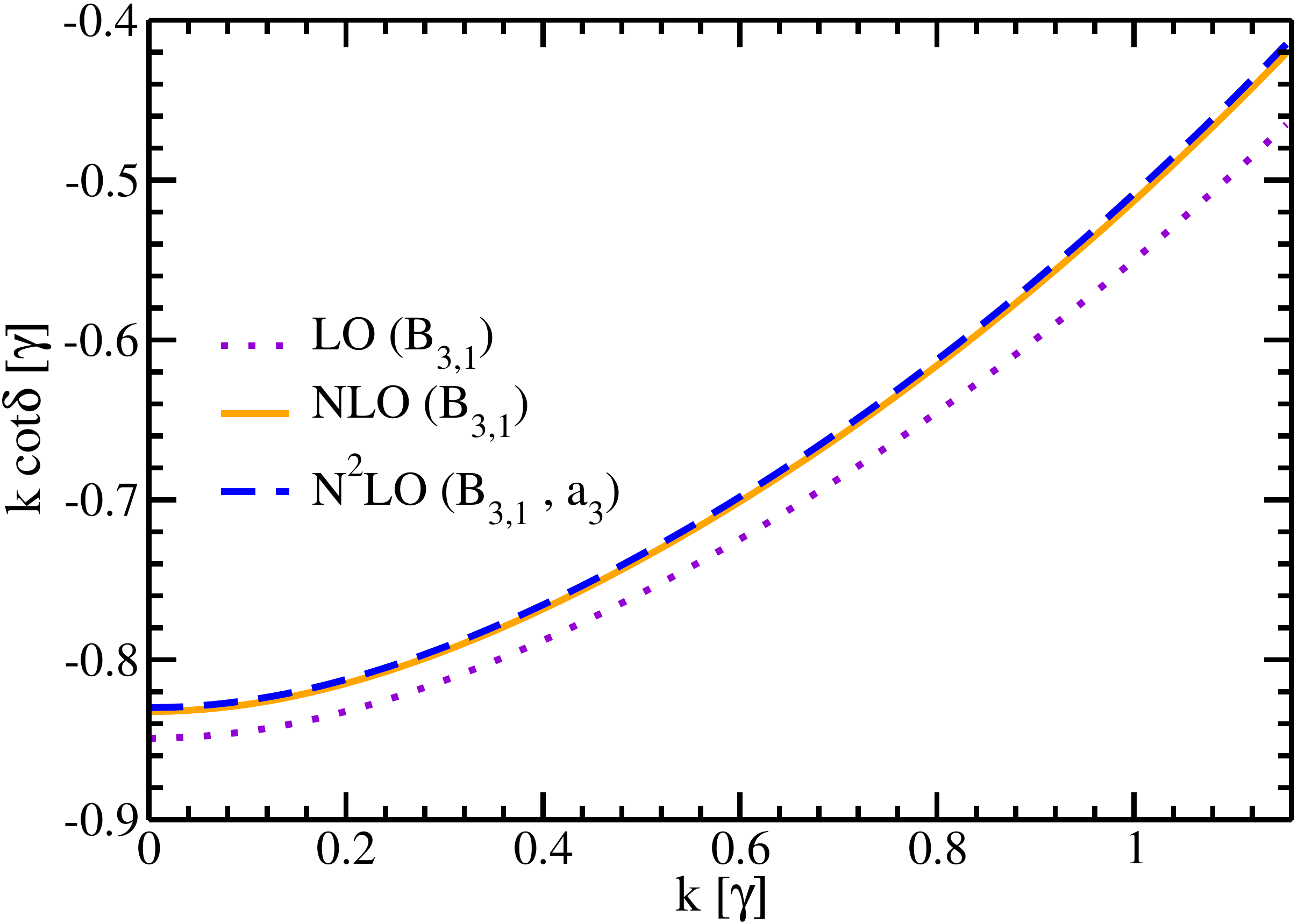}
}
\caption{$^4$He atom-dimer scattering phase shifts: $B_t^{(1)}=1.738 B_d$ is fixed at LO (dotted line), and also NLO (solid line). Both $B_t^{(1)}$ and $a_{ad}=1.205\gamma^{-1}$ are fixed at N$^2$LO (dashed line). $k\cot\delta$ is in units of the $^4$He dimer binding momentum.}
\label{pic:cot2}
\end{figure}

\begin{figure}[tbpic:cot3]
\centerline{
\includegraphics[width=12cm,angle=0,clip=true]{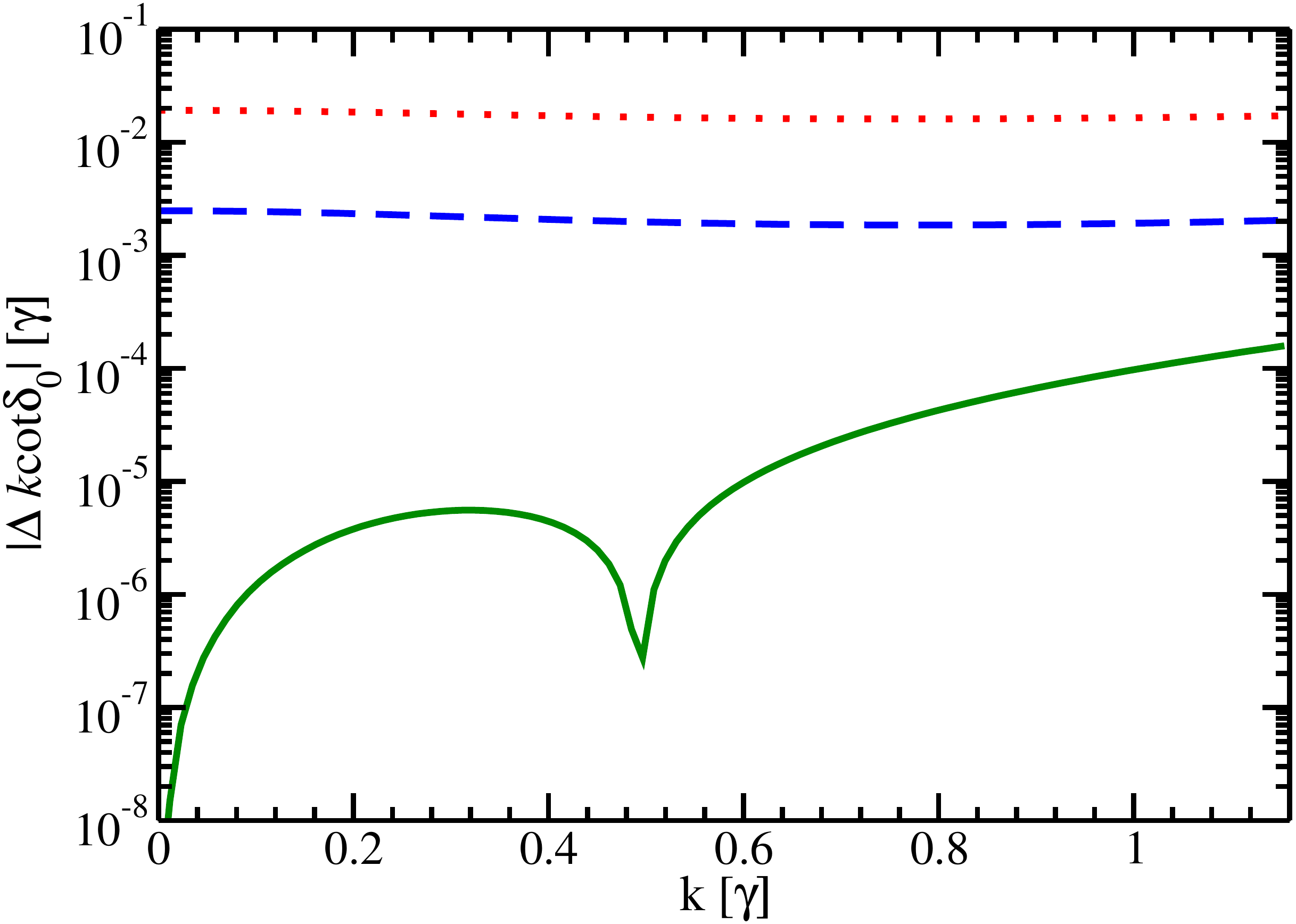}
}
\caption{Difference between $^4$He atom-dimer scattering phase shifts calculated in two renormalization schemes, which are illustrated in Figs.~\ref{pic:cot1} and \ref{pic:cot2} respectively. The difference is calculated at LO (red dotted line), NLO (blue dashed line) and N$^2$LO (green solid line). $|\Delta(k\cot\delta)|$ is in units of the $^4$He dimer binding momentum.}
\label{pic:cot3}
\end{figure}

Eq.~(\ref{eq:Bt0pred}) suggests that, at best, the SREFT expansion for the trimer ground state is poorly convergent. The internal momentum of the state is near the convergence boundary of the EFT expansion, making it a questionable venue for the application of SREFT. In contrast, a low-momentum observable like $r_{ad}$ seems to converge very well. In order to further demonstrate the usefulness of SREFT for low-energy atom-dimer interactions we calculated the $^4$He atom-dimer scattering phase shifts. The results for
$k\cot\delta$ shown in Fig.~\ref{pic:cot1} reproduce $a_{ad}$ at all orders, and $B_t^{(1)}$ at N$^2$LO. They should be compared to Fig.~\ref{pic:cot2} where $B_t^{(1)}$ is fixed at all orders, and $a_{ad}$ at N$^2$LO. In either renormalization scheme (Fig.~\ref{pic:cot1} or Fig.~\ref{pic:cot2}), the atom-dimer phase shift converges from LO to N$^2$LO. 
The difference of the phase shifts in these two renormalization schemes is shown in Fig.~\ref{pic:cot3} at each order, with the absolute value of this difference, $|\Delta \,(k\cot\delta)|$, plotted as a function of the on-shell momentum $k$. $|\Delta \,(k\cot\delta)|$ is $< 2\%$ at LO, and decreases to less than $0.3\%$ at NLO. The N$^2$LO $|\Delta \,(k\cot\delta)|$ stays below $2\times10^{-4}$ for $0<k<\frac{2\gamma}{\sqrt{3}}$. Based on this renormalization-scheme difference, and the convergence pattern of the phase shifts themselves, we deduce that N$^3$LO effects from the $l=0$ atom-atom amplitude will alter the phase shifts by $<0.1$\%  throughout the kinematic range shown. The first atom-atom partial wave with $l \neq 0$ that is allowed by bosonic symmetry is $l=2$, and this amplitude does not affect the atom-dimer scattering amplitude until N$^4$LO~\cite{Gao1998}. Therefore the phase-shift predictions shown here should be accurate to better than $0.2$\%.  

Refs.~\cite{Motovilov:1999iz,Roudnev2003} give results of atom-dimer phase-shift calculations for the TTY potential. However, calculations by Motovilov {\it et al.}~\cite{Motovilov:1999iz} correspond to an incorrect value of the atom-dimer scattering length, $a_{ad} \gamma=1.362$, a value which was amended to $a_{ad} \gamma=1.200(5)$ in later work by the same group of authors~\cite{Kolganova:2004}. Platter and Phillips used the earlier value, and compared their partially-resummed SREFT calculation to the phase shifts from Ref.~\cite{Motovilov:1999iz}. Semi-quantitative agreement was found. We employ the correct input $a_{ad} \gamma=1.205$ calculated by Roudnev~\cite{Roudnev2003}, however, our predicted phase shifts differ from results in Ref.~\cite{Roudnev2003} by about $3\%$. Since contributions from higher-partial-wave atom-atom interactions only enter at N$^4$LO~\cite{Gao1998}, the origin of this discrepancy is worthy of further investigation. 

High-precision computations with gaussian soft-core atom-atom potentials have recently been performed both below~\cite{RomeroRedondo11} and above~\cite{Garrido12} the three-atom threshold. The soft-core potential in these calculations is constructed to reproduce solely the S-wave parameters in $^4$He-$^4$He systems (binding energy and scattering length). Such atom-atom potentials are supplemented by an energy-independent three-atom force by the same group of authors in Ref.~\cite{Kievsky:2011ut} in order to obtain a better description of three-atom systems. Since the soft-core potential includes only short-distance interactions, calculations with such potentials are possiblily equivalent to SREFT calculations. Detailed comparison of SREFT results and the calculations of Refs.~\cite{RomeroRedondo11,Garrido12,Kievsky:2011ut} could therefore be quite illuminating.

\section{Conclusion}
\label{sec:conclusion}

We calculated the atom-dimer scattering amplitude up to N$^2$LO in the SREFT framework, i.e. in a perturbative expansion in $r_0/a$. This amplitude determines three-body observables, such as three-body binding energies and atom-dimer phase shifts. We showed the asymptotic behavior of LO K-matrices for half-on-shell and fully-off-shell cases. This allowed us to infer that, in order to achieve proper renormalization at N$^2$LO, the three-body force at that order must contain an energy-dependent part---an inference confirmed by explicit numerical calculation.
This means that N$^2$LO SREFT calculations in the three-body system require a second three-body input, a finding that disagrees with the results of a partially-resummed N$^2$LO calculation by Platter and Phillips~\cite{Platter:2006ev}, where the theory was not renormalized in a manner consistent with the perturbative expansion in $r_0/a$ and $r_0 k$.

Our perturbative N$^2$LO calculation can be straightforwardly extended to few-body systems in nuclear physics, such as the triton or s-wave two-neutron halo nuclei~\cite{CanhamHammer1,CanhamHammer2}, once spin and isospin degrees of freedom are included. For example, the partially-resummed N$^2$LO calculation of Ref.~\cite{Platter:2006ev} was extended to the three-nucleon system in Ref.~\cite{Platter:2006}. Ref.~\cite{Platter:2006} found a good result for the triton binding energy, but this was obtained without the energy-dependent three-nucleon contact term whose inclusion we are advocating here. Bedaque {\it et al.} studied three-nucleon observables in SREFT at N$^2$LO, and---motivated by the power counting developed in Ref.~\cite{Bedaque:1998km}---included an energy-dependent three-nucleon contact term which played the same role as the $h_{22}$ piece of the atom-dimer contact interaction~\cite{Bedaque:2002yg}.
The calculation of Ref.~\cite{Bedaque:2002yg} also involved partial resummation, which they argued was equivalent to a perturbative calculation, up to N$^2$LO accuracy, 
for momentum-space cutoffs $\sim 1/r_0$. However, this partial resummation meant that they could not fully demonstrate proper renormalization of their SREFT result at N$^2$LO. Indeed, the only rigorous way to demonstrate correct renormalization of the three-nucleon problem is to follow an analysis similar to that presented here for three-boson systems. Given the formal similarities between the integral equations for three-nucleon system and that for three bosons, we anticipate that the conclusions of Ref.~\cite{Bedaque:2002yg} (see also Ref.~\cite{BB05}) regarding the necessity of an energy-dependent three-nucleon interaction for renormalization at N$^2$LO will be affirmed, once a strict, perturbative N$^2$LO analysis is carried out. 

We have carried out such an analysis for a system of three Helium-4 atoms, calculating trimer binding energies and atom-dimer scattering parameters ($a_{ad}$, $r_{ad}$ and $k\cot\delta$) up to N$^2$LO and comparing with results from realistic TTY potentials~\cite{Roudnev2000, Roudnev2003} and the partially-resummed EFT result of Ref.~\cite{Platter:2006ev}. The convergence of the SREFT expansion is poor for the ground-state Helium-4 trimer, since the typical momentum in that state is such that $k r_0 \sim 1$. In the three-nucleon system the triton plays a role analogous to that of this ground-state Helium-4 trimer. Thus, slow convergence for the triton in a true N$^2$LO SREFT computation (c.f. Ref.~\cite{Platter:2006}) might be expected, because of short-distance effects that scale with $k^2 r_0^2$. In contrast, the SREFT expansion converges rapidly for low-energy atom-dimer scattering, which facilitates very accurate predictions for the atom-dimer phase shifts in computations with the TTY potential. This 
establishes definitive N$^2$LO predictions for these phase shifts, superseding the results of Ref.~\cite{Platter:2006ev}. Once improved calculations for TTY-potential Helium-4-atom-dimer phase shifts become available it will be interesting to compare them to the full N$^2$LO SREFT results computed here. 

\acknowledgments{We are indebted to Lucas Platter for many useful conversations, much encouragement regarding this research, and for his valuable comments on the manuscript. We also thank Shung-ichi Ando for useful comments on the manuscript. We are grateful for the support of the US Department of Energy under contract no. DE-FG02-93ER40756. This work was also supported in part by both 
the Natural Sciences and Engineering Research Council (NSERC) and by 
the
National Research Council of Canada.}

\appendix
\addappheadtotoc

\section{Asymptotic expansion of half-on-shell $\tilde{K}_0(k,p;E)$}
\label{app:asymptotic-K0}
The half-on-shell K-matrix $\tilde{K}_0(k,p;E)$ obeys Eq.~\eqref{eq:K0} with the
on-shell-incoming momentum $k$, off-shell-outgoing momentum $p$ and the 3-body
energy $E = 3k^2/(4m) -\gamma^2/m$. $\tilde{K}_0(k,p;E)$ is cutoff independent after
renormalization. Its result is thus unchanged by use of a cutoff $\Lambda$ or a different one $\Lambda'>\Lambda$:
\begin{eqnarray}
\label{eq:K0-cut1}
\tilde{K}_0(k,p;E) &=& M(k,p;E) +\frac{2}{\pi}\int^{\Lambda'}_0 dq\
\frac{q^2}{-\gamma+\sqrt{3q^2/4-mE}} M(q,p;E) \tilde{K}_0(k,q;E)
\nn
&=& M(k,p;E)+\frac{2}{\pi}\int^{\Lambda}_0 dq\
\frac{q^2}{-\gamma+\sqrt{3q^2/4-mE}} M(q,p;E) \tilde{K}_0(k,q;E).
\end{eqnarray}
By substituting the kernel $M(q,p;E)$ of Eq.~\eqref{eq:kernel-M} into Eq.~\eqref{eq:K0-cut1}, we cancel its common part on the two sides of the equation:
\begin{eqnarray}
\label{eq:K0-cut2}
&& \frac{2}{\pi}\int^{\Lambda'}_\Lambda dq\
\frac{q^2}{-\gamma+\sqrt{3q^2/4-mE}}\
\frac{1}{qp}\ln\left(\frac{q^2+p^2+qp-mE}{q^2+p^2-qp-mE}\right)
\tilde{K}_0(k,q;E)
\nn
&=&\frac{4H_0(\Lambda)}{\pi\Lambda^2} \int^{\Lambda}_0 dq\, \frac{q^2
\tilde{K}_0(k,q;E)}{-\gamma+\sqrt{3q^2/4-mE}}\, -\
\frac{4H_0(\Lambda')}{\pi\Lambda'^2} \int^{\Lambda'}_0 dq\, \frac{q^2
\tilde{K}_0(k,q;E)}{-\gamma+\sqrt{3q^2/4-mE}}
\end{eqnarray}
As the second integral on the right side of Eq.~\eqref{eq:K0-cut2} vanishes in the
limit $\Lambda' \to \infty$, we have
\begin{eqnarray}
\label{eq:K0-cut3}
&& \frac{2}{\pi}\int^{\infty}_\Lambda dq\, \frac{q^2}{-\gamma+\sqrt{3q^2/4-mE}}\
\frac{1}{qp}\ln\left(\frac{q^2+p^2+qp-mE}{q^2+p^2-qp-mE}\right)
\tilde{K}_0(k,q;E)
\nn
&=&\frac{4H_0(\Lambda)}{\pi\Lambda^2} \int^{\Lambda}_0 dq\, \frac{q^2
\tilde{K}_0(k,q;E)}{-\gamma+\sqrt{3q^2/4-mE}}.
\end{eqnarray}

Therefore, the effect from the three-body force will cancel the cutoff-dependence in
Eq.~\eqref{eq:K0}, which indicates that
\begin{eqnarray}
\label{eq:K0-incut}
\tilde{K}_0(k,p;E) &=& \frac{1}{kp}\ln
\left(\frac{k^2+p^2+kp-mE}{k^2+p^2-kp-mE}\right)
\nn
&&\hspace{-10mm} +\frac{2}{\pi} \int^\infty_0\frac{dq}{p} \frac{q}{-\gamma+\sqrt{3q^2/4-mE}}
\ln \left(\frac{q^2+p^2+qp-mE}{q^2+p^2-qp-mE}\right) \tilde{K}_0(k,q;E)~,
\end{eqnarray}
where the ultraviolet regularization in Eq.~\eqref{eq:K0-incut} can then be taken to infinity after the renormalization. In fact, Eqs.~\eqref{eq:K0-cut3} and \eqref{eq:K0-incut} are rigorously valid only if $\tilde{K}_0$'s asymptotic behavior in the region $q>\Lambda$ exactly follows Eq.~\eqref{eq:K0-asymptotic1}. However, this condition cannot be obtained numerically using a hard-cutoff regularization. With a finite $\Lambda$, $\tilde{K}_0$'s asymptotic behavior at $p\sim \Lambda$ is distorted due to this stiff boundary. We cannot continuously extend the asymptotic form of $\tilde{K}_0$ to the region $p>\Lambda$ without considering these distortion effects.

A rigorous way to derive a cutoff-independent integral equation for $\tilde{K}_0$ without introducing the three-body force is to apply the subtractive renormalization scheme. For details of this method, see Ref.~\cite{Afnan:2003bs}.

The function $\tilde{K}_0(k,p;E)$ is bounded when $p\sim k,\gamma$; while it is in an
expansion of $k/p$ and $\gamma/p$ when $p\gg k,\gamma$. This statement is also true for each term in Eq.~\eqref{eq:K0-incut}.

 When $p\gg k,\gamma$, the kernel is expanded as
\begin{equation}
\frac{1}{kp} \ln\left(\frac{k^2+p^2+kp-mE}{k^2+p^2-kp-mE}\right) =
\frac{2}{p^2}+\frac{k^2}{6p^4}-\frac{2\gamma^2}{p^4}+\cdots,
\end{equation}
and the propagator is expanded as
\begin{equation}
\frac{q}{-\gamma+\sqrt{3q^2/4-mE}} = \frac{2}{\sqrt{3}} +\frac{4\gamma}{3q}
+\frac{8}{3\sqrt{3}}\frac{\gamma^2}{q^2} +\frac{4}{3\sqrt{3}}\frac{mE}{q^2}
+\cdots;
\end{equation}
while for arbitrary $q$, if $p \gg k,\gamma$, we have
\begin{equation}
\ln \frac{q^2+p^2+qp-mE}{q^2+p^2-qp-mE} 
=
 \ln \frac{q^2+p^2+qp}{q^2+p^2-qp}\, +
\frac{2mE}{p^2}\frac{q/p}{(q/p)^4+(q/p)^2+1}+\cdots.
\end{equation}

In the rest of this section, we simplify the notation for the
half-on-shell K-matrix to $\tilde{K}(p)\equiv \tilde{K}_0(k,p;E)$. i.e., its asymptotics at large $p$ is
expressed in Eq.~\eqref{eq:K0-asymptotic0} as
\begin{equation}
\label{eq:K0-asymptotic}
\tilde{K}_{>}(p) = p^{is_0-1} +\gamma D_1 p^{is_0-2}+\gamma^2 D_2 p^{is_0-3}
+mE\cdot C_1 p^{is_0-3},
\end{equation}
which was also derived by Bedaque {\it et al.} in~\cite{Bedaque:2002yg}. 
If we compare the inhomogeneous part in Eq.~\eqref{eq:K0-incut} with
$\tilde{K}_{>}(p)$ as regards each term in their $\gamma/p$ and $k/p$
expansion, we find that each term in the inhomogeneous part is $1/p$ order lower
than the corresponding term in $\tilde{K}_{>}(p)$. i.e. $2/p^2$ corresponds to $p^{is_0-1}$, and $k^2/p^4$ corresponds to $mE\, C_1\, p^{is_0-3}$, {\it etc.}. The inhomogeneous terms therefore
only affect the overall amplitude of the asymptotic form, not the phase. The phase is determined by the factors $\lbrace D_n\rbrace$ and $\lbrace C_n\rbrace$. Therefore, we drop the inhomogeneous part in calculating $\tilde{K}_0$'s asymptotics, and restore an overall amplitude---which is affected by the infrared physics of the inhomogeneous term---in the final calculation. After expanding each term in Eq.~\eqref{eq:K0-incut}, we have
\begin{eqnarray}
\label{eq:K0-asymptotics-a9}
\tilde{K}_{>}(p) &=& \frac{2}{\pi} \int^{\mu p}_0 \frac{dq}{p}
\frac{q}{-\gamma+\sqrt{3q^2/4-mE}}
\ln\frac{q^2+p^2+qp}{q^2+p^2-qp}\tilde{K}_{<}(q)
\nn
&&+\frac{2}{\pi} \int^{\infty}_{\mu p} \frac{dq}{p} \left[\frac{2}{\sqrt{3}}
+\frac{4\gamma}{3q} +\frac{8}{3\sqrt{3}} \frac{\gamma^2}{q^2}
+\frac{4}{3\sqrt{3}} \frac{mE}{q^2} +\cdots\right]
\ln\frac{q^2+p^2+qp}{q^2+p^2-qp}\tilde{K}_{>}(q)
\nn
&&+\frac{2}{\pi}\int^{\mu p}_0 \frac{dq}{p}\frac{q}{-\gamma+\sqrt{3q^2/4-mE}}
\frac{2mE}{p^2} \frac{q/p}{(q/p)^4+(q/p)^2+1} \tilde{K}_{<}(q)
\nn
&&+\frac{2}{\pi}\int^\infty_{\mu p} \frac{dq}{p}
\left[\frac{2}{\sqrt{3}}+\cdots\right] \frac{2mE}{p^2}
\frac{q/p}{(q/p)^4+(q/p)^2+1} \tilde{K}_{>}(q),
\end{eqnarray} 
where $\mu p$ separates the integration range into two parts: $\tilde{K}_{>}(q)$
obeys the asymptotic expansion at $q > \mu p$, and $\tilde{K}_{<}(q)$ is a
bounded function at $q < \mu p$. 

The bounded function $\tilde{K}_{<}(q)$
guarantees that integrals in the range $\int^{\mu p}_0$ are finite, and the resultant values of integrals in this region can also be expanded in powers of $\gamma/p$ and $mE/p^2$. At each order in this expansion in powers of $1/p$, the $\mu$-dependence of these parts must be canceled by $\mu$-dependence in the high-momentum integral at the corresponding order, because $\mu$ is arbitrarily chosen to separate the integral $\int^\infty_0$
in two parts. After canceling the $\mu$-dependence, the combination of terms from low- and high-momentum integrations at a given order will reproduce the corresponding term in Eq.~\eqref{eq:K0-asymptotic}.

Therefore, we can find a series of functions $f_{is_0-n}(q)$,
\begin{equation}
f_{is_0-n}(q)=
\begin{cases}
		\mbox{bounded}, & \mbox{if }q<\mu p \\
		q^{is_0-n-1}, & \mbox{if }q>\mu p
\end{cases},
\end{equation}
each one of which matches a term in Eq.~\eqref{eq:K0-asymptotic}. $f_{is_0-n}$ at $q<\mu p$ is built to absorb the explicit $\mu$-dependence in Eq.~\eqref{eq:K0-asymptotics-a9}. Therefore, we can replace $\tilde{K}_{>}(q)$ by $f_{is_0-n}(q)$ in integrals in the high-momentum region, and extend the lower limit to $0$. After performing the integrals the result yields an expression for $\tilde{K}_{>}$, which can be matched, term-by-term, to Eq.~\eqref{eq:K0-asymptotic}.

The coefficients, $D_n$ and $C_n$ can be generated from these integrals, which are  related to Mellin transforms. For example:
\begin{subequations}
\begin{equation}
D_n = \left(\frac{2}{\sqrt{3}}\right)^n \frac{I(is_0-n)}{\prod^n_{k=1}
[1-I(is_0-k)]},
\end{equation}
\begin{equation}
C_1 = \frac{ \frac{2}{3} I(is_0-2) +L(is_0)}{1-I(is_0-2)}.
\end{equation}
\end{subequations}
$I(s)$ and $L(s)$ are two types of Mellin transform, that are defined respectively as
\begin{subequations}
\begin{equation}
\label{eq:Is-appA}
I(s) \equiv \frac{4}{\sqrt{3}\pi} \int^\infty_0 dx\, \ln \frac{x^2+x+1}{x^2-x+1}\, f_s(x),
\end{equation}
\begin{equation}
\label{eq:Ls-appA}
L(s) \equiv \frac{8}{\sqrt{3}\pi} \int^\infty_0 dx \frac{x}{x^4+x^2+1} g_s(x),
\end{equation}
\end{subequations}
where $x$ denotes the momentum ratio, $x=q/p$. $f_s(x)$ in \eqref{eq:Is-appA} and \eqref{eq:Ls-appA} is a dimensionless version of the function $f_s(q)$: 
\begin{equation}
f_{s}(x)=
\begin{cases}
		\mbox{bounded}, & \mbox{if }x<\mu \\
		x^{s-1}, & \mbox{if }x>\mu
\end{cases},
\end{equation}
a definition used in the integrals calculated from the Mellin transform~\cite{Ji:2012}.

In the rest of Appendix~\ref{app:asymptotic-K0}, we will explain how we obtain $D_n$ and $C_n$ at each order in the $\gamma/p$ and $E/p^2$ expansion.

\subsection{$s_0$ in the leading-order calculation}

At leading order, we have
\begin{equation}
p^{is_0-1} = \frac{4}{\sqrt{3}\pi} \int^\infty_0 \frac{dq}{p}
\ln\frac{q^2+p^2+qp}{q^2+p^2-qp} f_{is_0}(q).
\end{equation}
By substituting $q=xp$, we obtain
\begin{equation}
\label{eq:mellin-f}
1=\frac{4}{\sqrt{3\pi}}\int^\infty_0 dx\, \ln \frac{x^2+x+1}{x^2-x+1} f_{is_0}(x).
\end{equation}
The integral on the right hand side of Eq.~\eqref{eq:mellin-f} is calculated in Ref.~\cite{Ji:2012}, which results in the analytic expression in Eq.~\eqref{eq:mellin-IL}. Therefore Eq.~\eqref{eq:mellin-f} determines the value of $s_0$ by
\begin{equation}
I(is_0)=1.
\end{equation}

\subsection{$D_n$'s}

By extracting terms proportional to $\gamma$ on both sides of
Eq.~\eqref{eq:K0-incut} (and dropping the inhomogeneous terms, as previously discussed), we arrive at
\begin{equation}
\gamma D_1 =
 \gamma\left(D_1+\frac{2}{\sqrt{3}}\right) \frac{4}{\sqrt{3}\pi}
\int^\infty_0 dx\, \ln \frac{x^2+x+1}{x^2-x+1} f_{is_0-1}(x).
\end{equation}
Therefore, we relate $D_1$ to the Mellin transform $I(is_0-1)$ by
\begin{equation}
D_1 = \frac{2}{\sqrt{3}} \frac{I(is_0-1)}{1-I(is_0-1)}.
\end{equation}

Similarly, we find
\begin{equation}
\gamma^2 D_2 = \gamma^2 \left(D_2+\frac{2}{\sqrt{3}}D_1
+\frac{4}{3}\right) I(is_0-2).
\end{equation}
Therefore, $D_2$ is expressed as
\begin{equation}
D_2 
=
\frac{4}{3}\frac{I(is_0-2)}{[1-I(is_0-1)][1-I(is_0-2)]}.
\end{equation}

This generalizes to~\cite{Ji:2012}:
\begin{equation}
\label{eq:Bn-sum}
D_n = \frac{I(is_0-n)}{1-I(is_0-n)} \left(\frac{2}{\sqrt{3}}\right)^{n}\
\sum^{n-1}_{l=0} \left(\frac{\sqrt{3}}{2}\right)^l D_l.
\end{equation}
By inserting the expression for $D_{n-1}$ back into Eq.~\eqref{eq:Bn-sum} for $D_{n}$, we find the recursion relation
\begin{equation}
\label{eq:Bn-Bn-1}
D_n =  \frac{2}{\sqrt{3}} \frac{I(is_0-n)}{I(is_0-n+1)\, [1-I(is_0-n)]}D_{n-1}.
\end{equation}
Using induction, we then derive an expression for $D_n$:
\begin{equation}
D_n = \left(\frac{2}{\sqrt{3}}\right)^n \frac{I(is_0-n)}{\prod^n_{k=1}
[1-I(is_0-k)]},
\end{equation}
where we have chosen the normalization $D_0=1$.

\subsection{$C_1$ at $\ensuremath{\mathcal{O}}(mE)$}

The coefficient $C_1$ at $\ensuremath{\mathcal{O}}(mE)$ is derived from
\begin{eqnarray}
mE\cdot C_1 p^{is_0-3} &=& \frac{2}{\pi} \int^\infty_0 \frac{dq}{p} \ln
\ln\frac{q^2+p^2+qp}{q^2+p^2-qp} \left[\frac{2}{\sqrt{3}} mE\, C_1 f_{is_0-2}(q)
+\frac{4}{3\sqrt{3}}\frac{mE}{q^2} f_{is_0}(q)\right]
\nn
&&+ \frac{2}{\pi}\int^\infty_0\frac{dq}{p} \frac{2}{\sqrt{3}}\frac{2mE}{p^2}
\frac{q/p}{(q/p)^4+(q/p)^2+1}f_{is_0}(q).
\end{eqnarray}
By substituting $x=q/p$, we find that the resulting $C_1$ is expressed as 
\begin{equation}
\label{eq:C1}
C_1 = \frac{ \frac{2}{3} I(is_0-2) +L(is_0)}{1-I(is_0-2)}.
\end{equation}
The Mellin transfrom $L(s)$ is computed in Ref.~\cite{Ji:2012}, with the final analytic expression given in Eq.~\eqref{eq:mellin-IL}.

\section{Relevant Integrals}
\label{app:integrals}

In this section, we define the integrals that are used in the derivation of analytic expressions for the N$^2$LO three-body forces $h_{20}$ and $h_{22}$.

Integrals that contain a single function are defined by
\begin{eqnarray}
\varPhi_{-1}(\Lambda) 
&\equiv& \frac{1}{\pi}\int^\Lambda dq\, \frac{\phi_0(q)}{q^2} 
= -\frac{1}{\pi\Lambda\sqrt{1+s_0^2}} \sin\left(s_0\ln\frac{\Lambda}{\bar{\Lambda}}+\arctan s_0\right),
\end{eqnarray}

\begin{eqnarray}
\varPsi_{-1}(\Lambda) 
&\equiv& \frac{1}{\pi}\int^\Lambda dq\, \frac{\psi_1(q)}{q^2} 
= -\frac{|C_1|}{\pi\Lambda\sqrt{1+s_0^2}} \sin\left(s_0\ln\frac{\Lambda}{\bar{\Lambda}}+\arg C_1+\arctan s_0\right),
\end{eqnarray}

\begin{equation}
\varPhi_{n} (\Lambda) 
\equiv \frac{1}{\pi}\int^\Lambda dq\, q^{n-1} \phi_0(q) 
= \frac{1}{\pi}\int^\Lambda dq\,  q^{n-1} \sin\left(s_0\ln\frac{q}{\bar{\Lambda}}\right)
= \frac{\Lambda^n}{\pi\sqrt{n^2+s_0^2}}\sin\left[s_0\ln
\frac{\Lambda}{\bar{\Lambda}}-\arctan\left(\frac{s_0}{n}\right)\right],
\end{equation}

\begin{equation}
\bar{\varPhi}_{n} (\Lambda) 
\equiv \frac{1}{\pi}\int^\Lambda dq\, q^{n-1} \bar{\phi}_0 (q)
= \frac{1}{\pi}\int^\Lambda dq\, q^{n-1} \cos\left(s_0\ln\frac{q}{\bar{\Lambda}}\right)
= \frac{\Lambda^n}{\pi\sqrt{n^2+s_0^2}}\cos\left[s_0\ln
\frac{\Lambda}{\bar{\Lambda}}-\arctan\left(\frac{s_0}{n}\right)\right],
\end{equation}
where the subscript denotes the power of $\Lambda$ present in the integrals for $\varPhi_n$ and $\bar{\varPhi}_n$, $n=1,2$.

Meanwhile, integrals that involve a product of functions are:
\begin{equation}
\mathcal{U}_{\phi \phi, 0}(\Lambda)  
\equiv \frac{1}{\pi} \int^\Lambda \frac{dq}{q}\, \phi_0^2(q)
= \frac{1}{2\pi} \left[\ln \Lambda-\frac{1}{2 s_0} \sin\left(2s_0 \ln \frac{\Lambda}{\bar{\Lambda}}\right)\right],
\end{equation}

\begin{equation}
\mathcal{U}_{\phi \psi, 0}(\Lambda)  
\equiv \frac{1}{\pi} \int^\Lambda \frac{dq}{q}\, \phi_0(q)\psi_1(q)
=  \frac{|C_1|}{2\pi} \left[\cos\left(\arg C_1\right) \ln \Lambda-\frac{1}{2 s_0} \sin\left(2s_0 \ln \frac{\Lambda}{\bar{\Lambda}}+\arg C_1\right)\right],
\end{equation}

\begin{equation}
\mathcal{U}_{\phi \phi, 1}(\Lambda)  
\equiv \frac{1}{\pi} \int^\Lambda dq\, \phi_0^2(q)
= \frac{\Lambda}{2\pi} \left[1-\frac{1}{\sqrt{1+4s_0^2}} \cos\left(2s_0 \ln \frac{\Lambda}{\bar{\Lambda}} -\arctan(2s_0)\right)\right],
\end{equation}

\begin{equation}
\mathcal{U}_{\phi \bar{\phi}, 1} (\Lambda) 
\equiv \frac{1}{\pi} \int^\Lambda dq\, \phi_0(q) \bar{\phi}_0(q)
= \frac{\Lambda}{2\pi\sqrt{1+4s_0^2}} \sin\left(2s_0 \ln \frac{\Lambda}{\bar{\Lambda}} -\arctan(2s_0)\right),
\end{equation}

\begin{equation}
\mathcal{U}_{\phi \phi, 2} (\Lambda) 
\equiv \frac{1}{\pi}\int^\Lambda dq\, q \phi_0^2(q)=
 \frac{\Lambda^2}{4\pi} \left[ 1 - \frac{1}{\sqrt{1+s_0^2}} \cos\left(2s_0 \ln\frac{\Lambda}{\bar{\Lambda}} -\arctan s_0 \right)\right],
\end{equation}

\begin{equation}
\mathcal{U}_{\phi \bar{\phi}, 2} (\Lambda) 
\equiv \frac{1}{\pi}\int^\Lambda dq\, q\phi_0(q) \bar{\phi}_0(q)
= \frac{\Lambda}{4\pi\sqrt{1+s_0^2}} \sin\left(2s_0 \ln \frac{\Lambda}{\bar{\Lambda}} -\arctan s_0\right).
\end{equation}

Lastly, four double integrals used in Sec.~\ref{sec:renormalization} are
\begin{eqnarray} 
\mathcal{W}_{\phi(\bar{\phi}),1+1} (\Lambda) 
\equiv \frac{1}{\pi} \int^\Lambda dq\, \phi_0(q)\, \bar{\varPhi}_{1}(q)=\frac{\Lambda^2}{4\pi^2(1+s_0^2)} \left[s_0+ \sin\left( 2s_0 \ln\frac{\Lambda}{\bar{\Lambda}} -2\arctan s_0 \right) \right],\nn
\end{eqnarray}

\begin{eqnarray}
&& \mathcal{W}_{\phi\phi(\bar{\phi}),1+1}(\Lambda) 
\equiv \frac{1}{\pi} \int^\Lambda dq\, \phi_0^2(q)\, \bar{\varPhi}_{1}(q)\nn
&& \qquad = -\frac{\Lambda^2}{4\pi^2 \sqrt{1+s_0^2}\sqrt{4+s_0^2}}
\left[
\frac{\sqrt{4+s_0^2}}{\sqrt{4+9s_0^2}} \cos\left( 3s_0\ln\frac{\Lambda}{\bar{\Lambda}} -\arctan s_0 -\arctan \frac{3s_0}{2}\right)
\right.
\nn
&& \qquad \left.
-2\cos\left(s_0\ln\frac{\Lambda}{\bar{\Lambda}}-\arctan s_0 -\arctan\frac{s_0}{2}\right)
+\cos\left(s_0\ln\frac{\Lambda}{\bar{\Lambda}}+\arctan s_0 -\arctan\frac{s_0}{2}\right)
\right],
\nn
\end{eqnarray}

\begin{eqnarray}
 \mathcal{W}_{\phi(\phi\bar{\phi}),1+1}(\Lambda) 
&\equiv& \frac{1}{\pi} \int^\Lambda dq\, \phi_0(q)\, \mathcal{U}_{\phi \bar{\phi}, 1}(q)\nn
&=& \frac{\Lambda^2}{4\pi^2 \sqrt{1+4s_0^2}\sqrt{4+s_0^2}}
\left[
\cos\left(s_0\ln\frac{\Lambda}{\bar{\Lambda}}-\arctan (2s_0) -\arctan\frac{s_0}{2}\right)
\right.
\nn
&&\qquad \left.
-\frac{\sqrt{4+s_0^2}}{\sqrt{4+9s_0^2}} \cos\left( 3s_0\ln\frac{\Lambda}{\bar{\Lambda}} -\arctan (2s_0) -\arctan \frac{3s_0}{2}\right)
\right],
\end{eqnarray}

\begin{eqnarray}
\mathcal{W}_{\phi\phi(\phi\bar{\phi}),1+1} (\Lambda) 
&\equiv& \frac{1}{\pi} \int^\Lambda dq\, \phi_0^2(q)\, \mathcal{U}_{\phi \bar{\phi}, 1}(q)\nn
&=& \frac{\Lambda^2}{16\pi^2(1+4s_0^2)} \left[
2s_0 - \sin\left(4s_0\ln\frac{\Lambda}{\bar{\Lambda}} -2\arctan (2s_0)\right)
\right.
\nn
&&\left. + 
\frac{2\sqrt{1+4s_0^2}}{\sqrt{1+s_0^2}}
\sin\left(2s_0\ln\frac{\Lambda}{\bar{\Lambda}} -\arctan (2s_0) -\arctan s_0\right)
\right].
\end{eqnarray}

\end{document}